# Breaking the curse of dimensionality in regression

Yinchu Zhu and Jelena Bradic
University of Oregon and University of California, San Diego

**Abstract**

Models with many signals, high-dimensional models, often impose structures on the signal strengths. The common assumption is that only a few signals are strong and most of the signals are zero or close (collectively) to zero. However, such a requirement might not be valid in many real-life applications. In this article, we are interested in conducting large-scale inference in models that might have signals of mixed strengths. The key challenge is that the signals that are not under testing might be collectively non-negligible (although individually small) and cannot be accurately learned. This article develops a new class of tests that arise from a moment matching formulation. A virtue of these moment-matching statistics is their ability to borrow strength across features, adapt to the sparsity size and exert adjustment for testing growing number of hypothesis. GRoup-level Inference of Parameter, *GRIP*, test harvests effective sparsity structures with hypothesis formulation for an efficient multiple testing procedure. Simulated data showcase that GRIPs error control is far better than the alternative methods. We develop a minimax theory, demonstrating optimality of GRIP for a broad range of models, including those where the model is a mixture of a sparse and high-dimensional dense signals.

## 1 Introduction

The emergence of high-dimensional data, such as the gene expression values in microarray and the single nucleotide polymorphism data, brings challenges to many traditional statistical methods and theory. One important aspect of the high-dimensional data under the regression setting is that the number of covariates greatly exceeds the sample size. For example, in microarray data, the number of genes ($p$) is in the order of thousands whereas the sample size ($n$) is much less, usually less than 50. This is the so called "large-$p$, small-$n$" paradigm, which translates to a regime of asymptotics where $p \to \infty$ much faster than $n$. Inference in regression setting for large $p$, small $n$ settings, have been recently developed. Sparsity assumption on the model signals has had a significant role in achieving optimal inference – Cai and Guo (2015); Javanmard and Montanari (2015); Cai and Guo (2016) found minimax results quantifying the direct effect of the size of the sparsity.

In this article, we develop a test statistic that is able to quantify the simultaneous effect of a growing number of signals in a general high-dimensional linear model framework, allowing for a broad-ranging parameter structure. To be specific, let $\boldsymbol{\delta}^* = (\boldsymbol{\beta}^*, \boldsymbol{\gamma}^*) \in \mathbb{R}^p$ denote the model parameter containing $p$ signals. Given a sample of size $n$, the objective is to conduct inference on $d$ out of $p$ signals, i.e., testing hypotheses on a $d$-dimensional component $\boldsymbol{\beta}^*$.

$$H_0 : \boldsymbol{\beta}^* = \boldsymbol{\beta}_0 \qquad \text{versus} \qquad H_1 : \boldsymbol{\beta}^* \neq \boldsymbol{\beta}_0 \tag{1}$$

where $d$, $p$ and $p - d$ can be much larger than $n$. Multivariate testing (1) is a very relevant problem in practice and yet, it has only been studied on a case-by-case basis; all of the existing methods are asymptotically exact only under the strict sparsity assumptions requiring $\|\boldsymbol{\delta}^*\|_0 \ll \sqrt{n}$.

However, exactly (or even approximately) sparse models are hardly appropriate for many modern scientific studies, see Hall et al. (2014); Ward (2009); Jin and Ke (2014); Pritchard (2001). One such example would include testing of brain-connectivity patterns as there is new evidence highlighting the lack of sparsity in functional networks. Gaussian graphical models have been found to be good at recovering the main brain networks from fMRI data. Nevertheless, recent work in neuroscience has shown that the structural wiring of the brain doesn't correspond to a sparse network (Markov et al., 2013), thus questioning the underlying assumption of sparsity often used to estimate brain network connectivity. In such setting, we are interested in determining which edges (e.g. autistic) are present and we want to provide confidence-intervals on our results.

The importance of allowing deviations from strictly sparse signals is two-fold. First, the null hypothesis can directly rule out the sparse assumption of the model signals. In practice, the hypothesized value $\boldsymbol{\beta}_0$ can be with non-zero entries, possibly larger than $n$ implying that the model is not strictly sparse under the null hypothesis. To fix ideas, consider the problem of testing the specification of parameterizing $\boldsymbol{\beta}^* = g(b^*)$, where $g(\cdot)$ is a known parametric function and $b^*$ is a scalar, say $\boldsymbol{\beta}^* = (b^*, ..., b^*)^\top$. A natural approach is to construct a confidence set for $b^*$ by inverting a test for $\boldsymbol{\beta}^* = g(b)$ for each $b$ and see whether the confidence set is empty. Hence, it reduces to the problem of (1) with $\boldsymbol{\beta}_0 = g(b_0)$ for a given $b_0 \in \mathbb{R}$. However, $\boldsymbol{\beta}_0$ is not sparse whenever $b_0 \neq 0$, and the null model, even with sparse $\boldsymbol{\gamma}^*$ is not even approximately sparse; in fact it belongs to a class of hybrid models where the signal is a composition of sparse and dense structures (see Chernozhukov et al. (2015) for example).

Second, the nuisance parameter, $\boldsymbol{\gamma}^*$, might not be sparse. This is motivated by the latest need in biology to identify significant and large sets of genes (the sets of genes representing biological pathways in the cell, or sets of genes whose DNA sequences are close together on the cell's chromosomes), which are associated with certain clinical outcome, rather than identifying only a restrictive set of individual genes (e.g. Subramanian et al. (2005); Trapnell et al. (2013); Sumazin et al. (2016)). As the dimension of a gene-set ranges from a few to thousands, and the gene sets can overlap as they share common genes, there are both high dimensionality and multiplicity in gene-set testing. Testing such hypotheses is a necessity in determining the effects of covariates on certain disease related outcome. Hence, our work represents a step towards fully discriminating the relation between covariates and changes in the response variable in models where potentially all $p$ covariates are associated with the outcome of the disease.

## 1.1 Related Work

In recent years, the number of papers considering high-dimensional inference has grown rapidly. In a large number of existing work the sparsity level of the regression model, $s = \|\boldsymbol{\delta}^*\|_0$, is a constant or grows slowly with $n$. To name a few Van de Geer et al. (2014), Zhang and Zhang (2014), Javanmard and Montanari (2014a) (JM), Ning and Liu (2014) (NL), Belloni et al. (2014) allow the sparsity to grow at the rate $o(\sqrt{n}/\log p)$ and design tests for testing univariate parameters in high-dimensional sparse models. For problems of univariate testing in linear models Janson et al. (2015); Zhu and Bradic (2016a,b) or two-sample setting Zhu and Bradic (2016c) make progress in removing sparsity assumption. However, a more challenging setting of group and/or multivariate testing in such generality has not been successfully developed. Zhang and Cheng (2016) (ZC) and Dezeure et al. (2016) define the simultaneous tests by designing appropriate bootstrap procedures. However, they assume sparsity to be of the order of $o(\sqrt{n}/(\log p)^3)$ and $o(\sqrt{n}/(\log p)^{3/2})$, respectively. Different from them we allow $s$ to grow faster than the sample size, thus introducing a useful and more broad alternative to the existing work. Our work is closely related to Chernozhukov et al. (2015); Belloni et al. (2014, 2015a,b) who utilize Neyman's orthogonal score method to design efficient moment equations. However, the underlying assumption therein is that $s = o(\sqrt{n}/(\log p)^{3/2})$. We decompose feature correlation, embed the hypothesis of interest into the moment condition and provide a test that does not rely on consistent parameter estimation. With this we are able to construct a double-robust test that is robust to both model sparsity and the presence of high-correlation among the features, albeit only one robust departure is allowed at any given time. This paper also fits into the literature of methods rigorously controlling false discovery rate in high-dimensional setting (e.g. Storey et al. (2004); Barber and Candès (2015); G'Sell et al. (2016)), and those papers testing many moment inequalities (e.g. Chernozhukov et al. (2013b); Fan et al. (2015); Bugni et al. (2016); Romano et al. (2014), among others). We establish a new minimax optimality result of independent interest and show that in certain hybrid high-dimensional models (Chernozhukov et al., 2015) our test is optimal for the control of the false discovery rate while allowing the number of tests and the model parameters to be much larger than the sample size. Due to the latency of the moment estimators and their growing number, the mathematical details required to obtain these results are involved and new to the literature.

## 1.2 Organization

The remainder of the paper is organized as follows. In Section 2, we introduce the model setup, the assumptions and the definitions of the moment equations of interest. We then proceed and describe an estimation and bootstrap scheme used for testing. In Section 4, we present the theoretical guarantees of the proposed scheme. We show asymptotically exact Type I error control and showcase minimax optimality of our test. Second 3 contains extensions from a linear model setting to the very broad nonparametric setting with possibly dependent errors. We demonstrate the finite sample performance of our test by Monte Carlo simulations in Section 5, where we also compare existing state of the art inferential methods. In Section 6, we propose a general bootstrap methodology, which is of independent interest. All proofs are deferred to the Supplement.

## 2 Large Scale Learning

Although the proposed model applied more broadly, we begin with a Gaussian setting where in the high-dimensional linear model, where the pairs of observations $\{(y_i, w_i)\}_{i=1}^n$, $y_i \in \mathbb{R}$, $w_i \in \mathbb{R}^p$ follow the model of the form

$$y_i = w_i^\top \boldsymbol{\delta}^* + \varepsilon_i, \qquad i = 1, \ldots, n, \tag{2}$$

the model errors are such that $\varepsilon_1, \ldots, \varepsilon_n$ are independent and identically distributed (i.i.d.) Gaussian random variables with mean zero and variance $0 < \sigma_\varepsilon^2 < \infty$. In the above $\boldsymbol{\delta}^* \in \mathbb{R}^p$ is a high-dimensional parameter of interest and $p \gg n$. We consider a random design case, similar to that of Janson et al. (2015), where $\boldsymbol{\Sigma}_W^{-1/2} w_i$ are independent standard Gaussian random variables with $\boldsymbol{\Sigma}_W = \mathbb{E}[w_i^\top w_i]$ satisfying $C_{\min} \leq \sigma_{\min}(\boldsymbol{\Sigma}_W) \leq \sigma_{\max}(\boldsymbol{\Sigma}_W) \leq C_{\max}$. Moreover, $\varepsilon_i$ are uncorrelated of $w_i$. Further extensions, with additional technical details, can be easily established; both design and error distributions can be allowed to have exponential-type tails.

A natural measure of the simultaneous effects of a number of covariates onto the response can be measured by a hypothesis (1) where $\boldsymbol{\beta}_0 = (\beta_{0,(1)}, \cdots, \beta_{0,(d)})^\top \in \mathbb{R}^d$ is potentially high-dimensional in the sense that $\log d = \mathcal{O}(n^a)$ for some $a > 0$. Here $d$ may be equal to $p$, but is not required; with $d \to \infty$ we denote large-scale inference as multiple testing problem of interest. Namely, we may be interested in only an effect of a specific set of genes that does not include all of the genes collected for observation. Thus, the unknown parameter of interests $\boldsymbol{\delta}$ can be decomposed as $\boldsymbol{\delta} = (\boldsymbol{\gamma}, \boldsymbol{\beta})$ with $\boldsymbol{\gamma} \in \mathbb{R}^{p-d}$ being a nuisance parameter and $\boldsymbol{\beta} \in \mathbb{R}^d$ the parameter of interest. Similarly, following this decomposition, we decompose the covariates $w_i = (x_i, z_i) \in \mathbb{R}^p$ where $x_1, \ldots, x_n \in \mathbb{R}^{p-d}$ and $z_1, \ldots, z_n \in \mathbb{R}^d$.

**Remark 1.** *We note that related papers (e.g. Bühlmann and van de Geer (2015); Zhang and Zhang (2014); Belloni et al. (2014); Van de Geer et al. (2014)) define the same problem of interest but assume that the nuisance parameter is strictly sparse, i.e. $\|\boldsymbol{\delta}^*\|_0 \ll n$, $\|\boldsymbol{\beta}^*\|_0 \ll n$, $\|\boldsymbol{\gamma}^*\|_0 \ll n$. These assumptions bear nice theoretical property: parameters of interest are well defined and estimable (see for example the minimax rates of estimation Cai and Guo (2016); Cai et al. (2016); Yuan and Zhou (2016)). Yet we argue that models defined in this way are less useful in real scientific applications of main interest in the modern scientific studies, where sparsity cannot be verified.*

## 2.1 Challenges

We showcase in this section the limitations of the existing methods. We consider a simple example $\mathbf{Y} = \mathbf{W}\boldsymbol{\delta}^* + \varepsilon$, where $\boldsymbol{\delta}^* = (\boldsymbol{\beta}^*, \boldsymbol{\gamma}^*) \in \mathbb{R}^p$, where $\boldsymbol{\beta}^* \in \mathbb{R}^d$ and $\boldsymbol{\gamma}^* \in \mathbb{R}^{p-d}$. We set $d = p/2$, $\boldsymbol{\beta}^* = 0$ and $\boldsymbol{\gamma}^* = (n^{-1/2}, ..., n^{-1/2}, 0, ..., 0)$ with $s = \|\boldsymbol{\gamma}^*\|_0$. For simplicity, all the entries of $\mathbf{W}$ and $\varepsilon$ to be independent and identically distributed (i.i.d.) with standard normal distribution $N(0,1)$. The goal is to test $H_0 : \boldsymbol{\beta}^* = 0$. Notice that this is a true hypothesis.

We consider the test proposed in Zhang and Cheng (2016) based on the de-sparsified Lasso. It is not hard to show that if $\log p = o(n)$, then with probability approaching one, we have $\widehat{\boldsymbol{\delta}} = 0$ and $\widehat{\Theta} = \text{diag}(\sigma_{W,1}^{-2}, ..., \sigma_{W,p}^{-2})$, where $\widehat{\boldsymbol{\delta}}$ is the Lasso estimator for $\boldsymbol{\delta}^*$, $\widehat{\Theta}$ is the nodewise Lasso estimator for $(E\mathbf{W}^\top \mathbf{W}/n)^{-1}$ and $\sigma_{W,j}^2 = \mathbf{W}_{(j)}^\top \mathbf{W}_{(j)}/n$. Let $\mathbf{W}_{(j)}$ denoting the $j$-th column of the design matrix, i.e. $\mathbf{W}_{(j)} = (w_{1,j}, \cdots, w_{n,j})^\top$. Hence, with probability approaching one, the test statistic based on de-sparsified Lasso estimator for $\boldsymbol{\beta}^*$ is

$$\|\widetilde{\boldsymbol{\beta}}\|_\infty = \|n^{-1}\widehat{\Theta}_Z \mathbf{Z}^\top \mathbf{Y}\|_\infty,$$

where $\mathbf{Z} \in \mathbb{R}^{n \times d}$ is the first $d$ columns of $\mathbf{W}$ and $\widehat{\Theta}_Z = \text{diag}(\sigma_{W,1}^{-2}, ..., \sigma_{W,d}^{-2})$. Since the variance of $\varepsilon$ is known to be $I_n$, the bootstrap critical value is, with probability approaching one, equal to the $1 - \alpha$ quantile (conditional on the data) of

$$\|n^{-1}\widehat{\Theta}_Z \mathbf{Z}^\top \xi\|_\infty,$$

where $\xi \in \mathbb{R}^n$ is drawn from $N(0, I_n)$ independent of the data.

The above analysis allows us to simulate the large sample behavior of the test. The simulations are implemented with $n = 300$ and $p = 700$. Based on 10000 simulations, we plot the probability of rejection when the null hypothesis holds true. We consider significance level of 1%, 5% and 10%. The results are presented in Figure 1.

Figure 1: Rejection probability of the true hypothesis as a function of sparsity. We plot three significance levels, 1% (red), 5% (green) and 10% (blue). Dashed lines represent the nominal level considered. The x-axis represents sparsity of the parameter $\boldsymbol{\delta}^*$ and the y-axis represents rejection probability.

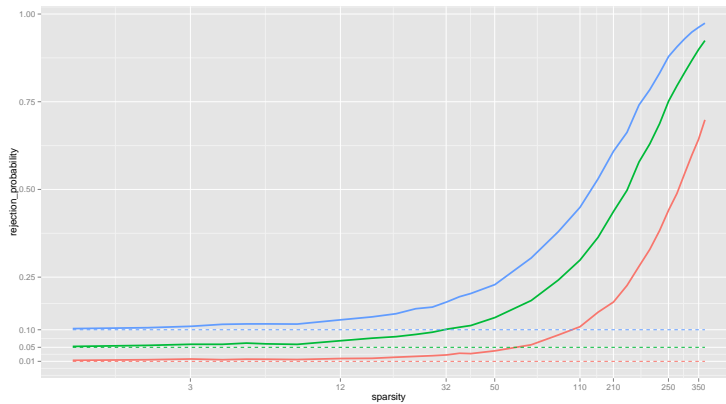

Notice that the probability of rejecting a true hypothesis can be larger than 0.9 if $s$ is large. This is due to the mixed signal strength. If the signal strength is sparse, i.e., $s \ll \sqrt{n}/\log p$, then the theory of ZC suggests that the size will be correct. However, when $s = n$, we have many weak signals that are collectively non-negligible, i.e., $\|\boldsymbol{\delta}^*\|_\infty = o(\sqrt{n^{-1} \log p})$ and $\|\boldsymbol{\delta}^*\|_2 = 1$. In this case, the existing methods do not suffice.

## 2.2 Global null

We begin by discussing a special case of testing the global null defined as

$$H_0 : \boldsymbol{\delta}^* = \boldsymbol{\delta}_0 \qquad \text{versus} \qquad H_1 : \|\boldsymbol{\Sigma}_W(\boldsymbol{\delta}^* - \boldsymbol{\delta}_0)\|_\infty > \bar{\tau}\sqrt{n^{-1}\log(p \vee n)}, \qquad (3)$$

for some $\bar{\tau} > 0$. In this setting $\boldsymbol{\delta}_0$ can be a vector of all zeros, in which case the problem is to test joint significance of all the parameters in the model. We traditionally use the F-test to test such null, however, it is well known that the F-test has low power and breaks down when the number of covariates in the model is close to the number of samples or exceeds the number of samples. In this section we explore testing of a null hypothesis against a high dimensional alternative. However, we would also like to allow a specification of $\boldsymbol{\delta}_0$ that allows it to be a $p$ dimensional vector of all ones or a vector of mixed values many of which are non-zero - problem typically much harder to solve.

For this end, we propose to consider a test statistic that measures maximum re-weighted correlation between the residuals under the null and the design matrix. Namely, we consider the test statistic $\|\mathbf{T}_n\|_\infty$, where $\mathbf{T}_n = (T_{n,1}, \cdots, T_{n,p})^\top$ and

$$T_{n,j} = n^{-1/2} \sum_{i=1}^n T_{i,j} := n^{-1/2} \sum_{i=1}^n \frac{w_{i,j}(y_i - w_i^\top \boldsymbol{\delta}_0)}{\|\mathbf{Y} - \mathbf{W}\boldsymbol{\delta}_0\|_2 \|\mathbf{W}_{(j)}\|_2/n}.$$

For the simple case of $\boldsymbol{\delta}_0 = \mathbf{0}$, typically employed in gene-wise studies for example, the test statistic takes the form

$$T_0 = \max_{1 \leq j \leq p} \frac{\sqrt{n}[\mathbf{W}_{(j)}^\top \mathbf{Y}]}{\|\mathbf{Y}\|_2 \|\mathbf{W}_{(j)}\|_2}.$$

If the columns of the design matrix $\mathbf{W}$ are normalized to have $l_2$ norm equal to 1, then the test takes the simpler form

$$T_0 = \max_{1 \leq j \leq p} \frac{\sqrt{n}[\mathbf{W}_{(j)}^\top \mathbf{Y}]}{\|\mathbf{Y}\|_2}.$$

We are interested in studying minimax optimality of the tests above where the alternative hypothesis are not restricted to only sparse vectors, but only the vectors whose maximal non-zero elements are strong enough - in particular, the vector can have $p$ non-zero elements. We refer to this as a dense alternative.

The case of sparse alternatives have been considered in Goeman et al. (2006) where the authors show optimality over an average of high-dimensional alternatives of an empirical Bayes test of the form $\mathbf{Y}^\top \mathbf{W} \mathbf{W}^\top \mathbf{Y}/\|\mathbf{Y}\|_2^2$. They also show that for the case of alternatives of the form $\|\boldsymbol{\delta}\|_2 \geq c$ their test is equivalent to the low-dimensional F-test. Lastly, Zhong and Chen (2011) discuss the same class of spherical alternatives and develop a test that is therein optimal even for $p \geq n$. However, class of dense alternatives poses significant theoretical challenges due to its exploding size.

Under the null hypothesis the test statistics is centered, i.e. $E[T_{n,j}] = 0$. It is usually not easy to compute the distribution of the maximal test statistic – the correlation between the elements together with the growing number of maxima make the computation challenging. In many cases however, we can find a reasonably good approximation to the distribution of $\|\mathbf{T}_n\|_\infty$ by considering a general multiplier bootstrap framework through a bootstrapped test statistics $\max_{1 \leq j \leq p} |\widetilde{T}_{n,j}|$. For that purpose, let $\xi_i$ be i.i.d. standard normal random variables that are independent of the observations $\{(x_i, z_i, y_i)\}_{i=1}^n$ and define

$$\widetilde{T}_{n,j} = n^{-1/2} \sum_{i=1}^n \xi_i \left( T_{i,j} - T_{n,j}/\sqrt{n} \right). \tag{4}$$

Additionally, we note that the method does not crucially depend on the normality of $\xi_i$'s. More elaborate distributions would suffice; for example Rademacher or Mammen distributions (see for example Mammen (1993)) would allow for additional robustness to the outliers in the model error. Let $\alpha$ be the predetermined nominal size of the test. Then, conditional on the original observations, compute

$$\mathcal{Q}(1 - \alpha, \|\widetilde{\mathbf{T}}_n\|_\infty) = \inf \left\{ x \in \mathbb{R} \;\Big|\; \mathbb{P}(\|\widetilde{\mathbf{T}}_n\|_\infty > x \mid \mathbf{X}, \mathbf{Z}, \mathbf{Y}) \leq 1 - \alpha \right\},$$

with $\widetilde{\mathbf{T}}_n = (\widetilde{T}_{n,1}, \cdots, \widetilde{T}_{n,d})^\top \in \mathbb{R}^d$ and $\mathbf{Y} = (y_1, \cdots, y_n)^\top \in \mathbb{R}^n$. Then, we reject $H_0$ in (1) if and only if

$$\|\mathbf{T}_n\|_\infty > \mathcal{Q}(1 - \alpha, \|\widetilde{\mathbf{T}}_n\|_\infty). \tag{5}$$

**Theorem 1.** *Consider the Gaussian model in which entries in $\Sigma_W^{-1/2} w_i$ and $\sigma_\varepsilon^{-1} \varepsilon_i$ are standard Gaussian random variables such that $\sigma_\varepsilon$ and all the eigenvalues of $\Sigma_W = \mathbb{E}[w_i^\top w_i]$ lie in the interval $[C_{\min}, C_{\max}]$ for some constants $C_{\min}, C_{\max} > 0$. Then the test (5) is an asymptotically exact test with $n, p \to \infty$ and $\log p = o(n^{1/8})$. If $\|\boldsymbol{\delta}^* - \boldsymbol{\delta}_0\|_2 = O(1)$ in (3), then this test is also minimax optimal for testing the problem (3).*

**Remark 2.** *Observe that in the Theorem 1 no sparsity assumptions are directly assumed, either in the model itself or the precision matrix of the design, and yet an ultra-high dimensional case with $p \gg n$ is allowed. Moreover, test is minimax optimal both against high-dimensional sparse alternatives where only a small number of beta are non-zero as well as more challenging setting of high-dimensional dense alternatives where all elements can be non-zero. In particular, the test is minimax optimal whenever p is extremely large and the non-zero coefficients (possibly p of them) of $\boldsymbol{\delta}$ are non-negligible. Achieving both theoretical guarantees seems unique in the existing literature on high-dimensional testing.*

### 2.3 From parametric null to many moments hypothesis

The presence of nuisance parameters in the model (2) complicates some of the issues that were described above. When nuisance parameters are present, the null hypothesis is not simple any more but composite and the problem of interest becomes more demanding to solve. We propose to explore Neyman orthogonalization and avoid penalization and estimation of the parameters under testing altogether. We transform the multivariate parametric hypothesis of interest into a sequence of growing number of moment equations, albeit univariate moment conditions. We also explicitly account for the dependence among features in the spirit of Neyman score orthogonalization. In particular, we consider the following $d$ simultaneous moment conditions

$$H_0 : \mathbb{E}\left[\psi(y_i - z_i^\top \boldsymbol{\beta}_0 - x_i^\top \boldsymbol{\gamma}^*)\phi_j(z_i, x_i)\right] = 0, \quad \forall j \in \{1, \cdots, m_d\}, \tag{6}$$

where $\phi_j$ and $\psi : \mathbb{R} \to \mathbb{R}$ are sequences of properly defined functions. The alternative hypothesis are defined as

$$H_1 : \mathbb{E}\left[\psi(y_i - z_i^\top \boldsymbol{\beta}_0 - x_i^\top \boldsymbol{\gamma}^*)\phi_j(z_i, x_i)\right] \neq 0, \quad \text{for some } j \in \{1, \cdots, m_d\}.$$

The power of the test depends heavily on the choice of the weights $\phi_j(z_i, x_i)$ and the function $\psi$. In particular, the test can be powerful against all possible alternatives if we consider a growing number of moment conditions; hence, $m_d = m(d) \to \infty$. Also, the more weighting functions we use the more likely will one of the moment conditions be violated under $H_1$. Thus, the weights $\phi_j(z_i, x_i)$ are carefully constructed random variables that are able to separate the null and the alternative hypothesis well enough. We also advocate for a choice that leads to the uncorrelated functions of the design vectors $x_i$. One particular choice considers a projection of $x_i$ onto $z_{i,(j)}$ defined as

$$\boldsymbol{\theta}^*_{(j)} = \boldsymbol{\Sigma}_X^{-1} \mathbb{E}[\mathbf{X}^\top \mathbf{Z}_{(j)}] \tag{7}$$

where $\boldsymbol{\Sigma}_X = \mathbb{E}[\mathbf{X}^\top \mathbf{X}] \in \mathbb{R}^{(p-d) \times (p-d)}$ is a variance covariance matrix of $x_i$'s, $\mathbf{X} = (x_1, ..., x_n)^\top \in \mathbb{R}^{n \times (p-d)}$ and $\mathbf{Z}_{(j)} = (z_{1,(j)}, ..., z_{n,(j)}) \in \mathbb{R}^n$. The residual of that projection, $\mathbf{u}_{(j)} = (u_{1,(j)}, ..., u_{n,(j)}) \in \mathbb{R}^n$ is defined as $u_{i,(j)} = \phi_j(z_i, x_i)$ with

$$\phi_j(z_i, x_i) := z_{i,(j)} - x_i^\top \boldsymbol{\theta}^*_{(j)}.$$

In the above display $\mathbf{u}_1, \ldots, \mathbf{u}_n$ are i.i.d. random vectors with $\mathbf{u}_i = (u_{i,(1)}, \cdots, u_{i,(d)})^\top \in \mathbb{R}^d$, and each Gaussian component $u_{i,(j)}$ has mean zero and variance $0 < \sigma_{u,(j)}^2 = \mathbb{E} u_{i,(j)}^2 < \infty$ and $m_d = d$. Different constructions of $\phi_j(z_i, x_i)$ are possible; see details in Section 3.

We will show that this choice of $\mathbf{u}_i$ leads to a great reduction in the number of assumptions needed for asymptotically valid inference. In particular, the introduction of $\boldsymbol{\theta}_{(j)}^*$ allows us to replace the model assumptions (such is strict sparsity in parameters) with the design assumptions (such is sparsity in the feature correlations); observe that the former cannot be checked whereas the later can be checked with an i.i.d. observations of the design. This is particular useful for practical scientific applications.

We proceed to provide an estimate of the moment defined in (6) – note that $\boldsymbol{\gamma}^*$ and $\boldsymbol{\theta}_{(j)}^*$ are unobservable and need to be estimated; however $\boldsymbol{\beta}^*$ does not need to. Whenever the nuisance parameters are of finite dimensions, a ridge regularized estimator can be employed; theory in this article can be extended easily for these cases. However, when they are of extremely high dimensions the problem becomes extremely difficult. Simple off-the-shelf method, like Lasso regularization or simple Dantzig selector cannot be effectively used for the purposes of inference; finite sample bias propagation and lack of asymptotic distribution prevent their use. Recent one-step estimates that rely on penalization of the moment condition itself and a bias correction mitigate the observed challenges, but remain highly computationally expensive. We develop novel theoretical arguments that allow a derivation of a limiting distribution of the test statistics without explicitly correcting for the bias. For that end, we use estimators that can provide effective control in the theoretical analysis (jumps of the martingales for example) and simultaneously control the size of the moments (6).

Although the moment conditions have recently been used for inference in high dimensions; see Chernozhukov et al. (2015); Belloni et al. (2017, 2014, 2015a,b), these methods rely on consistent estimation of the unknown parameters. However, when sparsity is in question, accurate estimation of the unknown high-dimensional model parameter is typically not guaranteed. As a major contribution of this paper, we construct tests using estimators that may or may not be consistent and show theoretically that valid inference can be achieved regardless of the quality of the estimator for the model parameter.

The proposed GRoup-level Inference of Parameters, *GRIP* for short, is summarized as follows.

Step 1 Estimate the unknown parameters $\boldsymbol{\gamma}$ and $\boldsymbol{\theta}_{(j)}$ for $j = 1, \cdots, d$. Estimation of the unknown parameter $\boldsymbol{\gamma}$ is done by imposing the null hypothesis directly into the optimization problem. We consider $\widehat{\boldsymbol{\gamma}}$ as follows

$$\begin{aligned}
\widehat{\boldsymbol{\gamma}} \in \quad & \underset{\boldsymbol{\gamma} \in \mathbb{R}^{p-d}}{\arg\min} \|\boldsymbol{\gamma}\|_1 \\
\text{s.t.} \quad & \|n^{-1}\mathbf{X}^\top \psi(\mathbf{Y} - \mathbf{Z}\boldsymbol{\beta}_0 - \mathbf{X}\boldsymbol{\gamma})\|_\infty \leq \eta_\gamma \\
& n^{-1}(\mathbf{Y} - \mathbf{Z}\boldsymbol{\beta}_0)^\top \psi(\mathbf{Y} - \mathbf{Z}\boldsymbol{\beta}_0 - \mathbf{X}\boldsymbol{\gamma}) \geq \bar{\eta}_\gamma \\
& \|\psi(\mathbf{Y} - \mathbf{Z}\boldsymbol{\beta}_0 - \mathbf{X}\boldsymbol{\gamma})\|_\infty \leq \mu_\gamma,
\end{aligned} \qquad (8)$$

where $\eta_\gamma, \bar{\eta}_\gamma$ and $\mu_\gamma$, are positive tuning parameters. In the display above $\mathbf{Z} = (\mathbf{Z}_{(1)}, ..., \mathbf{Z}_{(d)}) \in \mathbb{R}^{n \times d}$. This estimator directly depends on the null through $\mathbf{Z}\boldsymbol{\beta}_0$ and gives flexibility when estimating dense models; namely, the third constraint restricts the growth of the estimated residuals in the same spirit as classical robust estimators whereas the second constraint allows estimation of the variance of the model error, preventing fast convergence to zero for example. The choice of the function $\psi$ can be made based on prior assessment of the distribution of the model error – namely, models with heavier tails require more penalization in the last constraint than those with sub-gaussian error for example. Thus, it suffices to fine one such function for which the true model error satisfies all three constraints simultaneously. We discuss a simple case of identity function in depth; however, we illustrate other choices in Section 3.

Estimators of $\boldsymbol{\theta}_{(j)}$ are computed in the similar manner. Namely, we consider

$$
\begin{aligned}
\widehat{\boldsymbol{\theta}}_{(j)} &\in \underset{\boldsymbol{\theta}_{(j)} \in \mathbb{R}^{p-d}}{\arg\min} \|\boldsymbol{\theta}_{(j)}\|_1 \\
s.t. \quad & \|n^{-1}\mathbf{X}^\top(\mathbf{Z}_{(j)} - \mathbf{X}\boldsymbol{\theta}_{(j)})\|_\infty \leq \eta_{\theta,j} \\
& n^{-1}\mathbf{Z}_{(j)}^\top(\mathbf{Z}_{(j)} - \mathbf{X}\boldsymbol{\theta}_{(j)}) \geq \bar{\eta}_{\theta,j} \\
& \|\mathbf{Z}_{(j)} - \mathbf{X}\boldsymbol{\theta}_{(j)}\|_\infty \leq \mu_{\theta,j}
\end{aligned}
\tag{9}
$$

for suitable choices of positive tuning parameters $\eta_{\theta,j}, \bar{\eta}_{\theta,j}, \mu_{\theta,j}$. Observe that the above constraints are all linear in the unknown parameter and allow fast implementation.

**Step 2** Calculate the test statistic $\|\mathbf{T}_n\|_\infty = \max_{1 \leq j \leq p} |T_{n,j}|$, where $\mathbf{T}_n = (T_{n,1}, \cdots, T_{n,d})^\top \in \mathbb{R}^d$, and

$$T_{n,j} = n^{-1/2} \sum_{i=1}^n T_{i,j}, \text{ for} \tag{10}$$

$$T_{i,j} = \widehat{\sigma}_{u,j}^{-1}\widehat{\sigma}_\varepsilon^{-1}(z_{i,(j)} - x_i^\top \widehat{\boldsymbol{\theta}}_{(j)})^\top \psi(y_i - z_i^\top \boldsymbol{\beta}_0 - x_i^\top \widehat{\boldsymbol{\gamma}}), \text{ and} \tag{11}$$

$$\widehat{\sigma}_\varepsilon^2 = n^{-1}\|\psi(\mathbf{Y} - \mathbf{Z}\boldsymbol{\beta}_0 - \mathbf{X}\widehat{\boldsymbol{\gamma}})\|_2^2, \text{ and}$$

$$\widehat{\sigma}_{u,j}^2 = n^{-1}\|\mathbf{Z}_{(j)} - \mathbf{X}\widehat{\boldsymbol{\theta}}_{(j)}\|_2^2, \quad j = 1, \ldots, d.$$

Here, the function is evaluated on a vector coordinate wise.

The critical value for the test above is computed analogously as in Section 2.2. Namely, with

$$\widetilde{T}_{n,j} = n^{-1/2} \sum_{i=1}^n \xi_i \left(T_{i,j} - T_{n,j}/\sqrt{n}\right)$$

and the significance level $\alpha$, conditional on the original observations, we compute $\mathcal{Q}(1 - \alpha, \|\widetilde{\mathbf{T}}_n\|_\infty) = \inf \left\{ x \in \mathbb{R} \mid \mathbb{P}(\|\widetilde{\mathbf{T}}_n\|_\infty > x \mid \mathbf{X}, \mathbf{Z}, \mathbf{Y}) \leq 1 - \alpha \right\}$, with $\widetilde{\mathbf{T}}_n = (\widetilde{T}_{n,1}, \cdots, \widetilde{T}_{n,d})^\top \in \mathbb{R}^d$ and $\mathbf{Y} = (y_1, \cdots, y_n)^\top \in \mathbb{R}^n$. Then, we reject $H_0$ in (1) if and only if

$$\|\mathbf{T}_n\|_\infty > \mathcal{Q}(1 - \alpha, \|\widetilde{\mathbf{T}}_n\|_\infty).$$

We consider a regularization form that allows for both easy implementation and good theoretical properties. Estimators (9) are inspired by the recent work of Zhu and Bradic (2016a) for example. There the authors observe that $\bar{\eta}_\gamma$ and $\bar{\eta}_{\theta,j}$ should be of the order of $\|\mathbf{G}\|_2^2$ and $\|\mathbf{Z}_{(j)}\|_2^2$, respectively. In models where sparsity grows too fast, a naive overestimate of the variance of the model error may be warranted and is needed due to difficult tuning of the regularization parameters.

Note that the test statistic $\|\mathbf{T}_n\|_\infty$ considers the maximum over the estimated inner products and therefore allowing for the exploding number of simultaneous tests. Developing asymptotic distribution of such maxima is extremely difficult with correlation between the test statistics $T_{n,j}$ creating arduous technicalities. Moreover, observe that the estimators $\widehat{\boldsymbol{\theta}}_{(j)}$ and $\widehat{\boldsymbol{\gamma}}$ are regularized estimators and no de-biasing or refitting steps are needed, thus in turn, removing some of the computational costs of the existing methods. The multiplier bootstrap procedure is proposed in Steps 3-5 and allows for a very complicated dependence structure while approximating the null distribution successfully. Existing work on high-dimensional bootstrap largely focuses on the bootstrap of the linearized part of the test statistic with each components estimated consistently, whereas we allow larger errors in estimation of one of the underlying unknowns (hence one of them being not consistently estimated).

We also note that the implementation of the estimators above is simple and computationally efficient, as they can be framed into simple linear programming setting. We leave the discussion of practical choices of the tuning parameters for the Section 5.

## 3 Advancements beyond linear models

### 3.1 Weakly dependent data

In this section we illuminate the proposed methodology on a range of regression models different from the linear model considered in previous sections. We discuss dependence in the model errors first. Weak dependence is a common feature in many datasets. We consider the model
$$y_t = z_t^\top \boldsymbol{\beta}^* + x_t^\top \boldsymbol{\gamma}^* + \varepsilon_t \qquad \text{for } 1 \leq t \leq n, \tag{12}$$
where $\varepsilon_t$ independent of $x_t$ and $z_t$ such that $\mathbb{E}\varepsilon_t = 0$ and $\mathbb{E}\varepsilon_t^2 = \sigma^2$. Here, we assume that the data $\{(y_t, x_t, z_t)\}_{t=1}^n$ is $\beta$-mixing with exponential decay; see Bradley (2007) for the definition of $\beta$-mixing conditions. Many popular time series models including ARMA and GARCH satisfy this condition; see (Mokkadem, 1988; Carrasco and Chen, 2002).

We show that in order to test $\boldsymbol{\beta}_* = \boldsymbol{\beta}_0$ in this setup, it suffices to modify the bootstrap procedure to a allow for dependence in the model error; we define a block multiplier bootstrap for this purpose. We note that the robustness of the method is amplified here in that we do not need to change the way we estimate parameters in the model, although the model errors are highly dependent. This allows for fast and practically relevant alternative to many of the datasets that exhibit correlations and/or dependence.

The procedure is inspired by the Bernstein's blocking argument. The idea is to divide the sample into big and small blocks in an alternating order, i.e., a big block followed by a small block followed by a big block, etc. The strategy is to allow both big and small blocks to grow with $n$ such that the small blocks separate the big blocks and thus the big blocks can be viewed as approximately independent blocks. On the other hand, the small blocks only constitute a vanishing portion of the data and thus have almost no impact on the large-sample behavior of the statistics. This is a common proof technique in time series analysis but here we use it to explicitly construct the bootstrap procedure.

To describe the procedure, let $q_n, r_n \to \infty$ satisfy $q_n/r_n \to \infty$ and $n/q_n \to \infty$. Let $m$ be the integer part of $n/(q_n + r_n)$. We define the multipliers $\{\xi_t\}_{t=1}^n$ as follows:

(1) Draw $\{\xi_{1+(q_n+r_n)k}\}_{k=0}^{m-1}$ from i.i.d. $N(0,1)$;

(2) For $1 + (q_n + r_n)k \leq t \leq 1 + (q_n + r_n)k + q_n$, set $\xi_t = \xi_{1+(q_n+r_n)k}$;

(3) Set all the other $\xi_t$'s to zero.

In the above procedure, big blocks and small blocks contain $q_n$ and $r_n$ observations, respectively. The small blocks have $\xi_t = 0$ whereas the big blocks have weights drawn from $N(0,1)$. Notice that all the observations in a same big block have the same weight. Notice that the requirement for the choice of block size is very weak. For bootstrapping low-dimensional problems, the issue of optimal choice of block size has been studied; see Lahiri (2013). For high-dimensional problems, deriving the optimal choice of block size based on MSE or other criteria is much harder and unknown at the moment; thus, it is left for future research. We show that our procedure is still valid for weakly dependent data (see Corollary 1) as long as we modify the bootstrap procedure properly. The block multiplier bootstrap is very easy to implement in practice.

## 3.2 Nonlinear Models

In this section we consider a family of nonlinear regression models

$$y_i = f(x_i^\top \gamma^* + z_i^\top \beta^*) + \varepsilon_i, \qquad i = 1, \cdots, n$$

where $y_i$ are the response, $x_i \in \mathbb{R}^p$ and $z_i \in \mathbb{R}^d$ are the covariates and the error $\varepsilon_i$ is independent of the covariates. When $f$ is an identity function, the above model reduces to the linear model of Section 2. Moreover, a special case of this model is a semi-parametric high-dimensional model where $y_i = g(x_i^\top \gamma^*) + z_i^\top \beta^* + \varepsilon_i$ for some function $g$. As long as the function $f$ is twice differentiable, GRIP test extends easily to this setting.

GRIP test can be now defined as $T_n = \max_j |T_{n,j}|$ with

$$T_{n,j} = \frac{n^{-1/2}(\mathbf{Z}_{(j)} - \mathbf{X}\widehat{\boldsymbol{\theta}}_{(j)})^\top (\mathbf{Y} - f(\mathbf{X}\widehat{\boldsymbol{\gamma}} + \mathbf{Z}\boldsymbol{\beta}_0))}{\widehat{\sigma}_{u,j}\widehat{\sigma}_\varepsilon}$$

and estimates of the residual variances

$$\widehat{\sigma}_\varepsilon^2 = n^{-1}\|\mathbf{Y} - f(\mathbf{X}\widehat{\boldsymbol{\gamma}} + \mathbf{Z}\boldsymbol{\beta}_0)\|_2^2, \text{ and } \widehat{\sigma}_{u,j}^2 = n^{-1}\nu(\mathbf{X}\widehat{\boldsymbol{\gamma}} + \mathbf{Z}\boldsymbol{\beta}_0)\|\mathbf{Z}_{(j)} - \mathbf{X}\boldsymbol{\theta}_{(j)}\|_2^2.$$

Here,

$$\nu(x) = \{f'(x)\}^2 - (\mathbf{Y} - f(x))f''(x).$$

Estimates of the parameters of interest are now defined as

$$\begin{aligned}
\widehat{\boldsymbol{\gamma}} \in &\quad \underset{\boldsymbol{\gamma} \in \mathbb{R}^{p-d}}{\arg\min} \|\boldsymbol{\gamma}\|_1 \\
\text{s.t.} &\quad \left\| n^{-1} f'(\mathbf{X}\boldsymbol{\gamma} + \mathbf{Z}\boldsymbol{\beta}_0)\mathbf{X}^\top \left(\mathbf{Y} - f(\mathbf{X}\boldsymbol{\gamma} + \mathbf{Z}\boldsymbol{\beta}_0)\right)\right\|_\infty \leq \eta_\gamma \\
&\quad n^{-1}\mathbf{Y}^\top \left(\mathbf{Y} - f(\mathbf{X}\boldsymbol{\gamma} + \mathbf{Z}\boldsymbol{\beta}_0)\right) \geq \bar{\eta}_\gamma \\
&\quad \|\mathbf{Y} - f(\mathbf{X}\boldsymbol{\gamma} + \mathbf{Z}\boldsymbol{\beta}_0)\|_\infty \leq \mu_\gamma
\end{aligned} \qquad (13)$$

for the estimate of the unknown parameter $\boldsymbol{\gamma}^*$, whereas the estimate of the auxiliary variable $\boldsymbol{\theta}^*_{(j)}$ is now calibrated as follows

$$
\begin{aligned}
\widehat{\boldsymbol{\theta}}_{(j)} \in \quad & \arg\min_{\boldsymbol{\theta}_{(j)} \in \mathbb{R}^{p-d}} \|\boldsymbol{\theta}_{(j)}\|_1 \\
s.t. \quad & \|n^{-1}\nu(\mathbf{X}\widehat{\boldsymbol{\gamma}} + \mathbf{Z}\boldsymbol{\beta}_0)\mathbf{X}^\top(\mathbf{Z}_{(j)} - \mathbf{X}\boldsymbol{\theta}_{(j)})\|_\infty \leq \eta_{\theta,j} \\
& n^{-1}\nu(\mathbf{X}\widehat{\boldsymbol{\gamma}} + \mathbf{Z}\boldsymbol{\beta}_0)\mathbf{Z}^\top_{(j)}(\mathbf{Z}_{(j)} - \mathbf{X}\boldsymbol{\theta}_{(j)}) \geq \bar{\eta}_{\theta,j} \\
& \|\mathbf{Z}_{(j)} - \mathbf{X}\boldsymbol{\theta}_{(j)}\|_\infty \leq \mu_{\theta,j}
\end{aligned} \quad (14)
$$

Depending on the structure of the function $f$, the above optimization procedures may or may not be easily implementable. However, gradient descent algorithms can be applied through linearization of the gradient $f'$ and function $\nu$. Additionally, unlike linear models, non-linear models require an iterative scheme much in the spirit of reweighed least squares methods.

## 3.3 Generalized Additive Models

Lastly, we discuss an extremely wide class of models; a generalized additive model where

$$g(\mu(\mathbf{X})) = \alpha + \sum_{j=1}^{p} f_j(X_j), \qquad \mathbb{E}[f_j(X_j)] = 0$$

for $\mu(\mathbf{X}) = \mathbb{E}[\mathbf{Y}|\mathbf{X}]$ and the density of $\mathbf{Y}$ belongs to the exponential family, much like the generalized linear models. In the notation of generalized linear models $g(\cdot) = b'(\cdot)$. For simplicity we will assume that the responses are centered in such a way that $\alpha = 0$. It is worth pointing out that these class of models are extremely widespread and yet no regularization estimate exists. Here, we provide not only an estimator, but rather a testing statistics useful for a broad range of multivariate hypothesis tests.

Suppose that the test of interest is

$$H_0 : f_1 = f_{01}$$

for a pre-specified function $f_{01}$. More than one function can easily be tested by replacing 1 with $j$ and considering the maximum statistics. Such tests would be of great use in non-gaussian graphical models and in particular graphical models with both discrete and continuous observations. The test statistic of interest then becomes

$$T_{n,1} = \frac{n^{-1/2} \left( f_{01}(\mathbf{X}_1) - \sum_{k=2}^{p} \widehat{h}_k(\mathbf{X}_k) \right)^\top \left( \mathbf{Y} - b'(f_{01}(\mathbf{X}_1) + \sum_{k=2}^{p} \widehat{f}_k(\mathbf{X}_k)) \right)}{\widehat{\sigma}_{u,1}\widehat{\sigma}_\varepsilon}$$

with

$$\widehat{\sigma}^2_\varepsilon = n^{-1}\|\mathbf{Y} - b'(f_{01}(X_{i1}) + \sum_{k=2}^{p} \widehat{f}_k(X_{ik}))\|_2^2$$

and

$$\widehat{\sigma}^2_{u,1} = n^{-1}b''\left(f_{01}(\mathbf{X}_1) + \sum_{k=2}^{p} \widehat{f}_k(\mathbf{X}_k)\right) \|f_{01}(\mathbf{X}_1) - \sum_{k=2}^{p} \widehat{h}_k(\mathbf{X}_k)\|_2^2.$$

Here $f(\mathbf{X}_1) = (f(X_{11}), f(X_{21}), \ldots, f(X_{n1}))^\top \in \mathbb{R}^n$. With slight abuse in notation, let $\mathbf{b}_j$ denote a vector of $m_j$ basis functions. Then, we represent the estimator

$$\widehat{f}_j(X_j) = b_j(X_j)^\top \widehat{\boldsymbol{\gamma}}_j$$

Let $\mathbf{B}_j \in \mathbb{R}^{n \times m_j}$ be the matrix of evaluations of this function at the $n$ values $\{X_{ij}\}_{i=1}^n$ and assume without loss of generality that $\mathbf{B}_j$ has orthonormal columns. Let $f_{01}(X_1)$ denote the vector of size $n$ containing evaluations $f_{01}(X_{i1})$ as its coordinates.

Regarding the estimation of $\boldsymbol{\gamma}^*$ we adapt the estimator of Section 2 and propose the following estimator

$$
\begin{aligned}
(\widehat{\boldsymbol{\gamma}_2}, \cdots \widehat{\boldsymbol{\gamma}_p}) \in \quad & \underset{\boldsymbol{\gamma}_j \in \mathbb{R}^{p-d}}{\arg\min} \sum_{j=2}^p \|\boldsymbol{\gamma}_j\|_1 \\
\text{s.t.} \quad & \max_{2 \le j \le p} \left\| n^{-1} \mathbf{B}_j^\top \left( \mathbf{Y} - b'(\sum_{j=2}^p \mathbf{B}_j \boldsymbol{\gamma}_j + f_{01}(X_1)) \right) \right\|_\infty \le \eta_\gamma \\
& n^{-1} \mathbf{Y}^\top \left( \mathbf{Y} - b'(\sum_{j=2}^p \mathbf{B}_j \boldsymbol{\gamma}_j + f_{01}(X_1)) \right) \ge \bar{\eta}_\gamma \\
& \|\mathbf{Y} - b'(\sum_{j=2}^p \mathbf{B}_j \boldsymbol{\gamma}_j + f_{01}(X_1))\|_\infty \le \mu_\gamma \\
& \sum_{j=2}^p \sqrt{\boldsymbol{\gamma}_j^\top \mathbf{D}_j \boldsymbol{\gamma}_j} \le \delta_j
\end{aligned} \quad (15)
$$

Here, we also denote the diagonal penalty matrix by $\mathbf{D}_j = \text{diag}(d_{1j}, d_{2j}, \ldots, d_{nj})$. Assuming that the $b_j$ are ordered in increasing order of complexity, $\mathbf{D}_j$ has the property that $0 = d_{1j} \le d_{2j} \le d_{3j} \le \cdots d_{nj}$. Here $d_1$ and $d_2$ correspond to the constant and linear basis functions, respectively. The non-zero $d_j$ are associated with $u_j$ that are non-linear functions of $x$, with higher indexes corresponding to $u_j$ with greater numbers of zero-crossings.

Observe that whenever the $b'$ function is an identity the above optimization procedure can be broken down into $p$ individual optimization procedures much in line of those of Section 2; the only challenge is the last constraint that introduces a quadratic and not linear restrictions making the problems now convex optimizations instead of linear. With slight abuse in notation, let $\mathbf{b}_j$ denote a vector of $m_j$ basis functions possibly different from those used previously. Then, we represent the estimator

$$\widehat{h}_j(X_j) = b_j(X_j)^\top \widehat{\boldsymbol{\theta}}_j$$

with

$$
\begin{aligned}
(\widehat{\boldsymbol{\theta}}_2, \cdots \widehat{\boldsymbol{\theta}}_p) \in \quad & \underset{\boldsymbol{\gamma}_j \in \mathbb{R}^{p-d}}{\arg\min} \sum_{j=2}^p \|\boldsymbol{\theta}_j\|_1 \\
\text{s.t.} \quad & \max_{2 \le j \le p} \left\| n^{-1} \mathbf{B}_j^\top \left( f_{01}(X_1) - \sum_{j=2}^p \mathbf{B}_j \boldsymbol{\theta}_j \right) \right\|_\infty \le \eta_\gamma \\
& n^{-1} f_{01}(X_1)^\top \left( f_{01}(X_1) - \sum_{j=2}^p \mathbf{B}_j \boldsymbol{\theta}_j \right) \ge \bar{\eta}_\gamma \\
& \|f_{01}(X_1) - \sum_{j=2}^p \mathbf{B}_j \boldsymbol{\theta}_j\|_\infty \le \mu_\gamma \\
& \sum_{j=2}^p \sqrt{\boldsymbol{\theta}_j^\top \mathbf{D}_j \boldsymbol{\theta}_j} \le \delta_j.
\end{aligned} \quad (16)
$$

## 4 Asymptotic properties

### 4.1 Size properties

For the model specified by (2) we impose the following regularity conditions. In the rest of the document we denote with $K_1, C_{\min}, C_{\max}, \kappa \in (0, \infty)$ constants independent of $n$, $d$ and $p$.

**Condition 1.** *There exists a positive constant $K_1$ such that $\sigma_{u,j} \geq K_1$ for $j \in \{1, \cdots, d\}$. Moreover, $\|\boldsymbol{\gamma}^*\|_2 = \mathcal{O}(1)$, $\log d = o(n^{1/8})$ and $\log d = \mathcal{O}(\log(p-d))$. In addition, at least one of the following conditions holds:*

*(i) $\max_{1 \leq j \leq d} \|\boldsymbol{\theta}^*_{(j)}\|_0 = o\left(\left[[n \log(d \vee n) \log d]^{1/4} / \log(p-d)\right] \wedge \left[[n/\log(p-d)]^{1/3}/\log(dn)\right]\right)$*

*or*

*(ii) $\|\boldsymbol{\gamma}^*\|_0 = o\left(\left[[n \log(d \vee n)]^{1/4} / \log(p-d)\right] \wedge \left[n \log(p-d)/\log^3(dn)\right]\right).$*

Condition 1 is very mild and more general than most conditions commonly imposed by the existing work on high-dimensional inference. Observe that we only require sparsity of either $\boldsymbol{\theta}^*$ or $\boldsymbol{\gamma}^*$ and not both of them, whereas simultaneous methods of Zhang and Cheng (2016); Dezeure et al. (2016) require both of them to be sparse. However, as we remove one of the sparse components, our requirement on the other is slightly stronger than the existing $o(\sqrt{n}/[\log p \sqrt{\log d}])$. This can be thought of as a price to pay for allowing procedure to be more robust to the sparsity or existence thereof in the other component. It is worth pointing that the sparsity of $\boldsymbol{\theta}^*$ is an assumption on the design matrix of the model (2) and therefore can be checked at least approximately, whereas sparsity of $\boldsymbol{\gamma}^*$ is a model requirement for which current literature does not provide any tests on.

**Remark 3.** *The requirement of $\log(d)/\log(p-d) = \mathcal{O}(1)$ implies that $p-d$ grows at least polynomially fast, compared to the number of simultaneous components being tested, $d$. However, this is not restrictive as it still allows for $d \gg n$ and $d/p \to 1$.*

For simplicity of presentation, in the following we present the theory for the special case of identity function as $\psi$. However, results extend easily for many convex and even non-convex functions.

**Theorem 2.** *Let Condition 1 hold. Consider a choice of tuning parameters such that $\eta_\gamma, \eta_{\theta,j} \asymp \sqrt{n^{-1}\log(p-d)}$, $\mu_\gamma, \mu_{\theta,j} \asymp \sqrt{\log(dn)}$, $\bar{\eta}_\gamma \in (c_0, \sigma_\varepsilon^2 - c_0)$ and $\bar{\eta}_{\theta,j} \in (c_0, \sigma_{u,j}^2 - c_0)$ for a fixed constant $c_0 > 0$. Then, under $H_0$ in (1), the optimization problems in (9) are jointly (for all $j$) feasible with probability approaching one and*

$$\limsup_{n \to \infty} \sup_{\alpha \in (0,1)} \left| \mathbb{P}\left(\|\mathbf{T}_n\|_\infty > \mathcal{Q}(1-\alpha, \|\widetilde{\mathbf{T}}_n\|_\infty)\right) - \alpha \right| = 0,$$

*where $\mathbf{T}_n$ and $\widetilde{\mathbf{T}}_n$ are defined in (10) and (4).*

Under the common regularity condition of sparse $\boldsymbol{\theta}^*_{(j)}$'s, no assumption on $\|\boldsymbol{\gamma}^*\|_0$ is needed. Hence, Theorem 2 guarantees the validity and exactness of the test proposed in this paper even if $\|\boldsymbol{\gamma}^*\|_0 \asymp p$ and $d, p \gg n$. To the best of our knowledge, this is one of the first theoretical results on high-dimensional simultaneous inference under such generality. For inference problems with large $d$, allowing for non-sparse models can in fact be a necessity.

**Remark 4.** *In addition, the result above guarantees asymptotic exactness of the proposed GRIP test in a class of high-correlation sparse models where the feature correlation and therefore sparsity of the precision matrix is unrestricted; Javanmard and Montanari (2014a) have a robust test in this setting when the testing set is univariate. Lastly, models of hybrid nature in which the signal is the sum of sparse and dense components are also a special case of the result above. Such models are extremely difficult to handle as demonstrated in Chernozhukov et al. (2015) and are yet practically often important (see Qu and Shi (2016) for more details). GRIP test provides one of the first valid inferential methods where such hybrid structures are allowed.*

## 4.2 Minimax testing

In this subsection, we show that our test is asymptotically minimax optimal in the case of high-dimensional models.

Since the power of (asymptotically) minimax optimal tests is guaranteed over a class of data-generating processes (dgp's) that do not satisfy the null hypothesis, we first formally introduce this class of dgp's. Let $\boldsymbol{\Theta}^* = (\boldsymbol{\theta}^*_{(1)}, \cdots, \boldsymbol{\theta}^*_{(d)}) \in \mathbb{R}^{(p-d) \times d}$. We define $\lambda = (\boldsymbol{\beta}, \boldsymbol{\gamma}, \boldsymbol{\Theta}, \boldsymbol{\Sigma}_X, \boldsymbol{\Sigma}_u, \sigma_\varepsilon^2)$ and view the dgp's as being indexed by $\lambda$. The following regularity conditions are imposed on the dgp's under consideration in order to simplify the proof. Simplifications of these conditions are possible at the cost of greater complications in the proofs. Let $\boldsymbol{\pi} = \boldsymbol{\beta} - \boldsymbol{\beta}_0$ be the difference between the true and the hypothesized values.

**Condition 2.** *Parameter $\lambda = (\boldsymbol{\beta}, \boldsymbol{\gamma}, \boldsymbol{\Theta}, \boldsymbol{\Sigma}_X, \boldsymbol{\Sigma}_u, \sigma_\varepsilon^2)$ satisfies Condition 1 together with both (i) and (ii). In addition, $d \to \infty$ and vector $\boldsymbol{\pi}$ is such that $\|\boldsymbol{\Sigma}_X \boldsymbol{\Theta} \boldsymbol{\pi}\|_\infty = \mathcal{O}(\sqrt{n^{-1} \log(p - d)})$ and $\|\boldsymbol{\pi}\|_2 = \mathcal{O}(1)$.*

Notice that these regularity conditions are very weak and match largely the existing literature. For the case of sparse models, we allow both $\|\boldsymbol{\beta}_0\|_0$ and $\|\boldsymbol{\gamma}\|_0$ to be $o(\sqrt{n}/\log p)$, where with $\|\boldsymbol{\pi}\|_\infty = \mathcal{O}(\sqrt{n^{-1} \log d})$ and $\|\boldsymbol{\pi}\|_2 = \mathcal{O}(1)$ deviations from the null, the above condition is easily satisfied with $\|\boldsymbol{\pi}\|_2 = o(\|\boldsymbol{\pi}\|_\infty n^{1/4} \log^{-1/2} p)$. Moreover, observe that, under Condition 1, we have $\|\boldsymbol{\Sigma}_X \boldsymbol{\Theta} \boldsymbol{\pi}\|_\infty = \mathcal{O}(\sqrt{n^{-1} \log(p - d)})$, whenever $\|\boldsymbol{\pi}\|_0 = \mathcal{O}(1)$ and $\|\boldsymbol{\pi}\|_\infty = \mathcal{O}(\sqrt{n^{-1} \log d})$. Conditions much like these are present in the existing work on simultaneous testing (Zhang and Cheng, 2016; Dezeure et al., 2016).

**Remark 5.** *However, observe that Condition 2 allows models much broader than the exactly sparse models. Suppose that the variance of $(z_i, x_i)$ has a Toeplitz dependence structure implying that $\boldsymbol{\Theta}$ is a row-sparse matrix. It follows that the rows of $\boldsymbol{\Sigma}_X \boldsymbol{\Theta}$ have bounded $\ell_1$-norm. Hence, the Condition 2 is satisfied whenever $\|\boldsymbol{\pi}\|_\infty = O(\sqrt{(\log(p - d))/n})$ regardless of the sparse or dense structure of the vector $\boldsymbol{\pi}$. In particular, this allows vectors $\boldsymbol{\beta}$ and $\boldsymbol{\beta}_0$ to be extremely dense and high-dimensional simultaneously. We note that to the best of our knowledge existing literature provides no tests or the corresponding minimax optimality theory for such broad umbrella of models.*

In fact, with $d = p$, $\boldsymbol{\gamma}^* = 0$ and the above condition encompasses fully dense models where one is interested in testing all parameters of the model simultaneously. Recent work of Janson et al. (2015) considers similar models but univariate hypothesis testing of $l_2$ norm of $\boldsymbol{\beta}$ and allows only identity matrix for $\boldsymbol{\Sigma}_X$. The area of multivariate inference with dense structures is largely unexplored and demands new methodologies and theories.

In the following we define the set of dgp's that satisfy the null hypothesis

$$\mathcal{B}_0 = \{\lambda = (\boldsymbol{\beta}_0, \boldsymbol{\gamma}, \boldsymbol{\Theta}, \boldsymbol{\Sigma}_X, \boldsymbol{\Sigma}_u, \sigma_\varepsilon^2) \mid \lambda \text{ satisfies Condition 2}\},$$

and correspondingly, the set of dgp's that satisfy the alternative hypothesis:

$$\mathcal{B}_1(\tau) = \Big\{\lambda = (\boldsymbol{\beta}, \boldsymbol{\gamma}, \boldsymbol{\Theta}, \boldsymbol{\Sigma}_X, \boldsymbol{\Sigma}_u, \sigma_\varepsilon^2) \mid$$
$$\|\boldsymbol{\Sigma}_u(\boldsymbol{\beta}^* - \boldsymbol{\beta}_0)\|_\infty > \tau\sqrt{n^{-1}\log(d \vee n)} \text{ and } \lambda \text{ satisfies Condition 2}\Big\}.$$

We define the true value of the dgp parameter $\lambda^* = (\boldsymbol{\beta}^*, \boldsymbol{\gamma}^*, \boldsymbol{\Theta}^*, \boldsymbol{\Sigma}_X, \boldsymbol{\Sigma}_u, \sigma_\varepsilon^2)$. The goal is to test
$$H_0: \lambda^* \in \mathcal{B}_0 \quad \text{versus} \quad H_1: \lambda^* \in \mathcal{B}_1(\tau).$$

In this section we show our test has power approaching one against alternatives in $\mathcal{B}_1(\bar{\tau})$. Denote with $\mathbb{P}_\lambda$ and $\mathbb{E}_\lambda$ the probability and expectation under $\lambda$, respectively.

**Theorem 3.** *Under conditions illustrated above, there exists a constant $\bar{\tau} > 0$ such that*
$$\liminf_{n \to \infty} \inf_{\lambda \in \mathcal{B}_1(\bar{\tau})} \mathbb{P}_\lambda \left( \|\mathbf{T}_n\|_\infty > \mathcal{Q}(1-\alpha, \|\widetilde{\mathbf{T}}_n\|_\infty) \right) = 1 \ \forall \alpha \in (0,1).$$

This result says that, under weak conditions, our procedure for testing $\lambda^* \in \mathcal{B}_0$ against $\lambda^* \in \mathcal{B}_1(\bar{\tau})$ is asymptotically minimax optimal in that the asymptotic power on $\mathcal{B}_1(\bar{\tau})$ is guaranteed to be one.

Next, we derive an upper bound on the power of any valid procedure for this testing problem. We show that there exist two constants $\bar{\tau}, \underline{\tau} > 0$ such that any test that controls size has no power against alternatives in $\mathcal{B}_1(\underline{\tau})$.

**Theorem 4.** *Let the sample $(\mathbf{X}, \mathbf{Z}, \mathbf{Y})$ be jointly Gaussian. Then there exists a constant $\underline{\tau} > 0$ such that for any test $\phi_n : (\mathbb{R}^{p+1})^n \to [0,1]$ satisfying $\limsup_{n \to \infty} \sup_{\lambda \in \mathcal{B}_0} \mathbb{E}_\lambda \phi_n(\mathbf{X}, \mathbf{Z}, \mathbf{Y}) \leq \alpha$, we have*
$$\liminf_{n \to \infty} \inf_{\lambda \in \mathcal{B}_1(\underline{\tau})} \mathbb{E}_\lambda \phi_n(\mathbf{X}, \mathbf{Z}, \mathbf{Y}) \leq \alpha.$$

Theorem 4 rules out the existence of an asymptotically valid test that has power uniformly (in $\lambda$) against $\mathcal{B}_1(\underline{\tau})$. In other words, no test is guaranteed to detect alternatives satisfying $\|\boldsymbol{\Sigma}_u(\boldsymbol{\beta}^* - \boldsymbol{\beta}_0)\|_\infty \leq \underline{\tau}\sqrt{n^{-1}\log(d \vee n)}$.

**Remark 6.** *Note that Theorem 3 guarantees that our procedure have power approaching one against alternatives $\|\boldsymbol{\Sigma}_u(\boldsymbol{\beta}^* - \boldsymbol{\beta}_0)\|_\infty > \bar{\tau}\sqrt{n^{-1}\log(d \vee n)}$. Comparing the bound with Theorem 4, we see that the test we developed is asymptotically minimax optimal when $d = d(n) \to \infty$, $p = p(n) \to \infty$ as $n \to \infty$, under mild regularity conditions.*

The results in this section fill an important gap in the inferential theory of high-dimensional parameters. Existing results, such as Javanmard and Montanari (2014b), only address the testing problem of single entries of high-dimensional parameter and conclude that only deviations of the order at least $\mathcal{O}(n^{-1/2})$ are detectable. This result should not be expected to hold if the number of entries under testing, $d$, tends to infinity. We formally establish that, in this case, the null hypothesis is not testable against alternatives with deviations smaller than $\mathcal{O}(\sqrt{n^{-1}\log d})$. Together with Theorem 3 we see that GRIP test is minimax optimal for a wide class of regression models.

Further detailed comments provide comparisons with existing state of the art results.

1. Among the sparse models this result matches existing work; GRIP test achieves the same optimality rate as the work of Zhang and Cheng (2016) (see Theorem 2.4 therein). However, in this class of models GRIP offers a broader set of alternatives as $\boldsymbol{\beta}_0$ can be dense or sparse vector. In this setting, the alternative hypothesis is allowed to be different from the null in all elements, whereas existing work assumes that the alternative can only be different in a few, sparse, number of elements.

2. For models where sparsity is not exact, and a more reasonable hybrid model of signal composed of two vectors, sparse and dense, GRIP test achieves the same optimality rate while allowing $p \gg n$ and $d \gg n$. In this setting Theorem 4 represents a new result of independent interest that is perhaps unexpected; the dense but not too strong parameter does not affect detection threshold.

   A subset of models, of this kind were only recently studied in Chernozhukov et al. (2015) where estimation is considered under a stricter set of dimensionality assumptions, i.e., dense parameter length was considered growing at a slower rate than $n$. However, inference problems have not yet been studied for these class of high-dimensional models. Note that the method proposed here have an additional simplicity – it does not need to change between the ridge and the lasso type of estimation procedure. i.e. it adapts to the unknown structure using new estimators proposed in Section 2.

3. Additionally, our setup above encompasses a class of models where the signal is decomposed of three parts: strong signal, not-strong signal (below detection limit) and a number of zero components. Such models are practically extremely important. Estimation in such models was recently proposed in Qu and Shi (2016).

4. Lastly, our work breaks new ground in the area of strict false discovery rate control, as it showcases a possibility of testing simultaneous effect of all parameters, i.e., testing the global null, against both sparse and dense alternatives (where at least one coordinate is $\mathcal{O}(\sqrt{n^{-1} \log d})$ away from the true parameter vector $\boldsymbol{\beta}^*$) in models that are completely dense and high-dimensional at the same time (take $\boldsymbol{\beta}_0$ to be $p$-dimensional and $\boldsymbol{\gamma}^* = 0$).

   Such models are extremely important in practice. For example, consider the setting of genome-wide association study (GWAS), which typically refers to examination of associations between up to millions of genetic variants in the genome and certain traits of interest. According to GWAS catalog (Hindorff et al. (2009); http://www.genome.gov/gwastudies), as of October, 2013, more than 11,000 single-nucleotide polymorphisms (SNPs) have been reported to be associated with at least one trait/disease at the genome-wide significance level, many of which have been validated/replicated in further studies. However, these significantly associated SNPs only account for a small portion of the genetic factors underlying complex human traits/diseases (Manolio et al., 2009). One possible explanation for the missing heritability is that many SNPs jointly affect the phenotype, while the effect of each SNP is too weak to be detected at the genomewide significance level. Our test can make significant progress in this direction were one would be able to test joint effects of a large (and exploding) number of SNPs without requiring each one to be significant.

Moreover, previous theory easily lends a result as follows.

**Theorem 5.** *Let the assumptions of Theorem 2 hold. Suppose that the data is $\beta$-mixing with exponential decay. Let $\xi$ in (4) be replaced by the above block multipliers. Assume that $r_n \asymp n^c$ for some $c > 0$, $r_n/q_n = O(n^{-1/8}/\log^2 p)$ and $q_n = O(n^{3/4}/\log^{5/2}(pn))$. The test of Section 2 with new $\xi$ is asymptotically unbiased.*

Table 1: Size properties of the four competing methods in Model 1

| Sparsity of the model | Toeplitz Design $n = 200$, $p = 500$ $\mathbb{P}(\text{reject } H_0^{(J)} \mid H_0^{(J)})$ | | | |
|---|---|---|---|---|
| | ZC | JM | NL | GRIP |
| $s = 2$ | 0.02 | 0.04 | 0.03 | 0.02 |
| $s = 12$ | 0.24 | 0.32 | 0.09 | 0.05 |
| $s = 17$ | 0.22 | 0.24 | 0.16 | 0.04 |
| $s = 20$ | 0.22 | 0.18 | 0.19 | 0.03 |
| $s = 22$ | 0.29 | 0.24 | 0.18 | 0.05 |
| $s = 40$ | 0.34 | 0.19 | 0.41 | 0.01 |
| $s = 60$ | 0.45 | 0.21 | 0.50 | 0.02 |
| $s = 80$ | 0.60 | 0.29 | 0.66 | 0.04 |
| $s = 100$ | 0.59 | 0.30 | 0.68 | 0.04 |
| $s = 200$ | 0.58 | 0.23 | 0.49 | 0.03 |

Average size over 100 repetitions, where we see that irrespective of the size of the sparsity, existing methods fail to control Type I error.

## 5 Numerical Work

We begin by observing that the optimization problems (9) can be rephrased as linear programs. We follow Candes and Tao (2007) and choose the tuning parameters $\eta_\theta$ and $\eta_\gamma$ as the empirical maximum of $|\mathbf{X}\xi|_i$ over several realizations of $\xi \sim \mathcal{N}(0, \mathbb{I}_n)$; alternatively, one can choose these tuning parameters as in Section 4 of Chernozhukov et al. (2013a). Similarly, $\mu_\theta$ and $\mu_\gamma$ are chosen as the empirical maximum of $|\xi|_i$. We choose $\bar{\eta}_\gamma = (1-\lambda)\mathbf{G}^\top\mathbf{G}/n$, where $\lambda_\gamma \in (\boldsymbol{\gamma}^{*\top}\boldsymbol{\Sigma}\boldsymbol{\gamma}^*/\mathbb{E}[v_i^2], 1)$. Conceptually, $\lambda_\gamma$ is an upper bound for the regression $R^2$ and is more intuitive to choose as a tuning parameter. In the simulations below $\lambda_\gamma = 0.95$. We choose $\bar{\eta}_{\theta,j}$ in a similar way. We measure the performance of the proposed method through Type I and Type II error control. All the tests have a nominal size of 5%. All the results are based on 100 randomly generated samples.

### 5.1 Linear Gaussian Model

In order to compare performance we consider a simple linear model

$$y_i = x_i^\top \boldsymbol{\beta}^* + \varepsilon_i$$

with the sample size $n = 200$. The $\varepsilon_i$'s are generated as independent, standard Gaussian components. We fix the feature size to be $p = 500$. We compare the proposed method, GRIP, with three competing procedures: for ZC all the Lasso operations are replaced by the scaled Lasso with the universal tuning parameter; JM is implemented using two regularizations parameters. As suggested by JM, one equal to $4\widehat{\sigma}\sqrt{\log p/n}$ with $\widehat{\sigma}$ provided by the scaled Lasso and the other taken at a fixed value of $2\sqrt{\log p/n}$; the NL method is implemented by using the scaled Lasso with the universal tuning parameter whenever a Dantzig selector or Lasso is required.

We consider two models:

Table 2: Power properties of the four competing methods in Model 1

| Deviations from the null | Toeplitz Design $n = 200, p = 500$ Type II error | | | |
|---|---|---|---|---|
| | ZC | JM | NL | GRIP |
| $h = 0.8$ | 0.53 | 0.97 | 0.13 | 0.12 |
| $h = 1.6$ | 0.70 | 0.95 | 0.31 | 0.30 |
| $h = 2.4$ | 0.91 | 0.99 | 0.72 | 0.66 |
| $h = 3.2$ | 0.94 | 1.00 | 0.94 | 0.97 |
| $h = 4.0$ | 0.98 | 1.00 | 1.00 | 1.00 |

Average power over 100 repetitions.

- Model 1: $x_i \sim \mathcal{N}(0, \boldsymbol{\Sigma})$ with a Toeplitz covariance matrix $\boldsymbol{\Sigma}$: $\boldsymbol{\Sigma}_{ij} = (0.4)^{|i-j|}$. We generate $\boldsymbol{\beta}^* = 5a/\|a\|_2$, where $a = (a_1, \cdots, a_p)^\top \in \mathbb{R}^p$ is generated as follows. $a_j$ is generated from the uniform distribution on $(0, 1)$ if $j \leq 3s/2$ and $j/3$ is not an integer; otherwise, $a_j = 0$. Here, $s = \|\boldsymbol{\beta}^*\|_0$.

- Model 2: The setup is the same as in Model 1 except $\boldsymbol{\Sigma}$, which now satisfies that $\boldsymbol{\Sigma}_{ii} = 1$ and $\boldsymbol{\Sigma}_{ij} = 0.2s$ for $i \neq j$.

For $j \in J = \{4, 5, 7, 8, 10, 11\}$,
we test $H_0^{(J)} : \beta_j^* = \beta_{0,j} \ \forall j \in J$. We set $\beta_{0,j} = \beta_j^* + n^{-1/2}h$. The simulations with $h = 0$ correspond to size properties and those with $h \neq 0$ demonstrate the power properties.

The results are presented in Tables 1 and 2 (type I and power) for Model 1 and in Tables 3 and 4 (type I and power) for Model 2. First, we note that our procedure correctly keeps the empirical Type I error close to the nominal 5% level and reaches maximum power when the model parameters are far from the null hypothesis. Second, as essentially guaranteed by the established theory, the proposed method clearly outperforms other methods in terms of the size control – especially when the sparsity of the model increases. This is apparent in both examples, with the most prolific comparison can be seen in Table 3; there our method for simultaneous tests keeps nominal value, whereas the VBRD and JM tests with a nominal level of 5% rejects a true null hypothesis with probability close to 100 and NL being close to a random guess.

## 5.2 Linear Model with Heavy-tailed Designs

In the next example, we consider a settings that departs from normality assumptions. We consider the same simple linear model as above. Parameter choices are made by the same choices as in the Models 1-2 above: $n = 200, p = 500$.

- Model 3: The model is the same as in Model 2 except that entries of $\boldsymbol{\Sigma}^{-1/2}x_i$ are generated from a student $t$ distribution with 6 degrees of freedom, instead of a Gaussian distribution.

The results are presented in Tables 5 and 6 and clearly show that our approach works well under designs with heavy-tailed distributions. We can clearly observe that the proposed simultaneous test is the only one that successfully keeps Type I error near the nominal value, while achieving good power; other state-of-the-art are all close to random guesses.

Table 3: Size properties of the four competing methods in Model 2

| Sparsity of the model | Equicorrelation Design $n = 200$, $p = 500$ $\mathbb{P}(\text{reject} H_0^{(J)} \mid H_0^{(J)})$ | | | |
|---|---|---|---|---|
| | ZC | JM | NL | GRIP |
| $s = 1$ | 0.08 | 0.09 | 0.07 | 0.04 |
| $s = 2$ | 0.15 | 0.28 | 0.14 | 0.06 |
| $s = 12$ | 0.37 | 0.76 | 0.65 | 0.02 |
| $s = 17$ | 0.41 | 0.80 | 0.84 | 0.02 |
| $s = 20$ | 0.54 | 0.81 | 0.87 | 0.03 |
| $s = 22$ | 0.56 | 0.84 | 0.86 | 0.04 |
| $s = 40$ | 0.76 | 0.90 | 0.85 | 0.06 |
| $s = 60$ | 0.85 | 0.92 | 0.83 | 0.03 |
| $s = 80$ | 0.94 | 0.96 | 0.74 | 0.03 |
| $s = 100$ | 0.96 | 0.93 | 0.63 | 0.01 |
| $s = 200$ | 0.99 | 0.96 | 0.46 | 0.01 |

Average size over 100 repetitions, where we see that irrespective of the size of the sparsity, existing methods greatly fail to control Type I error – observed errors are very close to 50% making the tests useless in practice.

Table 4: Power properties of the four competing methods in Model 2

| Deviations from the null | Equicorrelation Design $n = 200$, $p = 500$ Type II error | | | |
|---|---|---|---|---|
| | ZC | JM | NL | GRIP |
| $h = 1.0$ | 0.59 | 0.64 | 0.94 | 0.03 |
| $h = 2.5$ | 0.93 | 0.96 | 1.00 | 0.20 |
| $h = 4.0$ | 1.00 | 1.00 | 1.00 | 0.70 |
| $h = 5.5$ | 1.00 | 1.00 | 1.00 | 0.95 |
| $h = 6.5$ | 1.00 | 1.00 | 1.00 | 0.98 |

Average power over 100 repetitions.

Table 5: Size properties of the four competing methods in Model 3

| Sparsity of the model | Student $t$ design $n = 200$, $p = 500$ $\mathbb{P}(\text{reject} H_0^{(J)} \mid H_0^{(J)})$ | | | |
|---|---|---|---|---|
| | ZC | JM | NL | GRIP |
| $s = 1$ | 0.04 | 0.04 | 0.04 | 0.03 |
| $s = 12$ | 0.18 | 0.48 | 0.11 | 0.03 |
| $s = 17$ | 0.20 | 0.42 | 0.13 | 0.02 |
| $s = 20$ | 0.13 | 0.34 | 0.13 | 0.02 |
| $s = 22$ | 0.20 | 0.40 | 0.16 | 0.02 |
| $s = 40$ | 0.30 | 0.30 | 0.39 | 0.01 |
| $s = 60$ | 0.44 | 0.36 | 0.47 | 0.04 |
| $s = 80$ | 0.62 | 0.36 | 0.60 | 0.03 |
| $s = 100$ | 0.62 | 0.31 | 0.56 | 0.01 |
| $s = 200$ | 0.52 | 0.25 | 0.50 | 0.03 |

Average size over 100 repetitions, where we see that irrespective of the size of the sparsity, existing methods greatly fail to control Type I error – observed errors are very close to 50% making the tests useless in practice.

Table 6: Power properties of the four competing methods in Model 3

| Deviations from the null | Student $t$ design $n = 200$, $p = 500$ Type II error | | | |
|---|---|---|---|---|
| | ZC | JM | NL | GRIP |
| $h = 1.0$ | 0.72 | 0.86 | 0.72 | 0.07 |
| $h = 2.5$ | 1.00 | 1.00 | 0.77 | 0.38 |
| $h = 4.0$ | 1.00 | 1.00 | 0.92 | 0.81 |
| $h = 5.5$ | 1.00 | 1.00 | 0.91 | 0.93 |
| $h = 6.5$ | 1.00 | 1.00 | 0.96 | 0.99 |

Average power over 100 repetitions.

# 6   Approximate Multiplier Bootstrap in High-Dimensions

Studying the theoretical properties of GRIP is not a straight-forward problem because the test statistic is the based on bootstrapping the sum of high-dimensional vectors $T_i = (T_{i,1}, ..., T_{i,d})^\top$, which are not independent (across $i$). Existing methods in high-dimensional bootstrap relies on the idea that the summands are independent and can be accurately estimated, see Chernozhukov et al. (2014, 2013a); Zhang and Cheng (2016). However, due to the inherent difficulty in estimating non-sparse high-dimensional parameters $\boldsymbol{\gamma}^*$, $\{T_i\}_{i=1}^n$ might not be consistently estimating any independent sequence and thus theoretical properties of GRIP cannot be established by simply applying existing result.

To address this problem, we propose a new framework of approximate multiplier bootstrap. This is a result of independent interest because it deals with a large class of problems, which includes the current problem (GRIP) as a special case. The key idea is to exploit conditional independence structures in the problem for deriving the theoretical properties.

We now start by illustrating the general setup of this new approach using the setup of Section 2. Observe that, the test statistics $T_{n,j}$ as defined in (10) can be written as $T_{n,j} = n^{-1/2} \sum_{i=1}^n w_{i,j} q_{i,j} + \Delta_{n,j}$, for two triangular arrays of random variables $w_{i,j} = (v_i - x_i \widehat{\gamma}) \widehat{\sigma}_\varepsilon^{-1}$ and $q_{i,j} = u_{i,(j)}/\sigma_{u,j}$ and a sequence $\Delta_{n,j} = n^{-1/2} \sum_{i=1}^n w_{i,j}(\widehat{q}_{i,j} - q_{i,j})$ with $\widehat{q}_{i,j} = \widehat{u}_{i,(j)}/\widehat{\sigma}_{u,j}$ and $\widehat{u}_{i,(j)} = z_{i,j} - x_i \widehat{\boldsymbol{\theta}}_{(j)}$. The decomposition (17) above, arises in many statistical problems, where the statistic of interest $\widehat{\boldsymbol{\Omega}}_n = (\widehat{\Omega}_{n,1}, \cdots, \widehat{\Omega}_{n,d})^\top \in \mathbb{R}^d$ can be represented as

$$\widehat{\Omega}_{n,j} = n^{-1/2} \sum_{i=1}^n w_{i,j} q_{i,j} + \Delta_{n,j}, \tag{17}$$

with the random variables $q_{i,j}$ and $\Delta_{n,j}$ being not observed and only a sequence of approximations $\{\widehat{q}_{i,j}\}$ are available. Moreover, typically, the bias term $\Delta_{n,j}$ doesn't have to be dependent on $\widehat{q}_{i,j}$. We would like to develop a multiplier bootstrap scheme that relies on pairs of observations

$$\{(w_{i,j}, \widehat{q}_{i,j})\}_{i=1}^n$$

and provides a good approximation for the distribution of $\max_{1 \leq j \leq d} |\widehat{\Omega}_{n,j}|$. We call such method the *approximate multiplier bootstrap*. Difficulties arise, as the approximating sequence $\{\widehat{q}_{i,j}\}$ is typically not independent of the sequence $w_{i,j}$ and the presence of $\Delta_{n,j}$ complicates the analysis.

The problem is very standard for a fixed $d$. However, in many high-dimensional or non-parametric problems, $d$ can be much larger than $n$ and, as a result, many of the classical tools, such as central limit theorem, are often inadequate. Moreover, the properties of $\widehat{\gamma}$ are not tractable whenever $\boldsymbol{\gamma}^*$ is non-sparse, thus leading to possible large values of $w_{i,j}$ of (17). However, the framework allows us to exploit the independence structure implied by the null hypothesis and hence avoid the formidable task of establishing properties of estimators of non-sparse high-dimensional parameters.

We now state some regularity conditions and then present a formal result on the validity of the multiplier bootstrap approximation. We also introduce the definition of sub-Gaussian and sub-exponential norms. For a random variable $X$, its sub-Gaussian norm is $\|X\|_{\psi_2} = \sup_{p \geq 1} p^{-1/2} (\mathbb{E}|X|^p)^{1/p}$ and its sub-exponential norm is $\|X\|_{\psi_1} = \sup_{p \geq 1} p^{-1} (\mathbb{E}|X|^p)^{1/p}$.

**Condition 3.** *The sequence $\{q_i\}_{i=1}^n$ is a sequence of independent random vectors with $q_i = (q_{i,1}, \cdots, q_{i,d})^\top \in \mathbb{R}^d$ such that $\mathbb{E} q_i = 0$, $\mathbb{E} q_{i,j}^2 = 1$ and $\max_{1 \leq i \leq n, 1 \leq j \leq d} \|q_{i,j}\|_{\psi_2} < L$ for some constant $L > 0$. Moreover, the sequence $\{w_{i,j}\}_{(i,j) \in \{1, \cdots, n\} \times \{1, \cdots, d\}}$ is independent of $\{q_i\}_{i=1}^n$ such that almost surely, $n^{-1} \sum_{i=1}^n w_{i,j}^2 = 1 \; \forall j \in \{1, \cdots, d\}$.*

An important feature in Condition 3 is that $w_{i_1, j_1}$ is allowed to have arbitrary dependence on $w_{i_2, j_2}$ for $(i_1, j_1) \neq (i_2, j_2)$. This flexibility turns out to be very useful for testing non-sparse high-dimensional models. Moreover, $q_{i, j_1}$ and $q_{i, j_2}$ can also be arbitrarily correlated for $j_1 \neq j_2$.

We show that a multiplier bootstrap scheme based on $\widehat{q}_{i,j}$ can be used to approximate the distribution of $\max_{1 \leq j \leq d} |\widehat{\Omega}_{n,j}|$, if both $\Delta_n$ and

$$r_n^2 = \max_{1 \leq j \leq d} n^{-1} \sum_{i=1}^n w_{i,j}^2 (\widehat{q}_{i,j} - q_{i,j})^2.$$

converge to zero fast enough. The formal result is summarized in the following theorem.

**Theorem 6.** *Suppose Condition 3 holds. Let $\{\xi_i\}_{i=1}^n$ be a sequence of independent standard normal random variables that is also independent of $w_{i,j}$, $q_{i,j}$ and $\widehat{q}_{i,j}$. Define $\mathbb{B}_n = (\mathbb{B}_{n,1}, \cdots, \mathbb{B}_{n,d})^\top \in \mathbb{R}^d$ with $\mathbb{B}_{n,j} = n^{-1/2} \sum_{i=1}^n \xi_i(w_{i,j}\widehat{q}_{i,j} - m_{n,j})$ and $m_{n,j} = n^{-1} \sum_{i=1}^n w_{i,j}\widehat{q}_{i,j}$
Assume*
$$\left(\bar{M}_n^2 \log^7(d \vee n) \bigvee \bar{M}_n^4 \log^2(d \vee n)\right) = o_P(n),$$

*with $\bar{M}_n = \max_{1 \leq i \leq n, 1 \leq j \leq d} |w_{i,j}|$. Moreover, assume that a sequence $\delta_n > 0$ satisfies $\delta_n \log d = o(1)$, $r_n = \mathcal{O}_P(\delta_n)$ and $\|\Delta_n\|_\infty = o_P(\delta_n^{1/2})$. Then,*
$$\limsup_{n \to \infty} \sup_{\eta \in (0,1)} \left|\mathbb{P}\left(\|\widehat{\Omega}_n\|_\infty > \mathcal{Q}(1-\eta, \|\mathbb{B}_n\|_\infty)\right) - \eta\right| = 0,$$

*where $\mathcal{Q}(\alpha, \|\mathbb{B}_n\|_\infty) = \inf\left\{x \in \mathbb{R} \mid \mathbb{P}(\|\mathbb{B}_n\|_\infty > x \mid \{(w_{i,j}, q_{i,j}, \widehat{q}_{i,j})\}_{1 \leq i \leq n, 1 \leq j \leq d}) \leq \alpha\right\}$ and $\widehat{\Omega}_n$ is defined in (17).*

The theory above although developed for Gaussian multipliers applies more widely; with some changes in the proof multipliers like those of Mammen (1993) can be allowed. However, studying their effect (in finite sample or otherwise) is beyond the scope of the current article.

## Discussion

We propose a new GRIP testing rule to tackle simultaneous hypothesis testing problems in high-dimensional settings while allowing exploding number of tests, dimensions and sparsity of the linear model. GRIP first augments the original hypothesis testing problem by leveraging correlations among the features and then achieves valid control over Type I error rate through an approximate multiplier bootstrap procedure. GRIP combines the adaptive and optimal feature correlation estimator with the essentially ill control estimator of what is essentially a misspecified model. Since features are split into useful and not-useful (for the purposes of testing), GRIP enjoys a flexible and optimal asymptotic results without requiring an initial model to be sparse. An array of simulation examples, supported developed theoretical findings.

To verify our methods' ability to generalize, we evaluate its performance different classifiers on the following two examples that have highly non-linear models. We consider two models for which penalized estimators do not exist – hence, illustrating the broad applicability of the proposed methodology. In particular, we consider Models 4 and 5 below. In each case a choice of $v_i = y_i$ was sufficient to guarantee all conditions.

- *Model 4:* (Nonlinear Single Index Model) The design matrix and $\boldsymbol{\beta}^*$ are generated as in the Model 1. However the model is now generated as
$$y_i = (x_i^\top \boldsymbol{\beta}^*) \exp\{\sin(x_i^\top \boldsymbol{\beta}^*)\} + \varepsilon_i \text{ with } \varepsilon_i \sim \mathcal{N}(0,1).$$

- *Model 5:* (Heckman Selection model) Consider the model
$$y_i^* = x_i^\top \boldsymbol{\beta}^* + \varepsilon_i$$
where $\boldsymbol{\beta}^* \in \mathbb{R}^p$ and $y_i^*$ is observed only if $w_i^\top \boldsymbol{\psi}^* + \xi_i > 0$ for $\boldsymbol{\psi}^* \in \mathbb{R}^{2p}$, that is,
$$y_i = y_i^* \mathbb{I}\{w_i^\top \boldsymbol{\psi}^* + \xi_i > 0\},$$

Table 7: Properties of the GRIP in Nonlinear Single Index Model

| Sparsity of the model | | Toeplitz Design $n = 200$, $p = 500$ | | | |
|---|---|---|---|---|---|
| | Type I | Power | | | |
| $s = 10$ | 0.07 | $0.47(h = 4.5)$ | $0.71(h = 5.5)$ | $0.90(h = 6.5)$ | $0.95(h = 7.5)$ |
| $s = 20$ | 0.05 | $0.50(h = 4.5)$ | $0.67(h = 5.5)$ | $0.85(h = 6.5)$ | $0.98(h = 7.5)$ |
| $s = 25$ | 0.06 | $0.38(h = 4.5)$ | $0.70(h = 5.5)$ | $0.94(h = 6.5)$ | $0.97(h = 7.5)$ |

Average size over 100 repetitions, where we see that for a model where there is not an efficient estimation scheme, GRIP performs valid inference.

where $\varepsilon_i \sim \mathcal{N}(0, 1)$, $w_i = (x_i^\top, \zeta_i^\top)^\top$, $\zeta_i \sim \mathcal{N}(0, \mathbb{I}_p)$, $\boldsymbol{\psi}^* = (\boldsymbol{\beta}^{*\top}, \boldsymbol{\pi}^{*\top})^\top$ and $\xi_i \sim \chi^2(1) - 1$. Moreover, the design for $x_i$ and $\boldsymbol{\beta}^*$ is generated as in the Example 1. Additionally, the signal $\boldsymbol{\pi}^* \in \mathbb{R}^p$ is drawn from $\mathcal{N}(0, \mathbb{I}_p)$ and then normalized to have $\|\boldsymbol{\pi}^*\|_2 = 7$.

Table 8: Properties of the GRIP in Heckman Selection model

| Sparsity of the model | | Toeplitz Design $n = 200$, $p = 500$ | | | |
|---|---|---|---|---|---|
| | Type I | Power | | | |
| $s = 10$ | 0.07 | $0.47(h = 4.5)$ | $0.71(h = 5.5)$ | $0.91(h = 6.5)$ | $0.92(h = 7.5)$ |
| $s = 20$ | 0.05 | $0.50(h = 4.5)$ | $0.67(h = 5.5)$ | $0.89(h = 6.5)$ | $0.99(h = 7.5)$ |
| $s = 25$ | 0.06 | $0.38(h = 4.5)$ | $0.70(h = 5.5)$ | $0.95(h = 6.5)$ | $0.98(h = 7.5)$ |

Average size over 100 repetitions, where we see that for a model where there is not an efficient estimation scheme, GRIP performs valid inference.

Results of these two experiments are summarized in Tables 7 and 8, respectively. For these nonlinear models, our method has good size properties and its power reaches power 1 for alternatives that are far from the null hypothesis. In contrast, other methods do not apply; there even an estimation in high-dimension is not well defined.

In Section A, we prove the theoretical results in Section 4 by casting the problem into the framework of approximate bootstrap summarized in Section 6. In Section B, we provide the proof for the approximate bootstrap in high-dimensions as well as the auxiliary technical lemmas.

## A  Proofs of the main results

*Proof of Theorem 2.* In this proof, we invoke Theorem 6. We phrase the problem under the notation of Theorem 6 and verify the assumptions of Theorem 6. The exact statements depend on whether Condition 1(i) or Condition 1(ii) is imposed. Hence, we discuss both scenarios separately.

Let $\mathbf{G} = \mathbf{Y} - \mathbf{Z}\boldsymbol{\beta}_0$ and define $(g_1, ..., g_n)^\top = \mathbf{G}$. Let $s_{*,\gamma} = \|\boldsymbol{\gamma}^*\|_0$ and $s_{*,\theta} = \max_{1 \leq j \leq d} \|\boldsymbol{\theta}^*_{(j)}\|_0$. By Theorem 6 in Rudelson and Zhou (2013) and the rate condition regarding $s_{*,\gamma}$ and $s_{*,\theta}$, there exists a constant $\kappa > 0$, such that $\mathbb{P}(\mathcal{D}_n(s_{*,\theta}, \kappa)) \to 1$ under Condition 1(i) and $\mathbb{P}(\mathcal{D}_n(s_{*,\gamma}, \kappa)) \to 1$ under Condition 1(ii), where for $s \geq 1$,

$$\mathcal{D}_n(s, \kappa) = \left\{ \min_{J_0 \subseteq \{1, \cdots, p\}, |J_0| \leq s} \min_{\boldsymbol{\delta} \neq 0, \|\boldsymbol{\delta}_{J_0^c}\|_1 \leq \|\boldsymbol{\delta}_{J_0}\|_1} \frac{\|\mathbf{X}\boldsymbol{\delta}\|_2}{\sqrt{n}\|\boldsymbol{\delta}_{J_0}\|_2} > \kappa \right\}. \tag{18}$$

Define the event $\mathcal{M} = \left\{ \boldsymbol{\theta}^*_{(j)} \text{ and } \boldsymbol{\gamma}^* \text{ are feasible for (9) } \forall j \in \{1, \cdots, d\} \right\}$. By Lemma 6, $\mathbb{P}(\mathcal{M}) \to 1$. Hence,

$$\begin{cases} \mathbb{P}(\mathcal{M} \bigcap \mathcal{D}_n(s_{*,\theta}, \kappa)) \to 1 & \text{under Condition 1(i)} \\ \mathbb{P}(\mathcal{M} \bigcap \mathcal{D}_n(s_{*,\gamma}, \kappa)) \to 1 & \text{under Condition 1(ii)}, \end{cases} \tag{19}$$

In either Condition 1(i) or Condition 1(ii), we will show that $T_{n,j}$ and $\widetilde{T}_{n,j}$ (defined in (10) and (4)) can be written as

$$\begin{cases} T_{n,j} = n^{-1/2} \sum_{i=1}^n w_{i,j} q_{i,j} + \Delta_{n,j} \\ \widetilde{T}_{n,j} = n^{-1/2} \sum_{i=1}^n \xi_i (w_{i,j} \widehat{q}_{i,j} - m_{n,j}) & \text{with } m_{n,j} = n^{-1} \sum_{i=1}^n w_{i,j} \widehat{q}_{i,j}, \end{cases} \tag{20}$$

where the definitions of $w_{i,j}$'s, $q_{i,j}$'s, $\widehat{q}_{i,j}$'s and $\Delta_{n,j}$'s depend on whether Condition 1 (i) or (ii) is imposed. Under this framework, Theorem 6 implies that it suffices to check the following conditions:

(a) $w_{i,j}$'s are independent of $q_{i,j}$'s and $n^{-1} \sum_{i=1}^n w_{i,j}^2 = 1 \; \forall j \in \{1, \cdots, d\}$.

(b) $\{q_i\}_{i=1}^n \; q_i = (q_{i,1}, \cdots, q_{i,j})^\top$ be a sequence of independent random vectors in $\mathbb{R}^d$ with $\mathbb{E} q_i = 0$, $\mathbb{E} q_{i,j}^2 = 1$ and $q_{i,j}$ sub-Gaussian with uniformly bounded sub-Gaussian norm.

(c) $\left( \bar{M}_n^2 \log^7(d \vee n) \bigvee \bar{M}_n^4 \log^2(d \vee n) \right) = o_P(n)$, where $\bar{M}_n = \max_{1 \leq j \leq d, 1 \leq i \leq n} |w_{i,j}|$.

(d) there exists a sequence $\delta_n > 0$ such that $\delta_n \log d = o(1)$, $\max_{1 \leq j \leq d} n^{-1} \sum_{i=1}^n w_{i,j}^2 (\widehat{q}_{i,j} - q_{i,j})^2 = \mathcal{O}_P(\delta_n^2)$ and $\|\Delta_n\|_\infty = o_P(\delta_n^{1/2})$.

In the rest of the proof, we verify the framework in (20) with formal definitions of $w_{i,j}$'s, $q_{i,j}$'s, $\widehat{q}_{i,j}$'s and $\Delta_{n,j}$'s and then check conditions (a)-(d). This is done for both Condition 1(i) and (ii).

**Case 1:** Condition 1(i) is imposed.

In this case, we define $w_{i,j} = (g_i - x_i^\top \widehat{\gamma})/\widehat{\sigma}_\varepsilon$, $q_{i,j} = u_{i,(j)}/\sigma_{u,j}$, $\widehat{q}_{i,j} = \widehat{u}_{i,(j)}/\widehat{\sigma}_{u,j}$ and

$$\Delta_{n,j} = n^{-1/2}(\mathbf{G} - \mathbf{X}\widehat{\gamma})^\top \mathbf{X}(\boldsymbol{\theta}^*_{(j)} - \widehat{\boldsymbol{\theta}}_{(j)})/(\widehat{\sigma}_\varepsilon \widehat{\sigma}_u),$$

where

$$\widehat{u}_{i,(j)} = z_{i,(j)} - x_i^\top \widehat{\boldsymbol{\theta}}_{(j)} = u_{i,(j)} + x_i^\top (\boldsymbol{\theta}^*_{(j)} - \widehat{\boldsymbol{\theta}}_{(j)}).$$

Notice that, with these definitions, (20) is satisfied and we only need to verify conditions (a)-(d) listed above.

Notice that $w_{i,j}$ does not depend on $j$ and, by the definition of $\widehat{\sigma}_\varepsilon$, $n^{-1}\sum_{i=1}^n w_{i,j}^2 = 1$. Also notice that both $x_i$ and $\varepsilon_i$ are uncorrelated with $\mathbf{u}_i = (u_{i,(1)}, ... u_{i,(d)})^\top$. Hence, under the null hypothesis of $\boldsymbol{\beta}^* = \boldsymbol{\beta}_0$, $g_i = x_i^\top \boldsymbol{\gamma}^* + \varepsilon_i$ is uncorrelated with $\mathbf{u}_i$. Since $\widehat{\gamma}$ and $\widehat{\sigma}_\varepsilon$ are computed using only $\mathbf{G}$ and $\mathbf{X}$, (a) follows. Notice that (b) holds by the properties of $u_{i,(j)}$'s. We proceed to verify (c) and (d).

By Lemma 5, on the event $\mathcal{M} \bigcap \mathcal{D}_n(s_{*,\theta}, \kappa)$,

$$\widehat{\sigma}_\varepsilon = \sqrt{n^{-1}(\mathbf{G} - \mathbf{X}\widehat{\gamma})^\top(\mathbf{G} - \mathbf{X}\widehat{\gamma})} \geq \bar{\eta}_\gamma/\sqrt{n^{-1}\mathbf{G}^\top\mathbf{G}}.$$

Since under the null hypothesis of $\boldsymbol{\beta}^* = \boldsymbol{\beta}_0$,

$$g_i = x_i^\top \boldsymbol{\gamma}^* + \varepsilon_i,$$

we have that $\mathbb{E}g_i^2 = (\boldsymbol{\gamma}^*)^\top \boldsymbol{\Sigma} \boldsymbol{\gamma}^* + \sigma_\varepsilon^2 = \mathcal{O}(1)$ (due to $\|\boldsymbol{\gamma}^*\|_2 = \mathcal{O}(1)$). The law of large numbers implies that $n^{-1}\mathbf{G}^\top\mathbf{G} = \mathbb{E}g_i^2 + o_P(1) = \mathcal{O}_P(1)$ and hence,

$$\widehat{\sigma}_\varepsilon^{-1} \leq \sqrt{n^{-1}\mathbf{G}^\top\mathbf{G}}/\bar{\eta}_\gamma = \mathcal{O}_P(1). \tag{21}$$

Notice that, on $\mathcal{M}$, $\max_{1 \leq i \leq n} |g_i - x_i^\top \widehat{\gamma}| \leq \mu_\gamma = \mathcal{O}(\sqrt{\log(dn)})$ and thus

$$\bar{M}_n = \max_{1 \leq i \leq n} |(g_i - x_i^\top \widehat{\gamma})/\widehat{\sigma}_\varepsilon| = \mathcal{O}_P(\sqrt{\log(dn)}). \tag{22}$$

This, along with the rate conditions in statement of theorem, implies (c).

From Lemma 4 applied to $\widehat{\boldsymbol{\theta}}_{(j)}$, on the event $\mathcal{M} \bigcap \mathcal{D}_n(s_{*,\theta}, \kappa)$, we have for each $1 \leq j \leq d$,

$$\|\widehat{\boldsymbol{\theta}}_{(j)} - \boldsymbol{\theta}^*_{(j)}\|_1 \leq 8\eta_{\theta,j} s_{*,\theta} \kappa^{-2} \text{ and } n^{-1/2}\|\mathbf{X}(\widehat{\boldsymbol{\theta}}_{(j)} - \boldsymbol{\theta}^*_{(j)})\|_2 \leq 4\eta_{\theta,j}\sqrt{s_{*,\theta}}\kappa^{-1}. \tag{23}$$

Thus, on $\mathcal{M} \bigcap \mathcal{D}_n(s_{*,\theta}, \kappa)$, for each $1 \leq j \leq d$,

$$\left| n^{-1/2}(\mathbf{G} - \mathbf{X}\widehat{\gamma})^\top \mathbf{X}(\boldsymbol{\theta}^*_{(j)} - \widehat{\boldsymbol{\theta}}_{(j)}) \right| \leq n^{1/2}\|n^{-1}\mathbf{X}^\top(\mathbf{G} - \mathbf{X}\widehat{\gamma})\|_\infty \|\widehat{\boldsymbol{\theta}}_{(j)} - \boldsymbol{\theta}^*_{(j)}\|_1 \leq 8n^{1/2}\eta_{\theta,j}\eta_\gamma s_{*,\theta}\kappa^{-2} \tag{24}$$

and

$$\left| n^{-1/2}\|\mathbf{Z}_{(j)} - \mathbf{X}\widehat{\boldsymbol{\theta}}_{(j)}\|_2 - n^{-1/2}\|\mathbf{Z}_{(j)} - \mathbf{X}\boldsymbol{\theta}^*_{(j)}\|_2 \right| \leq n^{-1/2}\|\mathbf{X}(\widehat{\boldsymbol{\theta}}_{(j)} - \boldsymbol{\theta}^*_{(j)})\|_2 \leq 4\eta_{\theta,j}\sqrt{s_{*,\theta}}\kappa^{-1}. \tag{25}$$

Recall $\mathbf{u}_{(j)} = (u_{1,(j)}, \cdots, u_{n,(j)})^\top \in \mathbb{R}^n$. Since $\mathbf{u}_{(j)} = \mathbf{Z}_{(j)} - \mathbf{X}\boldsymbol{\theta}^*_{(j)}$ has entries with uniformly bounded sub-Gaussian norms, it follows, by Bernstein's inequality and the union bound, that

$$\max_{1 \leq j \leq d} \left| n^{-1} \|\mathbf{Z}_{(j)} - \mathbf{X}\boldsymbol{\theta}^*_{(j)}\|_2^2 - \sigma_{u,j}^2 \right| = \mathcal{O}_P(\sqrt{n^{-1} \log d}). \tag{26}$$

From (25) and (26), we have

$$\max_{1 \leq j \leq d} |\widehat{\sigma}_{u,j} - \sigma_{u,j}| = \max_{1 \leq j \leq d} \left| n^{-1/2} \|\mathbf{Z}_{(j)} - \mathbf{X}\widehat{\boldsymbol{\theta}}_{(j)}\|_2 - \sigma_{u,j} \right| \tag{27}$$

$$= \mathcal{O}_P(\sqrt{n^{-1} \log d} \vee (\max_{1 \leq j \leq d} \eta_{\theta,j} \sqrt{s_{*,\theta}})) = o_P(1). \tag{28}$$

This, combined with (24) and (21), implies that

$$\|\Delta_n\|_\infty = \max_{1 \leq j \leq d} \left| \frac{n^{-1/2}(\mathbf{G} - \mathbf{X}\widehat{\boldsymbol{\gamma}})^\top \mathbf{X}(\boldsymbol{\theta}^*_{(j)} - \widehat{\boldsymbol{\theta}}_{(j)})}{\widehat{\sigma}_\varepsilon \widehat{\sigma}_{u,j}} \right| \leq \mathcal{O}_P(n^{1/2} \eta_\gamma \max_{1 \leq j \leq d} \eta_{\theta,j} s_{*,\theta}) \tag{29}$$

Moreover,

$$\max_{1 \leq j \leq d} n^{-1} \sum_{i=1}^n w_{i,j}^2 (\widehat{q}_{i,j} - q_{i,j})^2$$

$$\overset{(i)}{\leq} \bar{M}_n^2 \max_{1 \leq j \leq d} n^{-1} \sum_{i=1}^n \left( \frac{\widehat{u}_{i,(j)}}{\widehat{\sigma}_{u,j}} - \frac{u_{i,(j)}}{\sigma_{u,j}} \right)^2$$

$$\overset{(ii)}{\leq} \bar{M}_n^2 \max_{1 \leq j \leq d} 2n^{-1} \sum_{i=1}^n \left( \frac{\widehat{u}_{i,(j)}}{\widehat{\sigma}_{u,j}} - \frac{\widehat{u}_{i,(j)}}{\sigma_{u,j}} \right)^2 + \bar{M}_n^2 \max_{1 \leq j \leq d} 2n^{-1} \sum_{i=1}^n \left( \frac{\widehat{u}_{i,(j)}}{\sigma_{u,j}} - \frac{u_{i,j}}{\sigma_{u,j}} \right)^2$$

$$\overset{(iii)}{=} \bar{M}_n^2 \max_{1 \leq j \leq d} 2(\widehat{\sigma}_{u,j} - \sigma_{u,j})^2 \sigma_{u,j}^{-2} + \bar{M}_n^2 \max_{1 \leq j \leq d} 2n^{-1} \|\mathbf{X}(\widehat{\boldsymbol{\theta}}_{(j)} - \boldsymbol{\theta}^*_{(j)})\|_2^2 \sigma_{u,j}^{-2}$$

$$\overset{(iv)}{=} \mathcal{O}_P((n^{-1} \log d) \vee (\max_{1 \leq j \leq d} \eta_{\theta,j}^2 s_{*,\theta})) \log(dn), \tag{30}$$

where $(i)$ holds from the Holder's inequality, $(ii)$ follows from the inequality $(a+b)^2 \leq 2a^2 + 2b^2$, $(iii)$ from $\widehat{\sigma}_{u,j}^2 = n^{-1} \sum_{i=1}^n \widehat{u}_{i,(j)}^2$ and

$$n^{-1} \sum_{i=1}^n (\widehat{u}_{i,(j)} - u_{i,(j)})^2 = n^{-1} \|\mathbf{X}(\widehat{\boldsymbol{\theta}}_{(j)} - \boldsymbol{\theta}^*_{(j)})\|_2^2$$

and $(iv)$ follows from (23), (28) and (22).

Based on (29), (30) and the rate conditions in the statement of the theorem, one can easily verify that (d) holds with

$$\delta_n = \sqrt{n^{-1} \log(d \vee n)[(\log d) \vee (s_{*,\theta} \log(p - d))]}.$$

This completes the proof for Case 1.

<u>**Case 2:**</u> Condition 1(ii) is imposed.

In this case, we define $w_{i,j} = (z_{i,(j)} - x_i^\top \widehat{\boldsymbol{\theta}}_{(j)})/\widehat{\sigma}_{u,j}$, $q_{i,j} = \varepsilon_i/\sigma_\varepsilon$, $\widehat{q}_{i,j} = \widehat{\varepsilon}_i/\widehat{\sigma}_\varepsilon$ and

$$\Delta_{n,j} = n^{-1/2}(\boldsymbol{\gamma} - \widehat{\boldsymbol{\gamma}})^\top \mathbf{X}^\top (\mathbf{Z}_{(j)} - \mathbf{X}\widehat{\boldsymbol{\theta}}_{(j)})/(\widehat{\sigma}_\varepsilon \widehat{\sigma}_u),$$

where $\widehat{\varepsilon}_i = v_i - x_i^\top \widehat{\boldsymbol{\gamma}}$. Notice that, with these definitions, (20) is satisfied and we only need to verify conditions (a)-(d) listed after (20).

Notice that, from the definition of $\widehat{\sigma}_{u,j}$, $n^{-1} \sum_{i=1}^n w_{i,j}^2 = 1 \; \forall j$. Since $\boldsymbol{\varepsilon}$ is uncorrelated with $(\{\mathbf{Z}_{(j)}\}_{j=1}^d, \mathbf{X})$ and $\{(\widehat{\boldsymbol{\theta}}_{(j)}, \widehat{\sigma}_{u,j})\}_{j=1}^d$ is computed using only $(\{\mathbf{Z}_{(j)}\}_{j=1}^d, \mathbf{X})$, (a) follows.

Notice that $q_{i,j}$ and $\widehat{q}_{i,j}$ do not depend on $j$ and (b) holds by the assumptions on $\varepsilon_i$. It remains to verify (c) and (d).

Since $z_{i,(j)}$ is centered Gaussian with uniformly bounded variance, the sub-exponential norm of $z_{i,(j)}^2$ is uniformly bounded. From Proposition 5.16 of Vershynin (2010) and the union bound,

$$\max_{1 \leq j \leq d} |n^{-1} \mathbf{Z}_{(j)}^\top \mathbf{Z}_{(j)} - \mathbb{E} z_{i,(j)}^2| = \mathcal{O}_P(\sqrt{n^{-1} \log d}) = o_P(1)$$

and thus $\max_{1 \leq j \leq d} n^{-1} \mathbf{Z}_{(j)}^\top \mathbf{Z}_{(j)} = \mathcal{O}_P(1)$. Therefore, on the event $\mathcal{M}$,

$$\max_{1 \leq j \leq d} \widehat{\sigma}_{u,j}^{-1} = \max_{1 \leq j \leq d} \left(n^{-1/2} \|\mathbf{Z}_{(j)} - \mathbf{X}\widehat{\boldsymbol{\theta}}_{(j)}\|_2\right)^{-1}$$

$$\overset{(i)}{\leq} \max_{1 \leq j \leq d} \left(n^{-1/2} \|\mathbf{Z}_{(j)}\|_2 / \bar{\eta}_{\theta,j}\right) \tag{31}$$

$$\leq \sqrt{\max_{1 \leq j \leq d} n^{-1} \mathbf{Z}_{(j)}^\top \mathbf{Z}_{(j)}} / (\min_{1 \leq j \leq} \bar{\eta}_{\theta,j}) = \mathcal{O}_P(1), \tag{32}$$

where $(i)$ follows by Lemma 5 (applied to $\widehat{\boldsymbol{\theta}}_{(j)}$ for each $j$). Notice that, on $\mathcal{M}$,

$$\max_{1 \leq i \leq n, 1 \leq j \leq d} |z_{i,(j)} - x_i^\top \widehat{\boldsymbol{\theta}}_{(j)}| \leq \max_{1 \leq j \leq d} \mu_{\theta,j} = \mathcal{O}(\sqrt{\log(dn)})$$

and thus, by (32),

$$\bar{M}_n = \max_{1 \leq i \leq n, 1 \leq j \leq d} |(z_{i,(j)} - x_i^\top \widehat{\boldsymbol{\theta}}_{(j)})/\widehat{\sigma}_{u,j}| = \mathcal{O}_P(\sqrt{\log(dn)}). \tag{33}$$

This, along with the rate conditions in statement of theorem, implies (c).

From Lemma 4 applied to $\boldsymbol{\gamma}^*$, on the event $\mathcal{M} \bigcap \mathcal{D}_n(s_{*,\gamma}, \kappa)$, we have

$$\|\widehat{\boldsymbol{\gamma}} - \boldsymbol{\gamma}^*\|_1 \leq 8\eta s_{*,\gamma} \kappa^{-2} \text{ and } n^{-1/2}\|\mathbf{X}(\widehat{\boldsymbol{\gamma}} - \boldsymbol{\gamma}^*)\|_2 \leq 4\eta_\gamma \sqrt{s_{*,\gamma}} \kappa^{-1}. \tag{34}$$

Thus, on $\mathcal{M} \bigcap \mathcal{D}_n(s_{*,\gamma}, \kappa)$,

$$\left| n^{-1/2}(\boldsymbol{\gamma}^* - \widehat{\boldsymbol{\gamma}})^\top \mathbf{X}^\top (\mathbf{Z}_{(j)} - \mathbf{X}\widehat{\boldsymbol{\theta}}_{(j)}) \right| \leq n^{1/2} \|n^{-1}\mathbf{X}^\top(\mathbf{Z}_{(j)} - \mathbf{X}\widehat{\boldsymbol{\theta}}_{(j)})\|_\infty \|\widehat{\boldsymbol{\gamma}} - \boldsymbol{\gamma}^*\|_1 \leq 8n^{1/2} \eta_{\theta,j} \eta_\gamma s_{*,\gamma} \kappa^{-2} \tag{35}$$

and

$$\left| n^{-1/2} \|\mathbf{G} - \mathbf{X}\widehat{\boldsymbol{\gamma}}\|_2 - n^{-1/2} \|\mathbf{G} - \mathbf{X}\boldsymbol{\gamma}^*\|_2 \right| \leq n^{-1/2} \|\mathbf{X}(\widehat{\boldsymbol{\gamma}} - \boldsymbol{\gamma}^*)\|_2 \leq 4\eta_\gamma \sqrt{s_{*,\gamma}} \kappa^{-1}. \tag{36}$$

Since, under the null hypothesis of $\boldsymbol{\beta}^* = \boldsymbol{\beta}_0$, $\boldsymbol{\varepsilon} = \mathbf{G} - \mathbf{X}\boldsymbol{\gamma}^*$, we apply (Lyapunov's) CLT and obtain $|n^{-1}\|\mathbf{G} - \mathbf{X}\boldsymbol{\gamma}^*\|_2^2 - \sigma_\varepsilon^2| = \mathcal{O}_P(n^{-1/2})$. Hence, by (36),

$$|\widehat{\sigma}_\varepsilon - \sigma_\varepsilon| = \mathcal{O}_P(n^{-1/2} \vee \eta_\gamma \sqrt{s_{*,\gamma}}) = o_P(1). \tag{37}$$

This, combined with (32) and (35), implies that

$$\|\Delta_n\|_\infty = \max_{1 \leq j \leq d} \left| \frac{n^{-1/2}(\boldsymbol{\gamma}^* - \widehat{\boldsymbol{\gamma}})^\top \mathbf{X}^\top (\mathbf{Z}_{(j)} - \mathbf{X}\widehat{\boldsymbol{\theta}}_{(j)})}{\widehat{\sigma}_\varepsilon \widehat{\sigma}_{u,j}} \right| \leq \mathcal{O}_P(n^{1/2} \eta_\gamma s_{*,\gamma} \max_{1 \leq j \leq d} \eta_{\theta,j}) \tag{38}$$

Moreover,

$$\max_{1 \leq j \leq d} n^{-1} \sum_{i=1}^n w_{i,j}^2 (\widehat{q}_{i,j} - q_{i,j})^2 \stackrel{(i)}{\leq} \bar{M}_n^2 n^{-1} \sum_{i=1}^n \left( \frac{\widehat{\varepsilon}_i}{\widehat{\sigma}_\varepsilon} - \frac{\varepsilon_i}{\sigma_\varepsilon} \right)^2$$

$$\stackrel{(ii)}{\leq} \bar{M}_n^2 2 n^{-1} \sum_{i=1}^n \left( \frac{\widehat{\varepsilon}_i}{\widehat{\sigma}_\varepsilon} - \frac{\widehat{\varepsilon}_i}{\sigma_\varepsilon} \right)^2 + \bar{M}_n^2 2 n^{-1} \sum_{i=1}^n \left( \frac{\widehat{\varepsilon}_i}{\sigma_\varepsilon} - \frac{\varepsilon_i}{\sigma_\varepsilon} \right)^2$$

$$\stackrel{(iii)}{=} \bar{M}_n^2 2(\widehat{\sigma}_\varepsilon - \sigma_\varepsilon)^2 \sigma_\varepsilon^{-2} + \bar{M}_n^2 2 n^{-1} \|\mathbf{X}(\widehat{\boldsymbol{\gamma}} - \boldsymbol{\gamma}^*)\|_2^2 \sigma_\varepsilon^{-2}$$

$$\stackrel{(iv)}{=} \mathcal{O}_P(n^{-1} \log(dn) [(s_{*,\gamma} \log(p-d)) \vee 1]), \tag{39}$$

where $(i)$ holds by Holder's inequality, $(ii)$ follows by the inequality $(a+b)^2 \leq 2a^2 + 2b^2$, $(iii)$ follows by $\widehat{\sigma}_\varepsilon^2 = n^{-1} \sum_{i=1}^n \widehat{\varepsilon}_i^2$ and

$$n^{-1} \sum_{i=1}^n (\widehat{\varepsilon}_i - \varepsilon_i)^2 = n^{-1} \|\mathbf{X}(\widehat{\boldsymbol{\gamma}} - \boldsymbol{\gamma}^*)\|_2^2$$

and $(iv)$ follows by (37), (34) and (33).

Based on (38), (39) and the rate conditions in the statement of the theorem, one can easily verify that (d) holds with

$$\delta_n = \sqrt{n^{-1} \log(dn) [(s_{*,\gamma} \log(p-d)) \vee 1]}.$$

This completes the proof for Case 2. $\square$

*Proof of Theorem 3.* For the majority of the proof, we lower bound the test statistic under the alternative. An upper bound of the critical value is provided by Lemma 10, which is proved later. With these two bounds, we prove the theorem.

Let $j^* \in \{1, \cdots, d\}$ satisfy $|\mathbb{E}\boldsymbol{\pi}^\top \mathbf{u}_i^\top u_{i,(j^*)}| = \max_{1 \leq j \leq d} |\mathbb{E}\boldsymbol{\pi}^\top \mathbf{u}_i^\top u_{i,(j)}| = \|\boldsymbol{\Sigma}_u \boldsymbol{\pi}\|_\infty$. Notice that

$$n^{-1/2}(\mathbf{G} - \mathbf{X}\widehat{\boldsymbol{\gamma}})^\top (\mathbf{Z}_{(j^*)} - \mathbf{X}\widehat{\boldsymbol{\theta}}_{(j^*)}) = \underbrace{n^{-1/2}(\mathbf{G} - \mathbf{X}\boldsymbol{\gamma}^*)^\top (\mathbf{Z}_{(j^*)} - \mathbf{X}\boldsymbol{\theta}^*_{(j^*)})}_{J_1}$$

$$+ \underbrace{n^{-1/2}(\mathbf{G} - \mathbf{X}\boldsymbol{\gamma}^*)^\top \mathbf{X}(\boldsymbol{\theta}^*_{(j^*)} - \widehat{\boldsymbol{\theta}}_{(j^*)})}_{J_2} + \underbrace{n^{-1/2}(\boldsymbol{\gamma}^* - \widehat{\boldsymbol{\gamma}})^\top \mathbf{X}^\top (\mathbf{Z}_{(j^*)} - \mathbf{X}\widehat{\boldsymbol{\theta}}_{(j^*)})}_{J_3}. \tag{40}$$

Moreover, under the alternative hypothesis of $\boldsymbol{\beta}^* = \boldsymbol{\beta}_0 + \boldsymbol{\pi}$, we have

$$(g_i - x_i^\top \boldsymbol{\gamma}^*)(z_{i,(j^*)} - x_i^\top \boldsymbol{\theta}^*_{(j^*)}) = \underbrace{(\varepsilon_i + x_i^\top \boldsymbol{\Theta}^* \boldsymbol{\pi} + \mathbf{u}_i^\top \boldsymbol{\pi})}_{J_{1,i}} u_{i,(j^*)}.$$

Since $\boldsymbol{\Sigma}_W$ has eigenvalues bounded away from zero and infinity, $\boldsymbol{\Sigma}_X$, $\boldsymbol{\Sigma}_u$ and $\boldsymbol{\Theta}$ also have singular values bounded away from zero and infinity. By $\|\boldsymbol{\pi}\|_2 = \mathcal{O}(1)$ and the bounded sub-Gaussian norms of $x_i$ and $\mathbf{u}_i$, $x_i^\top \boldsymbol{\Theta}^* \boldsymbol{\pi}$ and $\mathbf{u}_i^\top \boldsymbol{\pi}$ have sub-Gaussian norms bounded by some constant $K_1 > 0$. By the bounded sub-Gaussian norm of $\varepsilon_i$, we have that the sub-Gaussian norm of $J_{1,i}$ is bounded by some constant $K_2 > 0$. Using the sub-Gaussian properties of $u_{i,(j^*)}$ and Lemma 8, we have that the sub-exponential norm of $l_{n,i} := J_{1,i} u_{i,(j^*)}$ is bounded by some constant $K_3 > 0$. It follows by Proposition 5.16 of Vershynin (2010) that there exist constants $C_1, C_2 > 0$ such that $\forall x > 0$

$$\mathbb{P}\left(\left|n^{-1/2} \sum_{i=1}^n (l_{n,i} - \mathbb{E}l_{n,i})\right| > x\right) \leq 2\exp(-\min\{C_1 x^2, \sqrt{n} C_2 x\}). \tag{41}$$

Since $J_1 = n^{-1/2} \sum_{i=1}^n l_{n,i}$ and

$$|\mathbb{E}l_{n,i}| = |\mathbb{E}(\varepsilon_i + x_i^\top \boldsymbol{\Theta}^* \boldsymbol{\pi} + \mathbf{u}_i^\top \boldsymbol{\pi}) u_{i,(j^*)}| = \|\boldsymbol{\Sigma}_u \boldsymbol{\pi}\|_\infty,$$

it follows that $\forall x > 0$,

$$\begin{aligned}
\mathbb{P}\left(|J_1| > x\sqrt{\log d}\right) &\geq \mathbb{P}\left(\left|n^{1/2} \mathbb{E}l_{n,i}\right| - \left|n^{-1/2} \sum_{i=1}^n (l_{n,i} - \mathbb{E}l_{n,i})\right| > x\sqrt{\log d}\right) \\
&\geq \mathbb{I}\left\{\|\boldsymbol{\Sigma}_u \boldsymbol{\pi}\|_\infty > (x+1)\sqrt{n^{-1}\log d}\right\} - \mathbb{P}\left(\left|n^{-1/2} \sum_{i=1}^n (l_{n,i} - \mathbb{E}l_{n,i})\right| > \sqrt{n^{-1}\log d}\right) \\
&\stackrel{(i)}{\geq} \mathbb{I}\left\{\|\boldsymbol{\Sigma}_u \boldsymbol{\pi}\|_\infty > (x+1)\sqrt{n^{-1}\log d}\right\} - 2\exp(-\min\{C_1 n^{-1}\log d, \sqrt{n} C_2 \sqrt{n^{-1}\log d}\}) \\
&\geq \mathbb{I}\left\{\|\boldsymbol{\Sigma}_u \boldsymbol{\pi}\|_\infty > (x+1)\sqrt{n^{-1}\log d}\right\} - o(1),
\end{aligned}$$

where $(i)$ follows by (41). Then, we have that $\forall t_1, t_2 > 0$,

$$\begin{aligned}
\mathbb{P}\left(\|\mathbf{T}_n\|_\infty > t_1\sqrt{\log d}\right) &\geq \mathbb{P}\left(|T_{n,j^*}| > t_1\sqrt{\log d}\right) \\
&= \mathbb{P}\left(\widehat{\sigma}_\varepsilon^{-1} \widehat{\sigma}_{u,j^*}^{-1} |J_1 + J_2 + J_3| > t_1\sqrt{\log d}\right) \\
&\geq \mathbb{P}\left(\sigma_\varepsilon^{-1} \sigma_{u,j^*}^{-1} |J_1 + J_2 + J_3| > t_1 t_2 \sqrt{\log d}\right) - \mathbb{P}\left(\sigma_\varepsilon^{-1} \sigma_{u,j^*}^{-1} \widehat{\sigma}_\varepsilon \widehat{\sigma}_{u,j^*} > t_2\right) \\
&\geq \mathbb{P}\left(|J_1| > 5\sigma_\varepsilon \sigma_{u,j^*} t_1 t_2 \sqrt{\log d}\right) - \mathbb{P}\left(|J_2| > \sigma_\varepsilon \sigma_{u,j^*} t_1 t_2 \sqrt{\log d}\right) \\
&\quad - \mathbb{P}\left(|J_3| > \sigma_\varepsilon \sigma_{u,j^*} t_1 t_2 \sqrt{\log d}\right) - \mathbb{P}\left(\sigma_\varepsilon^{-1} \sigma_{u,j^*}^{-1} \widehat{\sigma}_\varepsilon \widehat{\sigma}_{u,j^*} > t_2\right) \\
&\geq \mathbb{I}\left\{\|\boldsymbol{\Sigma}_u \boldsymbol{\pi}\|_\infty > (5\sigma_\varepsilon \sigma_{u,j^*} t_1 t_2 + 1)\sqrt{n^{-1}\log d}\right\} - o(1) \\
&\quad - \mathbb{P}\left(|J_2| > \sigma_\varepsilon \sigma_{u,j^*} t_1 t_2 \sqrt{\log d}\right) - \mathbb{P}\left(|J_3| > \sigma_\varepsilon \sigma_{u,j^*} t_1 t_2 \sqrt{\log d}\right)
\end{aligned}$$

$$-\mathbb{P}\left(\sigma_\varepsilon^{-1}\sigma_{u,j^*}^{-1}\widehat{\sigma}_\varepsilon\widehat{\sigma}_{u,j^*} > t_2\right). \tag{42}$$

Define the event $\mathcal{M} = \left\{\boldsymbol{\theta}_{(j)}^* \text{ and } \boldsymbol{\gamma}^* \text{ are feasible for (9) } \forall j \in \{1,\cdots,d\}\right\}$. By Lemma 6, $\mathbb{P}(\mathcal{M}) \to 1$. By the same argument as in the beginning of the proof of Theorem 2, there exists a constant $\kappa > 0$ such that $\mathbb{P}(\mathcal{D}_n(s_*, \kappa)) \to 1$, where $s_{*,\gamma}$, $s_{*,\theta}$ and $\mathcal{D}_n(s,\kappa)$ are defined in the proof of Theorem 2 and $s_* = s_{*,\theta} \vee s_{*,\gamma}$. By Lemma 4 (applied to $\boldsymbol{\gamma}^*$ and $\boldsymbol{\theta}_{(j)}^*$), on $\mathcal{M} \bigcap \mathcal{D}_n(s_*, \kappa)$, we have

$$\|\widehat{\boldsymbol{\theta}}_{(j)} - \boldsymbol{\theta}_{(j)}^*\|_1 \le 8\eta_{\theta,j}s_{*,\theta}\kappa^{-2} \text{ and } n^{-1/2}\|X(\widehat{\boldsymbol{\theta}}_{(j)} - \boldsymbol{\theta}_{(j)}^*)\|_2 \le 4\eta_{\theta,j}\sqrt{s_{*,\theta}}\kappa^{-1} \tag{43}$$

$$\|\widehat{\boldsymbol{\gamma}}_{(j)} - \boldsymbol{\gamma}_{(j)}^*\|_1 \le 8\eta_\gamma s_{*,\gamma}\kappa^{-2} \text{ and } n^{-1/2}\|X(\widehat{\boldsymbol{\gamma}}_{(j)} - \boldsymbol{\gamma}_{(j)}^*)\|_2 \le 4\eta_\gamma\sqrt{s_{*,\gamma}}\kappa^{-1} \tag{44}$$

Hence, on the event $\mathcal{M} \bigcap \mathcal{D}_n(s_*, \kappa)$, we have

$$\begin{cases} |J_2| \le n^{-1/2}\|\mathbf{X}^\top(\mathbf{G} - \mathbf{X}\boldsymbol{\gamma}^*)\|_\infty \|\widehat{\boldsymbol{\theta}}_{(j^*)} - \boldsymbol{\theta}_{(j^*)}^*\|_1 \le 8\eta_\gamma \max_{1\le j \le d}\eta_{\theta,j}s_{*,\theta}\kappa^{-2} = o_P(1) \\ |J_3| \le n^{-1/2}\|\widehat{\boldsymbol{\gamma}} - \boldsymbol{\gamma}^*\|_1 \|\mathbf{X}^\top(\mathbf{Z}_{(j^*)} - \mathbf{X}\widehat{\boldsymbol{\theta}}_{(j^*)})\|_\infty \le 8\max_{1\le j \le d}\eta_{\theta,j}\eta_\gamma s_{*,\gamma}\kappa^{-2} = o_P(1) \end{cases} \tag{45}$$

Notice that $\bar{\sigma}_\varepsilon^2 := \mathbb{E}J_{1,i}^2 \in (K_4, K_5)$ for some constants $K_4, K_5 > 0$. One can use (43) and (44) and replicate the arguments for (28) and (37) in the proof of Theorem 2, obtaining

$$\max_{1\le j \le d} |\widehat{\sigma}_{u,j} - \sigma_{u,j}| = o_P(1) \text{ and } \widehat{\sigma}_\varepsilon = \bar{\sigma}_\varepsilon + o_P(1). \tag{46}$$

By Lemma 10, there exists a constant $C_* > 0$, such that

$$\mathbb{P}\left(\mathcal{Q}(1-\alpha, \|\widetilde{\mathbf{T}}_n\|_\infty) > 3\sqrt{2C_* \log d}\right) \to 0. \tag{47}$$

Let $K_6 > 1$ be a constant such that $\sigma_\varepsilon^{-1}\sigma_{u,j^*}^{-1}\bar{\sigma}_\varepsilon\sigma_{u,j^*} \le K_6$. Now we choose constants $t_1 = 4\sqrt{2C_*}$ and $t_2 = 2K_6$. By (42), (45) and (46) we obtain

$$\mathbb{P}\left(\|\mathbf{T}_n\|_\infty > 4\sqrt{2C_* \log d}\right) \ge \mathbb{1}\left\{\|\boldsymbol{\Sigma}_u\boldsymbol{\pi}\|_\infty > (40\sqrt{2C_*}K_6\sigma_\varepsilon\sigma_{u,j^*} + 1)\sqrt{n^{-1}\log d}\right\} - o(1).$$

This, combined with (47), implies that

$$\mathbb{P}\left(\|\mathbf{T}_n\|_\infty > \mathcal{Q}(1-\alpha, \|\widetilde{\mathbf{T}}_n\|_\infty)\right) \ge \mathbb{1}\left\{\|\boldsymbol{\Sigma}_u\boldsymbol{\pi}\|_\infty > (40\sqrt{2C_*}K_6\sigma_\varepsilon\sigma_{u,j^*} + 1)\sqrt{n^{-1}\log d}\right\} - o(1).$$

Let $K_7 > 0$ be a constant with $\sigma_\varepsilon \vee \sigma_{u,j^*} < K_7$. Then we can simply take $\bar{\tau} = 40K_6K_7^2\sqrt{2C_*}+1$. This completes the proof of the theorem. $\square$

*Proof of Theorem 4.* The basic idea of our proof is the following. We construct several alternative hypotheses in $\mathcal{B}_1(\underline{\tau})$ for some $\underline{\tau} > 0$ and show that the average power of any given test against these alternatives is at most equal to the nominal size. Hence, no test can have power uniformly against the alternatives in $\mathcal{B}_1(\underline{\tau})$.

In this proof, we write $\phi_n$ instead of $\phi_n(\mathbf{X}, \mathbf{Z}, \mathbf{Y})$ for notational simplicity. Let $\underline{\tau} > 0$ be a constant such that $K^2/2 > \underline{\tau}$, where $K > 0$ is an lower bound for $\sigma_\varepsilon$ and $\min_{1\le j \le d}\sigma_{u,j}$. Let $\boldsymbol{\beta}_{(j)} = \boldsymbol{\beta}_0 + c_j e_j$, where $e_j$ is the $j$th column of the $d \times d$ identity matrix and

$$c_j = 2^{-1/2}\sigma_{u,j}^{-1}\sigma_\varepsilon\sqrt{n^{-1}\log d}.$$

Consider $\boldsymbol{\gamma} = 0$, $\boldsymbol{\Theta} = 0$ and $\boldsymbol{\Sigma}_u = \text{diag}\{\sigma_{u,1}^2, \cdots, \sigma_{u,d}^2\}$. Define

$$\lambda_0 := (\boldsymbol{\beta}_0, \boldsymbol{\gamma}, \boldsymbol{\Theta}, \boldsymbol{\Sigma}, \boldsymbol{\Sigma}_u, \sigma_\varepsilon^2) \in \mathcal{B}_0 \quad \text{and} \quad \lambda_{(j)} := (\boldsymbol{\beta}_{(j)}, \boldsymbol{\gamma}, \boldsymbol{\Theta}, \boldsymbol{\Sigma}, \boldsymbol{\Sigma}_u, \sigma_\varepsilon^2) \in \mathcal{B}_1(\underline{\tau}) \, \forall j \in \{1, \cdots, d\}.$$

Let $\mathbb{E}_j$ and $\mathbb{P}_j$ denote the expectation and probability measure under $\lambda_{(j)}$, respectively. Similarly, let $\mathbb{E}_0$ and $\mathbb{P}_0$ denote the expectation and probability measure under $\lambda_0$, respectively. Since $\limsup_{n \to \infty} \mathbb{E}_0 \phi_n \leq \alpha$, we have

$$\liminf_{n \to \infty} \inf_{\lambda \in \mathcal{B}_1(\underline{\tau})} \mathbb{E}_\lambda \phi_n - \alpha$$

$$\leq \liminf_{n \to \infty} \left( d^{-1} \sum_{j=1}^d \mathbb{E}_j \phi_n - \mathbb{E}_0 \phi_n \right) \leq \limsup_{n \to \infty} \left| d^{-1} \sum_{j=1}^d (\mathbb{E}_j \phi_n - \mathbb{E}_0 \phi_n) \right|. \quad (48)$$

From the normality assumption, we have $(x_i^\top, \mathbf{u}_i^\top, \varepsilon_i)^\top \sim \mathcal{N}(0, \boldsymbol{\Omega}_*)$ with $\boldsymbol{\Omega}_* = \text{block diag}\{\boldsymbol{\Sigma}, \boldsymbol{\Sigma}_u, \sigma_\varepsilon^2\}$. Let $\{h_i\}_{i=1}^n$ denote the observed variables, where $h_i = (x_i^\top, z_i^\top, y_i)^\top \in \mathbb{R}^p$. Then, under $\mathbb{P}_0$, $h_i \sim \mathcal{N}(0, \boldsymbol{\Omega}_0)$ with p.d.f

$$(2\pi)^{-k/2} (\det \boldsymbol{\Omega}_0)^{-1/2} \exp(-h_i^\top \boldsymbol{\Omega}_0^{-1} h_i / 2),$$

where

$$\boldsymbol{\Omega}_0 = \begin{pmatrix} \boldsymbol{\Sigma} & 0 & 0 \\ 0 & \boldsymbol{\Sigma}_u & \boldsymbol{\Sigma}_u \boldsymbol{\beta}_0 \\ 0 & \boldsymbol{\beta}_0^\top \boldsymbol{\Sigma}_u & \boldsymbol{\beta}_0^\top \boldsymbol{\Sigma}_u \boldsymbol{\beta}_0 + \sigma_\varepsilon^2 \end{pmatrix}.$$

Moreover, under $\mathbb{P}_j$, $h_i \sim \mathcal{N}(0, \boldsymbol{\Omega}_j)$ with p.d.f

$$(2\pi)^{-k/2} (\det \boldsymbol{\Omega}_j)^{-1/2} \exp(-h_i^\top \boldsymbol{\Omega}_j^{-1} h_i / 2),$$

where

$$\boldsymbol{\Omega}_j = \begin{pmatrix} \boldsymbol{\Sigma} & 0 & 0 \\ 0 & \boldsymbol{\Sigma}_u & \boldsymbol{\Sigma}_u \boldsymbol{\beta}_{(j)} \\ 0 & \boldsymbol{\beta}_{(j)}^\top \boldsymbol{\Sigma}_u & \boldsymbol{\beta}_{(j)}^\top \boldsymbol{\Sigma}_u \boldsymbol{\beta}_{(j)} + \sigma_\varepsilon^2 \end{pmatrix}.$$

Hence, after straight-forward computations, we have that

$$\mathbb{E}_j \phi_n = \mathbb{E}_0 \frac{d\mathbb{P}_j}{d\mathbb{P}_0} \phi_n = \mathbb{E}_0 \phi_n \exp(S_j),$$

where

$$S_j := \sigma_\varepsilon^{-2} \sum_{i=1}^n \left[ c_j \varepsilon_i u_{i,(j)} - \frac{1}{2} c_j^2 u_{i,(j)}^2 \right].$$

Since $|\phi_n| \leq 1$ a.s., we have, by Lyapunov's inequality, that

$$\left| d^{-1} \sum_{j=1}^d (\mathbb{E}_j \phi_n - \mathbb{E}_0 \phi_n) \right| \leq \mathbb{E}_0 \left| d^{-1} \sum_{j=1}^d \exp(S_j) - 1 \right| \leq \sqrt{\mathbb{E}_0 \left( d^{-1} \sum_{j=1}^d \exp(S_j) - 1 \right)^2}. \quad (49)$$

Recall $\mathbf{u}_{(j)} = (u_{1,(j)}, \cdots, u_{n,(j)})^\top \in \mathbb{R}^n$ and notice that, conditional on $\mathbf{u} = (\mathbf{u}_{(1)}, \cdots, \mathbf{u}_{(d)}) \in \mathbb{R}^{n \times d}$,

$$S_j \sim \mathcal{N}\left(-\frac{1}{2}\sigma_\varepsilon^{-2} c_j^2 \sum_{i=1}^n u_{i,(j)}^2, \sigma_\varepsilon^{-2} c_j^2 \sum_{i=1}^n u_{i,(j)}^2\right) \qquad \forall j \in \{1, \cdots, d\}$$

$$S_{j_1} + S_{j_2} \sim \mathcal{N}\left(-\frac{1}{2}\sigma_\varepsilon^{-2} \sum_{i=1}^n (c_{j_1}^2 u_{i,(j_1)}^2 + c_{j_2}^2 u_{i,(j_2)}^2), \sigma_\varepsilon^{-2} \sum_{i=1}^n (c_{j_1} u_{i,(j_1)} + c_{j_2} u_{i,(j_2)})^2\right) \qquad \text{for } j_1 \neq j_2$$

Recall the moment generating functions of Gaussian distributions: for $X \sim \mathcal{N}(\mu, \sigma^2)$ and $t \in \mathbb{R}$, $\mathbb{E}\exp(tX) = \exp(t\mu + t^2\sigma^2/2)$. Hence,

$$\begin{cases} \mathbb{E}_0\left(\exp(S_j) \mid \mathbf{u}\right) = 1 & \forall j \in \{1, \cdots, d\} \\ \mathbb{E}_0\left(\exp(2S_j) \mid \mathbf{u}\right) = \exp(\sigma_\varepsilon^{-2} c_j^2 \sum_{i=1}^n u_{i,(j)}^2) & \forall j \in \{1, \cdots, d\} \\ \mathbb{E}_0\left(\exp(S_{j_1} + S_{j_2}) \mid \mathbf{u}\right) = \exp(\sigma_\varepsilon^{-2} c_{j_1} c_{j_2} \sum_{i=1}^n u_{i,(j_1)} u_{i,(j_2)}) & \text{for } j_1 \neq j_2 \end{cases} \qquad (50)$$

It follows that

$$\mathbb{E}_0\left[\left(d^{-1} \sum_{j=1}^d (\exp(S_j) - 1)\right)^2 \bigg| \mathbf{u}\right]$$

$$= \mathbb{E}_0\left[d^{-2} \sum_{j_1=1}^d \sum_{j_2=1}^d (\exp(S_{j_1}) - 1)(\exp(S_{j_2}) - 1) \bigg| \mathbf{u}\right]$$

$$= \mathbb{E}_0\left[d^{-2} \sum_{j_1,j_2=1, j_1 \neq j_2}^d (\exp(S_{j_1} + S_{j_2}) + 1 - \exp(S_{j_1}) - \exp(S_{j_2})) \bigg| \mathbf{u}\right]$$

$$+ \mathbb{E}_0\left[d^{-2} \sum_{j=1}^d (\exp(2S_j) - 2\exp(S_j) + 1) \bigg| \mathbf{u}\right]$$

$$= d^{-2} \sum_{j_1,j_2=1, j_1 \neq j_2}^d \left(\exp(\sigma_\varepsilon^{-2} c_{j_1} c_{j_2} \sum_{i=1}^n u_{i,(j_1)} u_{i,(j_2)}) - 1\right) + d^{-2} \sum_{j=1}^d \left(\exp(\sigma_\varepsilon^{-2} c_j^2 \sum_{i=1}^n u_{i,(j)}^2) - 1\right),$$
(51)

where the last line follows by (50).

Notice that for $j_1 \neq j_2$, $\mathbf{u}_{(j_1)}$ and $\mathbf{u}_{(j_2)}$ are independent since $\mathbf{\Sigma}_u$ is diagonal. Hence, conditional on $\mathbf{u}_{(j_1)}$,

$$\sigma_\varepsilon^{-2} c_{j_1} c_{j_2} \sum_{i=1}^n u_{i,(j_1)} u_{i,(j_2)}$$

is a zero-mean Gaussian random variable with variance $\sigma_\varepsilon^{-4} c_{j_1}^2 c_{j_2}^2 \sigma_{u,,j_2}^2 \sum_{i=1}^n u_{i,(j_1)}^2$. It follows that, for $j_1 \neq j_2$,

$$\mathbb{E}_0 \exp(\sigma_\varepsilon^{-2} c_{j_1} c_{j_2} \sum_{i=1}^n u_{i,(j_1)} u_{i,(j_2)}) \qquad (52)$$

$$
\begin{aligned}
&= \mathbb{E}_0 \left[ \mathbb{E}_0 \left( \exp(\sigma_\varepsilon^{-2} c_{j_1} c_{j_2} \sum_{i=1}^n u_{i,j_1} u_{i,j_2}) \mid \mathbf{u}_{(j_1)} \right) \right] \\
&= \mathbb{E}_0 \exp\left( \frac{1}{2} \sigma_\varepsilon^{-4} c_{j_1}^2 c_{j_2}^2 \sigma_{u,j_2}^2 \sum_{i=1}^n u_{i,(j_1)}^2 \right) \\
&\stackrel{(i)}{=} \mathbb{E}_0 \exp\left( (\frac{1}{8} n^{-2} \log^2 d)(\sum_{i=1}^n u_{i,(j_1)}^2 \sigma_{u,j_1}^{-2}) \right) \quad (53) \\
&\stackrel{(ii)}{=} (1 - \frac{1}{4} n^{-2} \log^2 d)^{-n/2}, \quad (54)
\end{aligned}
$$

where $(i)$ follows by the definition of $c_j$'s and $(ii)$ follows by the fact that $\sum_{i=1}^n u_{i,(j_1)}^2 \sigma_{u,j_1}^{-2} \sim \chi^2(n)$ and the moment generating function of chi-squared distributions. Recall that for $X \sim \chi^2(k)$ and $t < 1/2$, $\mathbb{E} \exp(tX) = (1-2t)^{-k/2}$. Similarly, we have

$$
\mathbb{E} \exp\left( \sigma_\varepsilon^{-2} c_j^2 \sum_{i=1}^n u_{i,(j)}^2 \right) = (1 - n^{-1} \log d)^{-n/2}. \quad (55)
$$

Now (51), (54) and (55) along with the law of iterated expectations imply that

$$
\mathbb{E}_0 \left( d^{-1} \sum_{j=1}^d (\exp(S_j) - 1) \right)^2 = \mathbb{E}_0 \left\{ \mathbb{E}_0 \left[ \left( d^{-1} \sum_{j=1}^d (\exp(S_j) - 1) \right)^2 \mid \mathbf{u} \right] \right\}
$$
$$
= \underbrace{d^{-1}(d-1) \left[ (1 - \frac{1}{4} n^{-2} \log^2 d)^{-n/2} - 1 \right]}_{A_1} + \underbrace{d^{-1}(1 - n^{-1} \log d)^{-n/2}}_{A_2} + d^{-1} \quad (56)
$$

To see that $A_1 = o(1)$, notice that $n^{-1} \log^2 d \to 0$,

$$
\log\left[ (1 - \frac{1}{4} n^{-2} \log^2 d)^{n^2 / \log^2 d} \right] \to -\frac{1}{4}
$$

and hence $\log\left[ (1 - \frac{1}{4} n^{-2} \log^2 d)^{-n/2} \right] = -\frac{1}{2}(n^{-1} \log^2 d) \log\left[ (1 - \frac{1}{4} n^{-2} \log^2 d)^{n^2/\log^2 d} \right] \to 0$. Since $d \to \infty$, we have

$$
\begin{aligned}
\log A_2 = \log\left[ d^{-1}(1 - n^{-1} \log d)^{-n/2} \right] &= -\log d + (\frac{1}{2} \log d) \left\{ \log\left[ (1 - n^{-1} \log d)^{-(n^{-1} \log d)^{-1}} \right] \right\} \\
&= -\log d + (\frac{1}{2} \log d)(1 + o(1)) \\
&\to -\infty.
\end{aligned}
$$

Now, (56), $A_1 = o(1)$ and $\log A_2 \to -\infty$ imply that $\mathbb{E}_0 \left( d^{-1} \sum_{j=1}^d (\exp(S_j) - 1) \right)^2 = o(1)$, which, by (48) and (49), implies the desired result. □

# B  Proofs of the remaining results

## B.1  Proofs for the High-Dimensional approximate bootstrap

We split the analysis into two major parts, with the first establishing a general result on approximate bootstrap and the second showing that many of the conditions established in the first part hold true.

**Preliminary Results**

The following result is the key argument for establishing the theoretical properties of the approximate multiplier bootstrap procedure. The setup is slightly more general than described in Section 6.

**Theorem 7.** *Consider* $\widehat{S}_n = S_n^\Psi + \Delta_n$ *and* $S_n^\Psi = n^{-1/2} \sum_{i=1}^n \Psi_i$ *with* $\Psi_i = (\Psi_{i,1}, \cdots, \Psi_{i,m_\Psi})^\top \in \mathbb{R}^{m_\Psi}$. *Define* $\widetilde{S}_n^\Psi = n^{-1/2} \sum_{i=1}^n (\Psi_i - \bar{\Psi}_n)\xi_i$ *and* $\bar{\Psi}_n = n^{-1} \sum_{i=1}^n \Psi_i$; *also define* $\widetilde{S}_n^{\widehat{\Psi}}$ *and* $\bar{\widehat{\Psi}}_n$ *similarly with* $\Psi_i$ *replaced by* $\widehat{\Psi}_i$. *Let* $\mathcal{F}_n$ *and* $\mathcal{G}_n$ *be* $\sigma$-*algebras such that the sequence* $\{\xi_i\}_{i=1}^n$ *is independent of* $\mathcal{G}_n$ *and* $\mathcal{G}_n$ *contains the* $\sigma$-*algebra generated by* $\mathcal{F}_n$ *and* $\{(\Psi_i, \widehat{\Psi}_i)\}_{i=1}^n$. *Suppose that the following hold:*

*(i)* $\mathbb{E}\left[\sup_{x \in \mathbb{R}} \left| \mathbb{P}\left(\|S_n^\Psi\|_\infty > x \mid \mathcal{F}_n\right) - \mathbb{P}\left(\|\widetilde{S}_n^\Psi\|_\infty > x \mid \mathcal{G}_n\right) \right| \right] \to 0$,

*(ii)* $\mathbb{E}\left[\sup_{x \in \mathbb{R}} \left| \mathbb{P}\left(\|\widetilde{S}_n^\Psi\|_\infty > x \mid \mathcal{G}_n\right) - \mathbb{P}\left(\|\widetilde{S}_n^{\widehat{\Psi}}\|_\infty > x \mid \mathcal{G}_n\right) \right| \right] \to 0$,

*(iii)* $\mathbb{P}\left(\min_{1 \leq j \leq m_\Psi} n^{-1} \sum_{i=1}^n (\Psi_{i,j} - \bar{\Psi}_{n,j})^2 > b\right) \to 1$, *for a constant* $b > 0$,

*(iv)* $\|\Delta_n\|_\infty = o_P(\varepsilon_n)$ *for a sequence of constants* $\varepsilon_n \to 0$, *such that* $\varepsilon_n \sqrt{\log m_\Psi} \to 0$.

*Then*

$$\limsup_{n \to \infty} \sup_{\eta \in (0,1)} \left| \mathbb{P}\left(\|\widehat{S}_n\|_\infty > \mathcal{Q}(1-\eta, \|\widetilde{S}_n^{\widehat{\Psi}}\|_\infty)\right) - \eta \right| = 0,$$

*where* $\mathcal{Q}(\alpha, \|\widetilde{S}_n^{\widehat{\Psi}}\|_\infty) = \inf\{x \in \mathbb{R} \mid \mathbb{P}(\|\widetilde{S}_n^{\widehat{\Psi}}\|_\infty > x \mid \mathcal{G}_n) \leq \alpha\}$.

*Proof of Theorem 7.* The main idea of the proof is based on the observation that $\|\Delta_n\|_\infty$ goes to zero faster than $\varepsilon_n$ while the probability of of the bootstrapped quantity lying in an interval of length $2\varepsilon_n$ goes to zero. Formally, let $\varepsilon_n$ be a sequence of constants satisfying (iv) in the statement of the theorem. We begin by introducing necessary notation. Let

$a_{n,1} := \sup_{x \in \mathbb{R}} \left| \mathbb{P}\left(\|S_n^\Psi\|_\infty > x \mid \mathcal{F}_n\right) - \mathbb{P}\left(\|\widetilde{S}_n^\Psi\|_\infty > x \mid \mathcal{G}_n\right) \right|$

and $a_{n,2} = \sup_{x \in \mathbb{R}} \left| \mathbb{P}\left(\|\widetilde{S}_n^\Psi\|_\infty > x \mid \mathcal{G}_n\right) - \mathbb{P}\left(\|\widetilde{S}_n^{\widehat{\Psi}}\|_\infty > x \mid \mathcal{G}_n\right) \right|$.

Observe that by the assumption (i) and (ii) we have $\mathbb{E} a_{n,1} \to 0$ and $\mathbb{E} a_{n,2} \to 0$. We follow by employing the triangular inequality guarantees,

$$\left| \mathbb{P}\left(\|S_n^\Psi\|_\infty \in (x - \varepsilon_n, x + \varepsilon_n] \mid \mathcal{F}_n\right) - \mathbb{P}\left(\|\widetilde{S}_n^\Psi\|_\infty \in (x - \varepsilon_n, x + \varepsilon_n] \mid \mathcal{G}_n\right) \right|$$
$$\leq \left| \mathbb{P}\left(\|S_n^\Psi\|_\infty > x - \varepsilon_n \mid \mathcal{F}_n\right) - \mathbb{P}\left(\|\widetilde{S}_n^\Psi\|_\infty > x - \varepsilon_n \mid \mathcal{G}_n\right) \right|$$
$$+ \left| \mathbb{P}\left(\|S_n^\Psi\|_\infty > x + \varepsilon_n \mid \mathcal{F}_n\right) - \mathbb{P}\left(\|\widetilde{S}_n^\Psi\|_\infty > x + \varepsilon_n \mid \mathcal{G}_n\right) \right| \leq 2a_{n,1}. \quad (57)$$

It follows that

$$\left| \mathbb{P}\left(\|S_n^{\widehat{\Psi}}\|_\infty > x \mid \mathcal{F}_n\right) - \mathbb{P}\left(\|\widetilde{S}_n^\Psi\|_\infty > x \mid \mathcal{G}_n\right) \right|$$
$$\overset{(i)}{\leq} \left| \mathbb{P}\left(\|S_n^{\widehat{\Psi}}\|_\infty > x \mid \mathcal{F}_n\right) - \mathbb{P}\left(\|S_n^\Psi\|_\infty > x \mid \mathcal{F}_n\right) \right| + a_{n,1}$$
$$\overset{(ii)}{\leq} \mathbb{P}\left(\|\Delta_n\|_\infty > \varepsilon_n \mid \mathcal{F}_n\right) + \mathbb{P}\left(\|S_n^\Psi\|_\infty \in (x - \varepsilon_n, x + \varepsilon_n] \mid \mathcal{F}_n\right) + a_{n,1}$$
$$\overset{(iii)}{\leq} \mathbb{P}\left(\|\Delta_n\|_\infty > \varepsilon_n \mid \mathcal{F}_n\right) + \mathbb{P}\left(\|\widetilde{S}_n^\Psi\|_\infty \in (x - \varepsilon_n, x + \varepsilon_n] \mid \mathcal{G}_n\right) + 3a_{n,1}, \quad (58)$$

where (i) follows by the triangular inequality and the definition of $a_{n,1}$, (ii) follows by a simple observation that for two random vectors $X$ and $Y$ and $\forall t, \varepsilon > 0$,

$$|\mathbb{P}(\|X\|_\infty > t) - \mathbb{P}(\|Y\|_\infty > t)| \leq \mathbb{P}(\|X - Y\|_\infty > \varepsilon) + \mathbb{P}(\|Y\|_\infty \in (t - \varepsilon, t + \varepsilon]). \quad (59)$$

and (iii) follows by (57).

Define the event
$$\mathcal{E}_n := \left\{ \min_{1 \leq j \leq m_\Psi} n^{-1} \sum_{i=1}^n (\Psi_{i,j} - \bar{\Psi}_{n,j})^2 \geq b \right\}.$$

Observe that by Nazarov's anti-concentration inequality (Lemma A.1 in Chernozhukov et al. (2014)), for a $p$-dimensional random vector $\mathbf{Y}$ with mean zero and such that $\min_{1 \leq j \leq m_\Psi} \mathbb{E}(Y_j^2) \geq b$ for some constant $b > 0$, there exists a constant $C_b > 0$ depending only on $b$, such that $\forall \varepsilon > 0$.
$$\sup_{x \in \mathbb{R}} \mathbb{P}(\|Y\|_\infty \in (x - \varepsilon, x + \varepsilon]) \leq C_b \varepsilon \sqrt{\log p}. \quad (60)$$

Together the fact that conditional on $\mathcal{G}_n$, $n^{-1/2} \sum_{i=1}^n (\Psi_i - \bar{\Psi}_n) \xi_i$ is a zero-mean Gaussian vector in $\mathbb{R}^{m_\Psi}$ whose $j$th component has variance equal to $n^{-1} \sum_{i=1}^n (\Psi_{i,j} - \bar{\Psi}_{n,j})^2$, we have

$$\sup_{x \in \mathbb{R}} \mathbb{P}\left(\|\widetilde{S}_n^\Psi\|_\infty \in (x - \varepsilon_n, x + \varepsilon_n] \mid \mathcal{G}_n\right) \leq \varepsilon_n C_b \sqrt{\log m_\Psi} + \mathbb{I}\{\mathcal{E}_n^c\}, \quad (61)$$

Then (58), (61) and the definition of $a_{n,2}$ imply that

$$\sup_{x \in \mathbb{R}} \left| \mathbb{P}\left(\|S_n^{\widehat{\Psi}}\|_\infty > x \mid \mathcal{F}_n\right) - \mathbb{P}\left(\|\widetilde{S}_n^{\widehat{\Psi}}\|_\infty > x \mid \mathcal{G}_n\right) \right|$$
$$\leq \sup_{x \in \mathbb{R}} \left| \mathbb{P}\left(\|S_n^{\widehat{\Psi}}\|_\infty > x \mid \mathcal{F}_n\right) - \mathbb{P}\left(\|\widetilde{S}_n^\Psi\|_\infty > x \mid \mathcal{G}_n\right) \right| + a_{n,2}$$
$$\leq \mathbb{P}(\|\Delta_n\|_\infty > \varepsilon_n \mid \mathcal{F}_n) + \sup_{x \in \mathbb{R}} \mathbb{P}\left(\|\widetilde{S}_n^\Psi\|_\infty \in (x - \varepsilon_n, x + \varepsilon_n] \mid \mathcal{G}_n\right) + 3a_{n,1} + a_{n,2}$$
$$\leq \mathbb{P}(\|\Delta_n\|_\infty > \varepsilon_n \mid \mathcal{F}_n) + \varepsilon_n C_b \sqrt{\log m_\Psi} + \mathbb{I}\{\mathcal{E}_n^c\} + 3a_{n,1} + a_{n,2}. \quad (62)$$

It follows that $\forall \delta > 0$

$$\mathbb{E}\left[\sup_{\eta \in (0,1)} \left| \mathbb{P}\left(\|S_n^{\widehat{\Psi}}\|_\infty > \mathcal{Q}(1 - \eta, \|\widetilde{S}_n^{\widehat{\Psi}}\|_\infty) \mid \mathcal{F}_n\right) - \eta \right| \right]$$
$$\stackrel{(i)}{\leq} \mathbb{E}\left[\delta + \mathbb{P}\left(\sup_{x \in \mathbb{R}} \left| \mathbb{P}\left(\|S_n^{\widehat{\Psi}}\|_\infty > x \mid \mathcal{F}_n\right) - \mathbb{P}\left(\|\widetilde{S}_n^{\widehat{\Psi}}\|_\infty > x \mid \mathcal{G}_n\right) \right| > \delta \bigg| \mathcal{F}_n\right)\right]$$
$$\stackrel{(ii)}{=} \delta + \mathbb{P}\left(\sup_{x \in \mathbb{R}} \left| \mathbb{P}\left(\|S_n^{\widehat{\Psi}}\|_\infty > x \mid \mathcal{F}_n\right) - \mathbb{P}\left(\|\widetilde{S}_n^{\widehat{\Psi}}\|_\infty > x \mid \mathcal{G}_n\right) \right| > \delta \right)$$
$$\stackrel{(iii)}{\leq} \delta + \mathbb{P}\left(\mathbb{P}(\|\Delta_n\|_\infty > \varepsilon_n \mid \mathcal{F}_n) + \varepsilon_n C_b \sqrt{\log m_\Psi} + 3a_{n,1} + a_{n,2} > \delta\right) + \mathbb{P}(\mathcal{E}_n^c)$$
$$\stackrel{(iv)}{\leq} \delta + \mathbb{P}(\mathbb{P}(\|\Delta_n\|_\infty > \varepsilon_n \mid \mathcal{F}_n) > \delta/5) + \mathbb{P}(a_{n,1} > \delta/15) + \mathbb{P}(a_{n,2} > \delta/5)$$
$$\quad + \mathbb{I}\left\{\varepsilon_n C_b \sqrt{\log m_\Psi} > \delta/5\right\} + \mathbb{P}(\mathcal{E}_n^c)$$
$$\stackrel{(v)}{\leq} \delta + \frac{5}{\delta}\mathbb{P}(\|\Delta_n\|_\infty > \varepsilon_n) + \frac{15}{\delta}\mathbb{E}[a_{n,1}] + \frac{5}{\delta}\mathbb{E}[a_{n,2}] + \mathbb{I}\left\{C_b \varepsilon_n \sqrt{\log m_\Psi} > \delta/5\right\} + \mathbb{P}(\mathcal{E}_n^c)$$

$$\stackrel{(vi)}{\leq} \delta + o(1),$$

where $(i)$ follows from Lemma 7, $(ii)$ from the law of iterated expectation, $(iii)$ from (62), $(iv)$ from the subadditivity of the probability measures, $(v)$ follows by the Chebyshev's inequality and the law of iterated expectation and $(vi)$ by Assumptions (i)-(iv) of the Theorem where $\varepsilon_n \sqrt{\log m_\Psi} \to 0$ and $\mathbb{E} a_{n,1} = \mathbb{E} a_{n,2} \to 0$.

Since $\delta$ is arbitrary, we have

$$\limsup_{n\to\infty} \mathbb{E}\left[ \sup_{\eta\in(0,1)} \left| \mathbb{P}\left( \|S_n^{\widehat{\Psi}}\|_\infty > \mathcal{Q}(1-\eta, \|\widetilde{S}_n^{\widehat{\Psi}}\|_\infty) \mid \mathcal{F}_n \right) - \eta \right| \right] = 0. \tag{63}$$

The desired result follows by noticing that

$$\sup_{\eta\in(0,1)} \left| \mathbb{P}\left( \|S_n^{\widehat{\Psi}}\|_\infty > \mathcal{Q}(1-\eta, \|\widetilde{S}_n^{\widehat{\Psi}}\|_\infty) \right) - \eta \right|$$
$$\leq \mathbb{E}\left[ \sup_{\eta\in(0,1)} \left| \mathbb{P}\left( \|S_n^{\widehat{\Psi}}\|_\infty > \mathcal{Q}(1-\eta, \|\widetilde{S}_n^{\widehat{\Psi}}\|_\infty) \mid \mathcal{F}_n \right) - \eta \right| \right].$$

$\square$

The next result shows sufficient conditions for the Condition (ii) of Theorem 7.

**Lemma 1.** Let $\Psi_i$, $\widehat{\Psi}_i$, $\widetilde{S}_n^\Psi$, $\widetilde{S}_n^{\widehat{\Psi}}$ and $\mathcal{G}_n$ be defined as in Theorem 7. Suppose that $\sigma_{n,*}^2 := \max_{1\leq j\leq m_\Psi} n^{-1} \sum_{i=1}^n (\widehat{\Psi}_{i,j} - \Psi_{i,j})^2 = \mathcal{O}_P(\delta_n^2)$ for a sequence of constants $\delta_n > 0$ that satisfies $\delta_n \log m_\Psi \to 0$. Assume that there exist a constant $b > 0$, such that

$$\mathbb{P}\left( \min_{1\leq j\leq m_\Psi} n^{-1} \sum_{i=1}^n (\Psi_{i,j} - \bar{\Psi}_{n,j})^2 > b \right) \to 1.$$

Then

$$\mathbb{E}\left[ \sup_{x\in\mathbb{R}} \left| \mathbb{P}\left( \|\widetilde{S}_n^\Psi\|_\infty > x \mid \mathcal{G}_n \right) - \mathbb{P}\left( \|\widetilde{S}_n^{\widehat{\Psi}}\|_\infty > x \mid \mathcal{G}_n \right) \right| \right] \to 0.$$

*Proof of Lemma 1.* Define the event $\mathcal{J}_n = \{\min_{1\leq j\leq m_\Psi} n^{-1} \sum_{i=1}^n (\Psi_{i,j} - \bar{\Psi}_{n,j})^2 > b\}$. By Equation (59),

$$\sup_{x\in\mathbb{R}} \left| \mathbb{P}\left( \|\widetilde{S}_n^\Psi\|_\infty > x \mid \mathcal{G}_n \right) - \mathbb{P}\left( \|\widetilde{S}_n^{\widehat{\Psi}}\|_\infty > x \mid \mathcal{G}_n \right) \right|$$
$$\leq \mathbb{P}\left( \|\widetilde{S}_n^{\widehat{\Psi}} - \widetilde{S}_n^\Psi\|_\infty > \sqrt{\delta_n} \mid \mathcal{G}_n \right) + \sup_{x\in\mathbb{R}} \mathbb{P}\left( \|\widetilde{S}_n^\Psi\|_\infty \in (x - \sqrt{\delta_n}, x + \sqrt{\delta_n}] \mid \mathcal{G}_n \right). \tag{64}$$

Notice that conditional on $\mathcal{G}_n$, $\widetilde{S}_n^\Psi$ is a Gaussian vector in $\mathbb{R}^{m_\Psi}$, whose $j$th entry has zero mean and variance $n^{-1} \sum_{i=1}^n (\Psi_{i,j} - \bar{\Psi}_{n,j})^2$. By the inequality in Equation (60), there exists a constant $C_b > 0$ depending only on $b$, such that

$$\sup_{x\in\mathbb{R}} \mathbb{P}\left( \|\widetilde{S}_n^\Psi\|_\infty \in (x - \sqrt{\delta_n}, x + \sqrt{\delta_n}] \mid \mathcal{G}_n \right) \leq C_b \sqrt{\delta_n \log m_\Psi} + \mathbb{I}\{\mathcal{J}_n^c\} \text{ a.s.} \tag{65}$$

Notice that the $j$th component of $\widetilde{S}_n^{\widehat{\Psi}} - \widetilde{S}_n^{\Psi}$ is

$$n^{-1/2} \sum_{i=1}^{n} (\widehat{\Psi}_{i,j} - \Psi_{i,j} + \bar{\Psi}_{n,j} - \bar{\widehat{\Psi}}_{n,j}) \xi_i,$$

which, conditional on $\mathcal{G}_n$, is a zero-mean Gaussian random variable with variance

$$n^{-1} \sum_{i=1}^{n} (\widehat{\Psi}_{i,j} - \Psi_{i,j} + \bar{\Psi}_{n,j} - \bar{\widehat{\Psi}}_{n,j})^2 = n^{-1} \sum_{i=1}^{n} (\widehat{\Psi}_{i,j} - \Psi_{i,j})^2 - (\bar{\Psi}_{n,j} - \bar{\widehat{\Psi}}_{n,j})^2 \leq n^{-1} \sum_{i=1}^{n} (\widehat{\Psi}_{i,j} - \Psi_{i,j})^2.$$

Observe that for $Z \sim \mathcal{N}(0, \sigma^2)$ and $x > 0$, $\mathbb{P}(|Z| > x) \leq K_1 \exp(-K_2 \sigma^{-2} x^2)$ for some universal constants $K_1, K_2 > 0$. This elementary fact, together with the union bound and the definition, implies that

$$\mathbb{P}\left( \|\widetilde{S}_n^{\widehat{\Psi}} - \widetilde{S}_n^{\Psi}\|_\infty > \sqrt{\delta_n} \mid \mathcal{G}_n \right) \leq K_1 \exp\left( \frac{(\sigma_{n,*}^2 \delta_n^{-2})(\delta_n \log m_\Psi) - K_2}{\sigma_{n,*}^2 \delta_n^{-1}} \right) = o_P(1), \quad (66)$$

where the last step follows by $K_2 > 0$, $\sigma_{n,*}^2 \delta_n^{-2} = \mathcal{O}_P(1)$, $\delta_n \log m_\Psi = o(1)$ and $\sigma_{n,*}^2 \delta_n^{-1} = \mathcal{O}_P(\delta_n) = o_P(1) > 0$. Hence,

$$\sup_{x \in \mathbb{R}} \left| \mathbb{P}\left( \|\widetilde{S}_n^{\Psi}\|_\infty > x \mid \mathcal{G}_n \right) - \mathbb{P}\left( \|\widetilde{S}_n^{\widehat{\Psi}}\|_\infty > x \mid \mathcal{G}_n \right) \right| \leq o_P(1) + C_b \sqrt{\delta_n \log m_\Psi} + \mathbb{I}\{\mathcal{J}_n^c\} = o_P(1),$$

where the inequality follows by (64), (65) and (66), and the last step follows by $P(\mathcal{J}_n^c) \to 0$ and $\delta_n \log m_\Psi = o(1)$. Since probabilities take values in $[0, 1]$, the left-hand side of the above equation is bounded and thus uniformly integrable. By Theorem 5.4 on p.220 of Gut (2013),

$$\mathbb{E}\left[ \sup_{x \in \mathbb{R}} \left| \mathbb{P}\left( \|\widetilde{S}_n^{\Psi}\|_\infty > x \mid \mathcal{G}_n \right) - \mathbb{P}\left( \|\widetilde{S}_n^{\widehat{\Psi}}\|_\infty > x \mid \mathcal{G}_n \right) \right| \right] \to 0.$$

$\square$

**Proof of the Theorem 6**

The following result is useful for verifying condition (i) of Theorem 7.

**Lemma 2.** *Let $\{w_{i,j}\}_{(i,j) \in \{1,\cdots,n\} \times \{1,\cdots,d\}}$ and $\{q_{i,j}\}_{(i,j) \in \{1,\cdots,n\} \times \{1,\cdots,d\}}$ be arrays of random variables in the setup of Theorem 6. Let $\mathcal{F}_n$ and $\mathcal{G}_n$ be $\sigma$-algebras, such that $\mathcal{F}_n = \sigma(\{w_{i,j}\})$ and $\mathcal{G}_n \supseteq \sigma(\mathcal{F}_n \bigcup \{q_{i,j}\})$. Moreover, define $\Psi_i = (\Psi_{i,1}, \cdots, \Psi_{i,d})^\top \in \mathbb{R}^d$ with $\Psi_{i,j} = w_{i,j} q_{i,j}$. If $n^{-1} \bar{M}_n^2 \log^7(dn) = o_P(1)$ with $\bar{M}_n = \max_{1 \leq i \leq n, 1 \leq j \leq d} |w_{i,j}|$, then*

$$\mathbb{E}\left[ \sup_{x \in \mathbb{R}} \left| \mathbb{P}\left( \|S_n^{\Psi}\|_\infty > x \mid \mathcal{F}_n \right) - \mathbb{P}\left( \|\widetilde{S}_n^{\Psi}\|_\infty > x \mid \mathcal{G}_n \right) \right| \right] \to 0,$$

*where $S_n^{\Psi}$ and $\widetilde{S}_n^{\Psi}$ are defined in Theorem 7.*

*Proof of Lemma 2.* Let $B_n = c_0 \max\{1, \bar{M}_n\}$, where $c_0 > 0$ is a constant to be chosen later. Notice that

$$n^{-1} \sum_{i=1}^{n} \mathbb{E}(\Psi_{i,j}^2 \mid \mathcal{F}_n) = n^{-1} \sum_{i=1}^{n} w_{i,j}^2 \mathbb{E} q_{i,j}^2 = n^{-1} \sum_{i=1}^{n} w_{i,j}^2 = 1. \quad (67)$$

Next, we check if conditions of Proposition 2.1 and Corollary 4.2 of Chernozhukov et al. (2014) hold. As $\{q_{i,j}\}$ is sub-Gaussian with bounded sub-Gaussian norm, there exists a constant $C_K > 0$ depending only on $K$, such that $\forall t \in \mathbb{R}$, $\mathbb{E}\exp(tq_{i,j}) \leq \exp(C_K t^2)$, $\max_{1 \leq j \leq d} \mathbb{E}|q_{i,j}|^3 < C_K$ and $\max_{1 \leq j \leq d} \mathbb{E}|q_{i,j}|^4 < C_K$. Hence, by (67) and the definition of $B_n$, we have

$$n^{-1} \sum_{i=1}^n \mathbb{E}(|\Psi_{i,j}|^3 \mid \mathcal{F}_n) B_n^{-1} \leq C_K \bar{M}_n n^{-1} \sum_{i=1}^n |w_{i,j}|^2 B_n^{-1} \leq C_K/c_0, \text{ and} \tag{68}$$

$$n^{-1} \sum_{i=1}^n \mathbb{E}(|\Psi_{i,j}|^4 \mid \mathcal{F}_n) B_n^{-2} \leq C_K \bar{M}_n^2 n^{-1} \sum_{i=1}^n |w_{i,j}|^2 B_n^{-2} \leq C_K/c_0^2 \tag{69}$$

Since $\forall t \in \mathbb{R}$, $\mathbb{E}\exp(tq_{i,j}) \leq \exp(C_K t^2)$, the definition of $B_n$ implies that

$$\mathbb{E}\left[\exp\left(|\Psi_{i,j}|/B_n\right) \mid \mathcal{F}_n\right] \leq \exp\left(C_K w_{i,j}^2 B_n^{-2}\right) \leq \exp\left(C_K/c_0\right). \tag{70}$$

Now, we choose the constant $c_0 > 0$, such that $C_K/c_0 \leq 1$, $C_K/c_0^2 \leq 1$ and $\exp(C_K/c_0) \leq 2$. This, together with (67), (68), (69) and (70), allows us to apply Proposition 2.1 of Chernozhukov et al. (2014) to the conditional probability measure $\mathbb{P}(\cdot \mid \mathcal{F}_n)$. Let $\{\Phi_i\}_{i=1}^n$ be a seqence of random elements in $\mathbb{R}^d$ such that conditional on $\mathcal{F}_n$, $\{\Phi_i\}_{i=1}^n$ is independent across $i$ and $\Phi_i \mid \mathcal{F}_n$ is Gaussian with mean zero and variance $\mathbb{E}(\Psi_i \Psi_i^\top \mid \mathcal{F}_n)$. It follows, by Proposition 2.1 of Chernozhukov et al. (2014), that

$$\sup_{x \in \mathbb{R}} \left| \mathbb{P}\left(\|S_n^\Psi\|_\infty > x \mid \mathcal{F}_n\right) - \mathbb{P}\left(\|S_n^\Phi\|_\infty > x \mid \mathcal{F}_n\right) \right| \leq C_1 D_n \text{ a.s.} \tag{71}$$

where $C_1 > 0$ is a universal constant (by (67)), $S_n^\Phi = n^{-1/2} \sum_{i=1}^n \Phi_i$ and $D_n = (n^{-1} B_n^2 \log^7(d \vee n))^{1/6}$. Moreover, from the Corollary 4.2 of Chernozhukov et al. (2014) applied to the probability measure $\mathbb{P}(\cdot \mid \mathcal{F}_n)$, we have that, for the sequence $\alpha_n = \min\{e^{-1}, n^{-1/2} d^{-1/2}\}$,

$$\mathbb{P}\left[\sup_{x \in \mathbb{R}} \left| \mathbb{P}\left(\|\widetilde{S}_n^\Psi\|_\infty > x \mid \mathcal{G}_n\right) - \mathbb{P}\left(\|S_n^\Phi\|_\infty > x \mid \mathcal{F}_n\right) \right| > C_2 D_n^{(\alpha_n)} \bigg| \mathcal{F}_n\right] \leq \alpha_n \text{ a.s,} \tag{72}$$

where $C_2 > 0$ is universal constant (by (67)) and

$$D_n^{(\alpha_n)} = (n^{-1} B_n^2 \log^5(dn) \log^2(\alpha_n^{-1}))^{1/6}.$$

Then, from (71), (72) and the law of iterated expectation, we have that $\forall \tau > 0$,

$$\mathbb{P}\left(\sup_{x \in \mathbb{R}} \left| \mathbb{P}\left(\|\widetilde{S}_n^\Psi\|_\infty > x \mid \mathcal{G}_n\right) - \mathbb{P}\left(\|S_n^\Psi\|_\infty > x \mid \mathcal{F}_n\right) \right| > \tau\right)$$
$$\leq \alpha_n + \mathbb{P}\left(C_1 D_n + C_2 D_n^{(\alpha_n)} > \tau\right) = o(1),$$

where the last step follows by $\alpha_n = o(1)$, $D_n = o(1)$ and $D_n^{(\alpha_n)} = o_P(1)$ (by the rate conditions in the assumption of the lemma). Since $\tau > 0$ is arbitrary, we have

$$\sup_{x \in \mathbb{R}} \left| \mathbb{P}\left(\|\widetilde{S}_n^\Psi\|_\infty > x \mid \mathcal{G}_n\right) - \mathbb{P}\left(\|S_n^\Psi\|_\infty > x \mid \mathcal{F}_n\right) \right| = o_P(1). \tag{73}$$

Since probabilities take values in $[0,1]$, the left-hand side of (73) is bounded and thus uniformly integrable. From the Theorem 5.4 on p.220 of Gut (2013),

$$\mathbb{E}\left[\sup_{x\in\mathbb{R}}\left|\mathbb{P}\left(\|\widetilde{S}_n^\Psi\|_\infty > x \mid \mathcal{G}_n\right) - \mathbb{P}\left(\|S_n^\Psi\|_\infty > x \mid \mathcal{F}_n\right)\right|\right] \to 0.$$

$\square$

The next lemma is helpful in verifying the assumptions of Lemma 1.

**Lemma 3.** *Consider the setting in Theorem 6. If $n^{-1/2}\bar{M}_n^2 \log(d \vee n) = \mathcal{O}_P(1)$, then, for any constant $a \in (0,1)$, we have*

$$\mathbb{P}\left(\min_{1\leq j\leq d} n^{-1}\sum_{i=1}^n (\Psi_{i,j} - \bar{\Psi}_{n,j})^2 > a\right) \to 1,$$

*where $\Psi_{i,j} = w_{i,j}q_{i,j}$ and $\bar{\Psi}_{n,j} = n^{-1}\sum_{i=1}^n \Psi_{i,j}$.*

*Proof of Lemma 3.* Since $q_{i,j}$'s have sub-Gaussian norms bounded by some constant $C_1 > 0$, it follows, by Equation (5.16) and Lemma 5.14 of Vershynin (2010), that there exists a constant $C_2 > 0$ depending only on $C_1$ such that $\forall (i,j)$, $q_{i,j}^2 - 1$ is sub-exponential with sub-exponential norm bounded by $C_2$. By Proposition 5.16 of Vershynin (2010) applied to the conditional probability measure $\mathbb{P}(\cdot \mid \mathcal{F}_n)$ and the union bound, there exist constants $C_3, C_4 > 0$ depending only on $C_2$ such that for any $t > 0$, almost surely

$$\mathbb{P}\left(\max_{1\leq j\leq d}\left|\sum_{i=1}^n w_{i,j}^2(q_{i,j}^2 - 1)\right| > n^{1/2}t \,\bigg|\, \mathcal{F}_n\right) \leq 2d\exp\left[-\min\left\{C_3 \frac{nt^2}{\sum_{i=1}^n w_{i,j}^4}, C_4 \frac{n^{1/2}t}{\bar{M}_n^2}\right\}\right].$$

Since $n^{-1}\sum_{i=1}^n w_{i,j}^2 = 1$, one can easily show that the right hand side above is upper bounded by

$$2d\exp\left[-\min\left\{C_3 \frac{t^2}{\bar{M}_n^2}, C_4 \frac{n^{1/2}t}{\bar{M}_n^2}\right\}\right].$$

Therefore, we can choose a sequence $\widetilde{t}_n = \mathcal{O}(\bar{M}_n^2 \log(d \vee n))$ such that

$$\mathbb{P}\left(\max_{1\leq j\leq d}\left|\sum_{i=1}^n w_{i,j}^2(q_{i,j}^2 - 1)\right| > n^{1/2}\widetilde{t}_n \,\bigg|\, \mathcal{F}_n\right) = o_P(1).$$

Since conditional probabilities are bound and hence uniformly integrable, Theorem 5.4 on p.220 of Gut (2013) implies

$$\mathbb{P}\left(\max_{1\leq j\leq d}\left|\sum_{i=1}^n w_{i,j}^2(q_{i,j}^2 - 1)\right| > n^{1/2}\widetilde{t}_n\right) \to 0.$$

Therefore,

$$\max_{1\leq j\leq d}\left|\sum_{i=1}^n \Psi_{i,j}^2 - w_{i,j}^2\right| = \max_{1\leq j\leq d}\left|\sum_{i=1}^n w_{i,j}^2(q_{i,j}^2 - 1)\right| = o_P(n^{1/2}\widetilde{t}_n) = o_P(n^{1/2}\bar{M}_n^2 \log(d \vee n)). \quad (74)$$

By Proposition 5.10 of Vershynin (2010) applied to the conditional probability measure $\mathbb{P}(\cdot \mid \mathcal{F}_n)$ and the union bound, there exists a constant $C_5 > 0$ depending only on $C_1$ such that for any $t > 0$, almost surely

$$\mathbb{P}\left(\max_{1 \leq j \leq d} \left|\sum_{i=1}^n w_{i,j} q_{i,j}\right| > n^{1/2} t \,\Big|\, \mathcal{F}_n\right) \leq \sum_{j=1}^d \exp\left(1 - C_5 \frac{nt^2}{\sum_{i=1}^n w_{i,j}^2}\right) = d \exp\left(1 - C_5 t^2\right),$$

where the last step follows by $n^{-1} \sum_{i=1}^n w_{i,j}^2 = 1$. Hence, we can choose $t = \mathcal{O}(\sqrt{\log(d \vee n)})$ such that the right-hand side of the above equation is $o_P(1)$. By the same argument as for (74), we have

$$\max_{1 \leq j \leq d} n|\bar{\Psi}_{n,j}| = \max_{1 \leq j \leq d} \left|\sum_{i=1}^n w_{i,j} q_{i,j}\right| = o_P(\sqrt{n \log(d \vee n)}). \tag{75}$$

It follows that

$$\min_{1 \leq j \leq d} n^{-1} \sum_{i=1}^n (\Psi_{i,j} - \bar{\Psi}_{n,j})^2 \geq \min_{1 \leq j \leq d} n^{-1} \sum_{i=1}^n \Psi_{i,j}^2 - \max_{1 \leq j \leq d} \bar{\Psi}_{n,j}^2$$

$$\geq \min_{1 \leq j \leq d} n^{-1} \sum_{i=1}^n w_{i,j}^2 - \max_{1 \leq j \leq d}\left[n^{-1}\sum_{i=1}^n (w_{i,j}^2 - \Psi_{i,j}^2)\right] - \max_{1 \leq j \leq d} \bar{\Psi}_{n,j}^2$$

$$= 1 - o_P(n^{-1/2} \bar{M}_n^2 \log(d \vee n)) - o_P(n^{-1} \log(d \vee n)),$$

where the last step follows by $n^{-1} \sum_{i=1}^n w_{i,j}^2 = 1$, (74) and (75). By assumption, $n^{-1/2} \bar{M}_n^2 \log(d \vee n) = \mathcal{O}_P(1)$. Since

$$\bar{M}_n^2 \geq (dn)^{-1} \sum_{j=1}^d \sum_{i=1}^n w_{i,jj}^2 = 1$$

and $n^{-1/2} \bar{M}_n^2 \log(d \vee n) = \mathcal{O}_P(1)$, we have $n^{-1} \log(d \vee n) = o(1)$. The result follows. □

Now we are ready to prove the main result of the Section 5.

*Proof of Theorem 6.* By Theorem 7, it suffices to show that (i)-(iv) therein hold. Notice that claim (iv) holds with $\varepsilon_n = \delta_n^{1/2}$. By Lemma 2, claim (i) holds. By Lemma 3, claim (iii) holds with $b = 1/2$. This and Lemma 1 imply claim (ii). □

### B.2 Auxiliary materials

The following result is useful in deriving the properties of our estimators. Its proof is similar to the proof of Theorem 7.1 of Bickel et al. (2009) and is thus omitted.

**Lemma 4.** *Let $Y \in \mathbb{R}^n$ and $X \in \mathbb{R}^{n \times p}$. Let $\widehat{\xi}$ be any vector satisfying $\|n^{-1} X^\top (Y - X\widehat{\xi})\|_\infty \leq \eta$. Suppose that there exists $\xi_*$ such that $\|n^{-1} X^\top (Y - X\xi_*)\|_\infty \leq \eta$ and $\|\widehat{\xi}\|_1 \leq \|\xi_*\|_1$. If $s_* = \|\xi_*\|_0$ and*

$$\min_{J_0 \subseteq \{1, \cdots, p\}, |J_0| \leq s_*} \min_{\delta \neq 0, \|\delta_{J_0^c}\|_1 \leq \|\delta_{J_0}\|_1} \frac{\|X\delta\|_2}{\sqrt{n} \|\delta_{J_0}\|_2} \geq \kappa, \tag{76}$$

*then $\|\delta\|_1 \leq 8\eta s_* \kappa^{-2}$ and $\delta^\top X^\top X \delta / n \leq 16 \eta^2 s_* \kappa^{-2}$, where $\delta = \widehat{\xi} - \xi_*$.*

**Lemma 5.** *If $n^{-1}\mathbf{G}^\top(\mathbf{G} - \mathbf{X}\widehat{\boldsymbol{\gamma}}) \geq \bar{\eta}$, then $n^{-1}(\mathbf{G} - \mathbf{X}\widehat{\boldsymbol{\gamma}})^\top(\mathbf{G} - \mathbf{X}\widehat{\boldsymbol{\gamma}}) \geq \bar{\eta}^2/(n^{-1}\mathbf{G}^\top\mathbf{G})$. Moreover, if $n^{-1}\mathbf{Z}_{(j)}^\top(\mathbf{Z}_{(j)} - \mathbf{X}\widehat{\boldsymbol{\theta}}) \geq \bar{\eta}$, then $n^{-1}(\mathbf{Z}_{(j)} - \mathbf{X}\widehat{\boldsymbol{\theta}})^\top(\mathbf{Z}_{(j)} - \mathbf{X}\widehat{\boldsymbol{\theta}}) \geq \bar{\eta}^2/(n^{-1}\mathbf{Z}_{(j)}^\top\mathbf{Z}_{(j)})$.*

*Proof of Lemma 5.* We only prove the result for $\widehat{\boldsymbol{\gamma}}$ because the result for $\widehat{\boldsymbol{\theta}}$ follows an analogous argument. Let $\widehat{\boldsymbol{\gamma}}$ satisfy $n^{-1}\mathbf{G}^\top(\mathbf{G} - \mathbf{X}\widehat{\boldsymbol{\gamma}}) \geq \bar{\eta}$. Then for any $a \geq 0$, we have

$$
\begin{aligned}
& n^{-1}(\mathbf{G} - \mathbf{X}\widehat{\boldsymbol{\gamma}})^\top(\mathbf{G} - \mathbf{X}\widehat{\boldsymbol{\gamma}}) \\
& \geq n^{-1}(\mathbf{G} - \mathbf{X}\widehat{\boldsymbol{\gamma}})^\top(\mathbf{G} - \mathbf{X}\widehat{\boldsymbol{\gamma}}) + a\left(\bar{\eta} - n^{-1}\mathbf{G}^\top(\mathbf{G} - \mathbf{X}\widehat{\boldsymbol{\gamma}})\right) \\
& \geq \min_{\boldsymbol{\gamma}} \left\{ n^{-1}(\mathbf{G} - \mathbf{X}\boldsymbol{\gamma})^\top(\mathbf{G} - \mathbf{X}\boldsymbol{\gamma}) + a\left(\bar{\eta} - n^{-1}\mathbf{G}^\top(\mathbf{G} - \mathbf{X}\boldsymbol{\gamma})\right) \right\} \\
& = a\bar{\eta} - \frac{1}{4}a^2 n^{-1}\mathbf{G}^\top\mathbf{G},
\end{aligned}
$$

where the last line follows by the first-order condition of a quadratic optimization. The desired result follows by minimizing the last right hand side with respect to $a$ and by choosing $a = 2\bar{\eta}/(n^{-1}\mathbf{G}^\top\mathbf{G})$. □

The following result shows that, with high probability, $\boldsymbol{\gamma}^*$ and $\boldsymbol{\theta}^*$ lie in the feasible region of (9), under the null hypothesis ($\boldsymbol{\pi} = 0$) or local alternatives ($\boldsymbol{\pi} \neq 0$).

**Lemma 6.** *Suppose that Condition 1 holds and $\boldsymbol{\beta}^* = \boldsymbol{\beta}_0 + \boldsymbol{\pi}$ such that*

$$\|\boldsymbol{\Sigma}_X \boldsymbol{\Theta} \boldsymbol{\pi}\|_\infty = \mathcal{O}(\sqrt{n^{-1}\log(p-d)})$$

*and $\|\boldsymbol{\pi}\|_2 = \mathcal{O}(1)$. Consider the optimization problems in (9). There exist constants $C_1, ..., C_7 > 0$ such that for $1 \leq j \leq d$, $\eta_\gamma, \eta_{\theta,j} \geq C_1\sqrt{n^{-1}\log p}$, and for*

$$n \geq [C_2 \log(p - d)] \vee C_3,$$

*$\bar{\eta}_{\theta,j} \leq 0.8\sigma_{u,j}^2$, $\bar{\eta}_\gamma \leq 0.8\sigma_\varepsilon^2$ and $\mu_\gamma, \mu_{\theta,j} > C_4\sqrt{\log(dn)}$, we have*

$$\mathbb{P}\left(\boldsymbol{\gamma}^* \text{ is feasible for } (9)\right) \geq 1 - 2/(p-d)^{C_5+1} - \frac{\exp(1)}{nd^2} - 2\exp\left(-[C_6 n\bar{\eta}_\gamma^2] \wedge [C_7 n\bar{\eta}_\gamma]\right)$$

*and*

$$\mathbb{P}\left(\bigcap_{j=1}^d \{\boldsymbol{\theta}_{(j)}^* \text{ is feasible for } (9)\}\right) \geq 1 - 2/(p-d) - \frac{\exp(1)}{n} - 2d\exp\left(-[C_6 n\bar{\eta}_\gamma^2] \wedge [C_7 n\bar{\eta}_\gamma]\right).$$

*Proof of Lemma 6.* First notice that,

$$\mathbf{G} - \mathbf{X}\boldsymbol{\gamma}^* = \mathbf{Z}\boldsymbol{\pi} + \boldsymbol{\varepsilon} = \mathbf{X}\boldsymbol{\Theta}^*\boldsymbol{\pi} + \mathbf{u}\boldsymbol{\pi} + \boldsymbol{\varepsilon}. \tag{77}$$

Since $\|\boldsymbol{\pi}\|_2 = \mathcal{O}(1)$ and $z_i$, $\varepsilon_i$ and $\mathbf{u}_i$ are centered Gaussian with bounded variance, it follows that there exists a constant $K_0 > 0$ such that $g_i - x_i^\top \boldsymbol{\gamma}^*$ is centered Gaussian with variance bounded by $K_0$. By Lemma 8 and the sub-Gaussian properties of $x_i$, there exists a constant $K_1 > 0$ such that each entry of $x_i(g_i - x_i^\top \boldsymbol{\gamma}^*)$ has sub-exponential norm bounded by $K_1$. By Proposition 5.16 of Vershynin (2010) and the union bound, there exist constants $K_2, K_3 > 0$ such that $\forall x > 0$,

$$\mathbb{P}\left(\|n^{-1}\mathbf{X}^\top(\mathbf{G} - \mathbf{X}\boldsymbol{\gamma}^*) - \mathbb{E}[x_i(g_i - x_i^\top \boldsymbol{\gamma}^*)]\|_\infty > x\sqrt{n^{-1}\log(p-d)}\right)$$
$$\leq 2(p-d)\exp\left[-[K_2 x^2 \log(p-d)] \wedge [K_3 x \sqrt{n\log(p-d)}]\right]. \tag{78}$$

Since $\log(d)/\log(p-d) = \mathcal{O}(1)$, there exists a constant $M > 0$ such that $d \leq (p-d)^M$. By (77) and Condition 2, we have $\|\mathbb{E}[x_i(g_i - x_i^\top \boldsymbol{\gamma}^*)]\|_\infty = \|\boldsymbol{\Sigma}_X \boldsymbol{\Theta}^* \boldsymbol{\pi}\|_\infty \leq K_4\sqrt{n^{-1}\log(p-d)}$ for some large enough constant $K_4 > \sqrt{(M+2)/K_2}$. This and the above display imply that $\forall \eta_\gamma > 2K_4\sqrt{n^{-1}\log(p-d)}$ and $\forall n \geq (K_2 K_4/K_3)^2 \log(p-d)$,

$$\mathbb{P}\left(\|n^{-1}\mathbf{X}^\top(\mathbf{G} - \mathbf{X}\boldsymbol{\gamma}^*)\|_\infty > \eta_\gamma\right)$$
$$\leq \mathbb{P}\left(\|n^{-1}\mathbf{X}^\top(\mathbf{G} - \mathbf{X}\boldsymbol{\gamma}^*) - \mathbb{E}[x_i(g_i - x_i^\top \boldsymbol{\gamma}^*)]\|_\infty > \eta_\gamma - K_4\sqrt{n^{-1}\log(p-d)}\right)$$
$$\leq \mathbb{P}\left(\|n^{-1}\mathbf{X}^\top(\mathbf{G} - \mathbf{X}\boldsymbol{\gamma}^*) - \mathbb{E}[x_i(g_i - x_i^\top \boldsymbol{\gamma}^*)]\|_\infty > K_4\sqrt{n^{-1}\log(p-d)}\right)$$
$$\stackrel{(i)}{\leq} 2(p-d)\exp\left[-[K_2 K_4^2 \log(p-d)] \wedge [K_3 K_4 \sqrt{n\log(p-d)}]\right]$$
$$\stackrel{(ii)}{\leq} 2/(p-d)^{M+1}, \tag{79}$$

where $(i)$ follows by (78) and $(ii)$ follows by $K_4 > \sqrt{2/K_2}$ and $n \geq (K_2 K_4/K_3)^2 \log(p-d)$.

Recall from previous analysis that $v_i - x_i^\top \boldsymbol{\gamma}^*$ is centered Gaussian with variance bounded by a constant $K_0$. It follows, by the union bound, that there exists a constant $K_5 > 0$ such that $\forall x > 0$,

$$\mathbb{P}(\|\mathbf{G} - \mathbf{X}\boldsymbol{\gamma}^*\|_\infty > x\sqrt{\log dn}) \leq n\exp(1 - K_5 x^2 \log(dn)). \tag{80}$$

By (77), $\|\boldsymbol{\gamma}^*\|_2 = \mathcal{O}(1)$ and $\|\boldsymbol{\Theta}^*\boldsymbol{\pi}\|_2 = \mathcal{O}(1)$, it follows that both $v_i$ and $v_i - x_i^\top \boldsymbol{\gamma}^*$ are centered Gaussian with bounded variance. By Lemma 8 and obtain that there exists a constant $K_6 > 0$ such that $v_i(v_i - x_i^\top \boldsymbol{\gamma}^*)$ has sub-exponential norm bounded by $K_6$. By Proposition 5.16 of Vershynin (2010), there exist constants $K_7, K_8 > 0$ such that $\forall x > 0$,

$$\mathbb{P}\left(\left|n^{-1}\mathbf{G}^\top(\mathbf{G} - \mathbf{X}\boldsymbol{\gamma}^*) - \mathbb{E}v_i(v_i - x_i^\top \boldsymbol{\gamma}^*)\right| > x\right) \leq 2\exp\left(-[K_7 n x^2] \wedge [K_8 n x]\right). \tag{81}$$

Again, by (77), we have $\mathbb{E}[v_i(v_i - x_i^\top \boldsymbol{\gamma}^*)] = \sigma_\varepsilon^2 + \boldsymbol{\pi}^\top(\boldsymbol{\Sigma}_u + \boldsymbol{\Theta}^{*\top}\boldsymbol{\Sigma}_X\boldsymbol{\Theta}^*)\boldsymbol{\pi} + \boldsymbol{\gamma}^{*\top}\boldsymbol{\Sigma}_X\boldsymbol{\Theta}^*\boldsymbol{\pi}$. Notice that

$$|\boldsymbol{\gamma}^{*\top}\boldsymbol{\Sigma}_X\boldsymbol{\Theta}^*\boldsymbol{\pi}| \leq \|\boldsymbol{\gamma}^*\|_1 \|\boldsymbol{\Sigma}_X\boldsymbol{\Theta}^*\boldsymbol{\pi}\|_\infty \tag{82}$$
$$\leq \|\boldsymbol{\gamma}^*\|_2 \|\boldsymbol{\gamma}^*\|_0 \|\boldsymbol{\Sigma}_X\boldsymbol{\Theta}^*\boldsymbol{\pi}\|_\infty = o\left(\left[n^{-1}\log(d \vee n)/\log^2(p-d)\right]^{1/4}\right), \tag{83}$$

where the last line follows by Conditions 1(ii) and 2. Hence,

$$\mathbb{E}[v_i(v_i - x_i^\top \boldsymbol{\gamma}^*)] = \sigma_\varepsilon^2 + \boldsymbol{\pi}^\top(\boldsymbol{\Sigma}_u + \boldsymbol{\Theta}^{*\top}\boldsymbol{\Sigma}_X\boldsymbol{\Theta}^*)\boldsymbol{\pi} + o(1).$$

Then there exists a constant $K_9 > 0$ such that for any $n \geq K_9$, $\mathbb{E}[v_i(v_i - x_i^\top \boldsymbol{\gamma}^*)] \geq 0.99\sigma_\varepsilon^2$. Therefore, $\forall n \geq K_9$ and $\forall \bar{\eta}_\gamma < 0.8\sigma_\varepsilon^2$, we have that

$$\mathbb{P}\left(\bar{\eta}_\gamma > n^{-1}\mathbf{G}^\top(\mathbf{G} - \mathbf{X}\boldsymbol{\gamma}^*)\right)$$
$$\leq \mathbb{P}\left(-0.1\bar{\eta}_\gamma > n^{-1}\mathbf{G}^\top(\mathbf{G} - \mathbf{X}\boldsymbol{\gamma}^*) - \mathbb{E}v_i(v_i - x_i^\top\boldsymbol{\gamma}^*)\right) + \mathbf{1}\{1.1\bar{\eta}_\gamma > \mathbb{E}v_i(v_i - x_i^\top\boldsymbol{\gamma}^*)\}$$
$$\stackrel{(i)}{=} \mathbb{P}\left(-0.1\bar{\eta}_\gamma > n^{-1}\mathbf{G}^\top(\mathbf{G} - \mathbf{X}\boldsymbol{\gamma}^*) - \mathbb{E}v_i(v_i - x_i^\top\boldsymbol{\gamma}^*)\right)$$
$$\leq \mathbb{P}\left(\left|n^{-1}\mathbf{G}^\top(\mathbf{G} - \mathbf{X}\boldsymbol{\gamma}^*) - \mathbb{E}v_i(v_i - x_i^\top\boldsymbol{\gamma}^*)\right| > 0.1\bar{\eta}_\gamma\right)$$
$$\stackrel{(ii)}{\leq} 2\exp\left(-[K_7 n\bar{\eta}_\gamma^2] \wedge [K_8 n\bar{\eta}_\gamma]\right), \tag{84}$$

where $(i)$ holds by $\bar{\eta}_\gamma < 0.8\sigma_\varepsilon^2$ and $\mathbb{E}[g_i(g_i - x_i^\top\boldsymbol{\gamma}^*)] \geq 0.99\sigma_\varepsilon^2$ and $(ii)$ follows by (81).

By (79), (80) and (84), it follows that for $\eta_\gamma > 2K_4\sqrt{n^{-1}\log p}$,
$$n \geq [(K_2 K_4/K_3)^2 \log(p-d)] \vee K_9,$$
$\bar{\eta}_\gamma < 0.8\sigma_\varepsilon^2$ and $\mu_\gamma > \sqrt{(2/K_5)\log(dn)}$, we have

$$\mathbb{P}\left(\boldsymbol{\gamma}^* \text{ is feasible for } (9)\right) \geq 1 - 2/(p-d)^{M+1} - \frac{\exp(1)}{nd^2} - 2\exp\left(-[K_7 n\bar{\eta}_\gamma^2] \wedge [K_8 n\bar{\eta}_\gamma]\right).$$

This proves the result for $\boldsymbol{\gamma}^*$. By the same reasoning, we can show that there exist positive constants $Q_1,..., Q_5$ such that for $1 \leq j \leq d$, $\eta_{\theta,j} > Q_1\sqrt{n^{-1}\log p}$, $n \geq [Q_2 \log(p-d)] \vee Q_3$, $\bar{\eta}_{\theta,j} < 0.8\sigma_{u,j}^2$ and $\mu_{\theta,j} > Q_3\sqrt{\log(dn)}$, we have

$$\mathbb{P}\left(\boldsymbol{\theta}_{(j)}^* \text{ is feasible for } (9)\right) \geq 1 - 2/(p-d)^{M+1} - \frac{\exp(1)}{nd^2} - 2\exp\left(-[Q_4 n\bar{\eta}_\gamma^2] \wedge [Q_5 n\bar{\eta}_\gamma]\right).$$

By the union bound, we have

$$\mathbb{P}\left(\bigcap_{j=1}^d \{\boldsymbol{\theta}_{(j)}^* \text{ is feasible for } (9)\}\right) \geq 1 - 2d/(p-d)^{M+1} - \frac{\exp(1)}{nd} - 2d\exp\left(-[Q_4 n\bar{\eta}_\gamma^2] \wedge [Q_5 n\bar{\eta}_\gamma]\right).$$

We finish the proof by noticing that $d \leq (p-d)^M$. $\square$

**Lemma 7.** *Let $\mathbf{X}$ and $\mathbf{Y}$ be two random vectors with cumulative distribution functions $F_X$ and $F_Y$, respectively. Then $\forall \varepsilon > 0$,*

$$\sup_{\alpha \in (0,1)} \left|\mathbb{P}\left(\|\mathbf{X}\|_\infty > F_Y^{-1}(1-\alpha)\right) - \alpha\right| \leq \varepsilon + \mathbb{P}\left(\sup_{x \in \mathbb{R}} |F_X(x) - F_Y(x)| > \varepsilon\right).$$

*Proof of Lemma 7.* Fix $\alpha \in (0,1)$ and notice that the following inequalities hold for all $\alpha \in (0,1)$. First, observe

$$\mathbb{P}\left(\|\mathbf{X}\|_\infty > F_Y^{-1}(1-\alpha)\right)$$
$$\leq \mathbb{P}\left(\|\mathbf{X}\|_\infty > F_Y^{-1}(1-\alpha) \text{ and } \sup_{x \in \mathbb{R}}|F_X(x) - F_Y(x)| \leq \varepsilon\right) + \mathbb{P}\left(\sup_{x \in \mathbb{R}}|F_X(x) - F_Y(x)| > \varepsilon\right)$$

$$
\begin{aligned}
&\overset{(i)}{\leq} \mathbb{P}\left(\|\mathbf{X}\|_\infty > F_X^{-1}(1-\alpha-\varepsilon)\right) + \mathbb{P}\left(\sup_{x\in\mathbb{R}} |F_X(x) - F_Y(x)| > \varepsilon\right) \\
&= \alpha + \varepsilon + \mathbb{P}\left(\sup_{x\in\mathbb{R}} |F_X(x) - F_Y(x)| > \varepsilon\right), \quad (85)
\end{aligned}
$$

where (i) follows from Lemma A.1(i) in Romano and Shaikh (2012) and the last line follows by the definition of $F_X(\cdot)$. Moreover,

$$
\begin{aligned}
&\mathbb{P}\left(\|\mathbf{X}\|_\infty > F_Y^{-1}(1-\alpha)\right) \\
&\geq \mathbb{P}\left(\|\mathbf{X}\|_\infty > F_Y^{-1}(1-\alpha) \text{ and } \sup_{x\in\mathbb{R}} |F_X(x) - F_Y(x)| \leq \varepsilon\right) \\
&\overset{(i)}{\geq} \mathbb{P}\left(\|\mathbf{X}\|_\infty > F_X^{-1}(1-\alpha+\varepsilon) \text{ and } \sup_{x\in\mathbb{R}} |F_X(x) - F_Y(x)| \leq \varepsilon\right) \\
&\overset{(ii)}{\geq} \mathbb{P}\left(\|\mathbf{X}\|_\infty > F_X^{-1}(1-\alpha+\varepsilon)\right) - \mathbb{P}\left(\sup_{x\in\mathbb{R}} |F_X(x) - F_Y(x)| > \varepsilon\right) \\
&= \alpha - \varepsilon - \mathbb{P}\left(\sup_{x\in\mathbb{R}} |F_X(x) - F_Y(x)| > \varepsilon\right) \quad (86)
\end{aligned}
$$

where (i) follows from Lemma A.1(ii) in Romano and Shaikh (2012), (ii) follows by the elementary result that holds for any two events $A$ and $B$,

$$\mathbb{P}(A\bigcap B) + \mathbb{P}(B^c) \geq \mathbb{P}(A\bigcap B) + \mathbb{P}(A\bigcap B^c) = \mathbb{P}(A),$$

and the last line follows by the definition of $F_X(\cdot)$. The desired result follows by (85) and (86). $\square$

**Lemma 8.** *Let $Z_1$ and $Z_2$ be random variables with bounded sub-Gaussian norms. Then $Z_1 Z_2$ has bounded sub-exponential norm.*

*Proof of Lemma 8.* Notice that, by the sub-Gaussian property, there exist constants $K_1, K_2 > 0$ such that $\forall z > 0$, $\mathbb{P}(|Z_1| > z) \leq K_1 \exp(-K_2 z^2)$ and $\mathbb{P}(|Z_2| > z) \leq K_1 \exp(-K_2 z^2)$. Hence, $\forall x > 0$,
$$\mathbb{P}(|Z_1 Z_2| > x) \leq \mathbb{P}(|Z_1| > \sqrt{x}) + \mathbb{P}(|Z_2| > \sqrt{x}) \leq 2K_1 \exp(-K_2 x).$$
Hence, $Z_1 Z_2$ is sub-exponential. $\square$

The following result can be viewed as a uniformly delta method for the function $f(x) = \sqrt{x}$.

**Lemma 9.** *Under the conditions of Theorem 3, there exists a constant $C_* > 0$ such that*

$$\mathbb{P}(\max_{1\leq j\leq d} \mathbb{E}(\widetilde{T}_{n,j}^2 \mid \mathbf{X}, \mathbf{Y}, \mathbf{Z}) > C_*) \to 0,$$

*where $\widetilde{T}_{n,j}$ is defined in (4).*

*Proof of Lemma 9.* Recall from (4) that

$$\widehat{H}_{i,j} = \widehat{\sigma}_\varepsilon^{-1} \widehat{\sigma}_{u,j}^{-1} (g_i - x_i^\top \widehat{\boldsymbol{\gamma}})(z_{i,j} - x_i^\top \widehat{\boldsymbol{\theta}}_{(j)})$$

and $\bar{H}_{n,j} = n^{-1}\sum_{i=1}^n \widehat{H}_{i,j}$. Then

$$\mathbb{E}(\widetilde{T}_{n,j}^2 \mid \mathbf{X}, \mathbf{Y}, \mathbf{Z}) = n^{-1}\sum_{i=1}^n (\widehat{H}_{i,j} - \bar{H}_{n,j})^2 = n^{-1}\sum_{i=1}^n \widehat{H}_{i,j}^2 - \bar{H}_{n,j}^2 \leq n^{-1}\sum_{i=1}^n \widehat{H}_{i,j}^2.$$

Let $H_{i,j} = \bar{\sigma}_\varepsilon^{-1}\sigma_{u,j}^{-1}(g_i - x_i^\top \boldsymbol{\gamma}^*)u_{i,j}$ with $\bar{\sigma}_\varepsilon^2 = \mathbb{E}(g_i - x_i^\top \boldsymbol{\gamma}^*)^2$. It suffices to show

(a) There exists a constant $K_* > 0$ such that $\forall C > K_*$, $\mathbb{P}(\max_{1\leq j \leq d} n^{-1}\sum_{i=1}^n H_{i,j}^2 > C) \to 0$.

(b) $\max_{1\leq j \leq d} n^{-1}\sum_{i=1}^n \widehat{H}_{i,j}^2 \leq o_P(1) + 4\max_{1\leq j \leq d} n^{-1}\sum_{i=1}^n H_{i,j}^2$.

We first show claim (a). Since $\boldsymbol{\beta}^* = \boldsymbol{\beta}_0 + \boldsymbol{\pi}$,

$$g_i - x_i^\top \boldsymbol{\gamma}_* = \boldsymbol{\pi}^\top z_i + \varepsilon_i = \boldsymbol{\pi}^\top ((\boldsymbol{\Theta}^*)^\top x_i + \mathbf{u}_i) + \varepsilon_i.$$

Since $\|\boldsymbol{\pi}\|_2$ and singular values of $\boldsymbol{\Theta}^*$ are bounded, the bounded sub-Gaussian norm of $(x_i, \mathbf{u}_i, \varepsilon_i)$ implies that there exist constants $K_1, K_2 > 0$ such that $K_1 < \bar{\sigma}_\varepsilon^2 < K_2$ and $g_i - x_i^\top \boldsymbol{\gamma}^*$ has a sub-Gaussian norm bounded by $K_2$. By the bounded sub-Gaussian norm of $u_{i,j}$ and Lemma 8, there exists a constant $K_3 > 0$ such that $\forall (i,j)$, $H_{i,j}$ has a sub-exponential norm bounded by $K_3$. Hence, there exist constants $K_4, K_5 > 0$ such that $\mathbb{P}(|\zeta_{i,j}| > z) \leq K_4 \exp(-K_5\sqrt{z})$ for any $z > 0$, where $\zeta_{i,j} = H_{i,j}^2 - \mathbb{E}H_{i,j}^2$. It follows by Theorem 1 of Merlevède et al. (2011) and the union bound that $\forall c > 0$

$$\mathbb{P}\left(\max_{1\leq j \leq d} n^{-1}\left|\sum_{i=1}^n \zeta_{i,j}\right| \geq c\right) \to 0.$$

Hence, $\max_{1\leq j \leq d} n^{-1}\sum_{i=1}^n (H_{i,j}^2 - \mathbb{E}H_{i,j}^2) = o_P(1)$. By the bounded sub-exponential norm of $H_{i,j}^2$, $\max_{1\leq j \leq d} \mathbb{E}H_{i,j}^2 = \mathcal{O}(1)$. Claim (a) follows.

It remains to verify claim (b). By essentially the same argument as for (30) and (39) in the proof of Theorem 2, we obtain

$$\max_{1\leq j \leq d} n^{-1}\sum_{i=1}^n (\widehat{H}_{i,j} - \widetilde{H}_{i,j})^2 = o_P(1) \quad \text{and} \quad \max_{1\leq j \leq d} n^{-1}\sum_{i=1}^n (\widetilde{H}_{i,j} - H_{i,j})^2 = o_P(1),$$

where $\widetilde{H}_{i,j} = \widehat{\sigma}_\varepsilon^{-1}\sigma_{u,j}^{-1}(g_i - x_i^\top \widehat{\boldsymbol{\gamma}})u_{i,j}$. Applying the elementary inequality $(a+b)^2 \leq 2a^2 + 2b^2$, we have

$$\widehat{H}_{i,j}^2 \leq 2(\widehat{H}_{i,j} - \widetilde{H}_{i,j})^2 + 2\widetilde{H}_{i,j}^2 \leq 2(\widehat{H}_{i,j} - \widetilde{H}_{i,j})^2 + 4(\widetilde{H}_{i,j} - H_{i,j})^2 + 4H_{i,j}^2.$$

This, combined with the display above, proves claim (b). $\square$

**Lemma 10.** *Under the conditions of Theorem 3, $\mathbb{P}\left(\|\widetilde{\mathbf{T}}_n\|_\infty > 3\sqrt{2C_* \log d}\right) \to 0$, where $C_* > 0$ is the constant defined in Lemma 9.*

*Proof of Lemma 10.* Notice that conditional on the data $(\mathbf{X}, \mathbf{Y}, \mathbf{Z})$, $\widetilde{\mathbf{T}}_n$ is a zero-mean Gaussian vector in $\mathbb{R}^d$. By Lemma 9, together with Proposition A.2.1 of van der Vaart and Wellner (1996), we have that $\forall x > 0$,

$$\mathbb{P}\left(\|\widetilde{\mathbf{T}}_n\|_\infty > \mathbb{E}\left(\|\widetilde{\mathbf{T}}_n\|_\infty \mid \mathbf{X}, \mathbf{Y}, \mathbf{Z}\right) + x \mid \mathbf{X}, \mathbf{Y}, \mathbf{Z}\right) \leq 2\exp\left(-\frac{x^2}{2C_*}\right) + \mathbb{I}\{A_n^c\} \ a.s, \qquad (87)$$

where the event
$$A_n = \{\max_{1\leq j\leq d} \mathbb{E}(\widetilde{T}_{n,j}^2 \mid \mathbf{X}, \mathbf{Y}, \mathbf{Z}) \leq C_*\}$$

and $C_* > 0$ is the constant defined in Lemma 9. Notice that for Gaussian variables $M_j \sim \mathcal{N}(0, \sigma_j^2)$ for $j = 1, \cdots, d$, an elementary inequality yields $\mathbb{E}\max_j M_j/\sigma_j \leq \sqrt{2\log d}$ and thus $\mathbb{E}\max_j M_j \leq \max_j \sigma_j \sqrt{2\log d}$. Therefore,

$$\mathbb{E}\left(\max_j \widetilde{T}_{n,j} \mid \mathbf{X}, \mathbf{Y}, \mathbf{Z}\right) \leq \sqrt{2\log d \max_{1\leq j\leq d} \mathbb{E}(\widetilde{T}_{n,j}^2 \mid \mathbf{X}, \mathbf{Y}, \mathbf{Z})} \text{ a.s.}$$

$$\mathbb{E}\left(\max_j (-\widetilde{T}_{n,j}) \mid \mathbf{X}, \mathbf{Y}, \mathbf{Z}\right) \leq \sqrt{2\log d \max_{1\leq j\leq d} \mathbb{E}(\widetilde{T}_{n,j}^2 \mid \mathbf{X}, \mathbf{Y}, \mathbf{Z})} \text{ a.s.}$$

and

$$\mathbb{E}\left(\|\widetilde{\mathbf{T}}_n\|_\infty \mid \mathbf{X}, \mathbf{Y}, \mathbf{Z}\right) \leq 2\sqrt{2\log d \max_{1\leq j\leq d} \mathbb{E}(\widetilde{T}_{n,j}^2 \mid \mathbf{X}, \mathbf{Y}, \mathbf{Z})} \text{ a.s.} \qquad (88)$$

It follows that

$$\begin{aligned}
\mathbb{P}\left(\|\widetilde{\mathbf{T}}_n\|_\infty > 3\sqrt{2C_*\log d} \mid \mathbf{X}, \mathbf{Y}, \mathbf{Z}\right) &\leq \mathbb{P}\left(\|\widetilde{\mathbf{T}}_n\|_\infty > \mathbb{E}\left(\|\widetilde{\mathbf{T}}_n\|_\infty \mid \mathbf{X}, \mathbf{Y}, \mathbf{Z}\right) + \sqrt{2C_*\log d} \mid \mathbf{X}, \mathbf{Y}, \mathbf{Z}\right) \\
&\quad + \mathbb{I}\left\{\mathbb{E}\left(\|\widetilde{\mathbf{T}}_n\|_\infty \mid \mathbf{X}, \mathbf{Y}, \mathbf{Z}\right) > 2\sqrt{2C_*\log d}\right\} \\
&\stackrel{(i)}{\leq} 2\exp\left(-\log d\right) + \mathbb{I}\{A_n^c\} \\
&\stackrel{(ii)}{=} o_P(1),
\end{aligned}$$

where $(i)$ follows by (87), (88) and the definition of $A_n$ and $(ii)$ follows by $d \to \infty$ and $\mathbb{P}(A_n) \to 1$. Since conditional probabilities are bounded and thus uniformly integrable, Theorem 5.4 on p.220 of Gut (2013) implies that $\mathbb{P}\left(\|\widetilde{\mathbf{T}}_n\|_\infty > 3\sqrt{2C_*\log d}\right) \to 0$. □

Bickel, P. J., Ritov, Y., and Tsybakov, A. B. (2009). Simultaneous analysis of lasso and dantzig selector. The Annals of Statistics, 37(4):1705–1732.

Bradley, R. C. (2007). Introduction to strong mixing conditions, volume 1. Kendrick Press Heber City.

Bugni, F. A., Caner, M., Bredahl Kock, A., and Lahiri, S. (2016). Inference in partially identified models with many moment inequalities using Lasso. ArXiv e-prints.

Bühlmann, P. and van de Geer, S. (2015). High-dimensional inference in misspecified linear models. Electron. J. Statist., 9(1):1449–1473.

Cai, T. T. and Guo, Z. (2015). Confidence intervals for high-dimensional linear regression: Minimax rates and adaptivity. arXiv preprint arXiv:1506.05539.

Cai, T. T. and Guo, Z. (2016). Accuracy Assessment for High-dimensional Linear Regression. arXiv preprint arXiv:1603.03474.

Cai, T. T., Ren, Z., and Zhou, H. H. (2016). Estimating structured high-dimensional covariance and precision matrices: Optimal rates and adaptive estimation. Electron. J. Statist., 10(1):1–59.

Candes, E. and Tao, T. (2007). The dantzig selector: Statistical estimation when p is much larger than n. Annals of Statistics, 35(6):2313–2351.

Carrasco, M. and Chen, X. (2002). Mixing and moment properties of various garch and stochastic volatility models. Econometric Theory, 18(1):17–39.

Chernozhukov, V., Chetverikov, D., and Kato, K. (2013a). Gaussian approximations and multiplier bootstrap for maxima of sums of high-dimensional random vectors. Ann. Statist., 41(6):2786–2819.

Chernozhukov, V., Chetverikov, D., and Kato, K. (2013b). Testing many moment inequalities. arXiv preprint arXiv:1312.7614.

Chernozhukov, V., Chetverikov, D., and Kato, K. (2014). Central limit theorems and bootstrap in high dimensions. arXiv preprint arXiv:1412.3661.

Chernozhukov, V., Hansen, C., and Liao, Y. (2015). A lava attack on the recovery of sums of dense and sparse signals. to appear in The Annals of Statistics.

Chernozhukov, V., Hansen, C., and Spindler, M. (2015). Valid post-selection and post-regularization inference: An elementary, general approach.

Dezeure, R., Bühlmann, P., and Zhang, C.-H. (2016). High-dimensional simultaneous inference with the bootstrap. ArXiv e-prints.

Fan, J., Liao, Y., and Yao, J. (2015). Power enhancement in high-dimensional cross-sectional tests. Econometrica, 83(4):1497–1541.
48